\documentclass[a4paper,11pt]{article}
\pdfoutput=1 
\usepackage[symbol]{footmisc} 

\usepackage{jheppub} 
 \usepackage{cancel}

\usepackage{feynmp}
\usepackage[compat=1.1.0]{tikz-feynhand}
\usepackage[compat=1.1.0]{tikz-feynman}
\usepackage{tikz}
\usetikzlibrary{external,shapes.geometric,arrows,arrows.meta}
\usepackage{color, colortbl}
\usepackage[T1]{fontenc}
\usepackage[utf8]{inputenc}
\usepackage{epsfig,latexsym}
\usepackage{amsmath}
\usepackage{adjustbox}
\usepackage{tensor}
\usepackage{graphicx}
\usepackage{subcaption}
\usepackage{verbatim}
\usepackage{mathrsfs}
\usepackage{amssymb}
\usepackage{multirow}
\usepackage{epsfig}
\usepackage{color,colordvi}
\usepackage{appendix}
\usepackage{slashed}
\usepackage{array}
\usepackage{cancel}
\usepackage{float}
\usepackage[font=footnotesize]{caption}
\usepackage{epsf}
\usepackage{amsmath}
\usepackage{physics}
\usepackage{MnSymbol}
\usepackage{rotating}
\usepackage{multirow}
\usepackage[normalem]{ulem}
\usepackage{tcolorbox}
\usepackage{bbold}
\usepackage{comment}
\usepackage{ulem}

\usepackage{tabularx}

\allowdisplaybreaks

\DeclareMathAlphabet{\mathpzc}{OT1}{pzc}{m}{it}

 \csname
@addtoreset\endcsname{equation}{section}

\newcolumntype{x}[1]{>{\centering\arraybackslash\hspace{0pt}}p{#1}}

\newcommand{\beq}{\begin{equation}}
\newcommand{\eeq}{\end{equation}}
\renewcommand{\[}{\left[}
\renewcommand{\]}{\right]}
\renewcommand{\(}{\left(}
\renewcommand{\)}{\right)}

\newcommand{\be}{\begin{eqnarray}}
\newcommand{\ee}{\end{eqnarray}}
\newcommand{\bea}{\begin{eqnarray}}
\newcommand{\eea}{\end{eqnarray}}
\newcommand{\bi}{\begin{itemize}}
\newcommand{\ei}{\end{itemize}}
\newcommand{\ben}{\begin{enumerate}}
\newcommand{\een}{\end{enumerate}}
\def\bes{\begin{equation*}}
\def\ees{\end{equation*}}
\def\bead{\begin{aligned}}
\def\eead{\end{aligned}}
\def\bmat{\left(\begin{matrix}}
\def\emat{\end{matrix}\right)}
\renewcommand{\Tr}{\text{Tr}}
\newcommand{\hc}{\text{h.c.}}
\newcommand{\thetaQCD}{\theta_{\text{QCD}}}
\newcommand{\wtilde}{\widetilde}

\def\Re{\text{Re}}
\def\Im{\text{Im}}

\def\cA{{\cal A}}
\def\cB{{\cal B}}

\def\cD{{\cal D}}

\def\cI{{\cal I}}

\def\cL{{\cal L}}

\def\cO{{\cal O}}

\def\cT{{\cal T}}

\def\s0{\slashed{0}}

\usepackage{multicol}
\usepackage{color}
\definecolor{verdes}{cmyk}{0.92,0,0.59,0.4}

\newcommand{\CPV}{$\Lambda_{\cancel{\rm CP}}$ }
\newcommand{\CPVm}{\Lambda_{\cancel{\rm CP}} }

\title{Small instanton-induced flavor invariants and \\ the axion potential}

\author[a]{Ravneet Bedi,}
\emailAdd{bedi0019@umn.edu}

\author[a]{Tony Gherghetta,}
\emailAdd{tgher@umn.edu}

\author[b,c,d]{Christophe Grojean,}
\emailAdd{christophe.grojean@desy.de}

\author[b]{Guilherme Guedes,}
\emailAdd{guilherme.guedes@desy.de}

\author[b,c]{Jonathan Kley,}
\emailAdd{jonathan.kley@desy.de}

\author[b]{Pham Ngoc Hoa Vuong}
\emailAdd{hoa.vuong@desy.de}

\affiliation[a]{School of Physics and Astronomy, University of Minnesota, Minneapolis, Minnesota 55455, USA}

\affiliation[b]{Deutsches Elektronen-Synchrotron DESY, Notkestr. 85, 22607 Hamburg, Germany}

\affiliation[c]{Institut für Physik, Humboldt-Universität zu Berlin, 12489 Berlin, Germany}

\affiliation[d]{Theoretical Physics Department, CERN, 1211 Geneva 23, Switzerland}

\abstract{Small instantons which increase the axion mass due to an appropriate modification of QCD at a UV scale $\Lambda_{\rm SI}$, can also enhance the effect of CP-violating operators to shift the axion potential minimum by an amount, $\theta_{\rm ind}$, proportional to the flavorful couplings in the SMEFT. Since physical observables must be flavor basis independent, we construct a basis of determinant-like flavor invariants that arise from instanton calculations containing the effects of dimension-six CP-odd operators at the scale $\CPVm$. This new basis provides a more reliable estimate of the shift $\theta_{\rm ind}$, that is severely constrained by neutron electric dipole moment experiments. In particular, for the case of four-quark, semi-leptonic and  gluon dipole operators, these invariants are then used to provide improved limits on the ratio of scales $\Lambda_{\rm SI}/\CPVm$ for different flavor scenarios. The CP-odd flavor invariants also provide a classification of the leading effects from Wilson coefficients, and as an example, we show that a semi-leptonic four-fermion operator is subdominant compared to the four-quark operators. More generally, the flavor invariants, together with an instanton NDA, can be used to more accurately estimate small instanton effects in the axion potential that arise from any SMEFT operator.
}

\begin{document} 
\begin{flushright}
CERN-TH-2024-022\\
DESY-24-019 \\
HU-EP-24/05 \\
UMN–TH–4313/24
\end{flushright}
\maketitle
\flushbottom
\newpage
\section{Introduction}\label{sec:Intro}

The Standard Model (SM) has provided a remarkably successful description of the elementary particles and non-gravitational interactions. However, one unresolved issue is CP violation which remains to be completely understood. 
In particular, there are two sources of CP violation in the Standard Model at the renormalizable level: a weak CP phase from the CKM matrix and the strong CP phase $\bar\theta=\theta_{\rm QCD}-{\rm arg}\,{\rm det}[Y_{\rm u} Y_{\rm d}]$, where $\theta_{\rm QCD}$ is the QCD vacuum angle and $Y_{\rm u,d}$ are the up, down Yukawa coupling matrices. Limits from not observing a neutron electric dipole moment (EDM) imply an upper bound $\bar\theta \lesssim 10^{-10}$~\cite{Abel:2020pzs}. This unexpectedly small value for $\bar\theta$ compares with the order-one weak CP phase and leads to the well-known strong CP problem. The strong CP problem occurs at the renormalizable level in the SM Lagrangian and remains rather stable under renormalization group flow where radiative corrections to $\bar{\theta}$, induced by the weak CP phase, first appear at seven 
loops~\cite{ELLIS1979141,Khriplovich:1993pf} (finite contributions appear at four loops although still highly suppressed~\cite{Khriplovich:1985jr}). 
However, the problem can be exacerbated by phases arising from higher-dimensional terms in the UV theory, which can cause order-one shifts in $\bar\theta$ thereby potentially invalidating solutions to the strong CP problem~\cite{Bigi:1990kz}. 

Such a solution to the strong CP problem is the Peccei--Quinn (PQ) mechanism~\cite{Peccei:1977hh} where a global $U(1)$ PQ symmetry is spontaneously broken giving rise to the axion~\cite{Weinberg:1977ma,Wilczek:1977pj}. The PQ symmetry is anomalous because it is explicitly broken by non-perturbative QCD effects, generating the axion potential with a minimum that exactly cancels $\bar\theta$. However, an underlying assumption of the PQ mechanism is that the explicit breaking of the global PQ symmetry must be dominated by non-perturbative QCD effects. Generically, shift-symmetry violating terms (which may be CP-violating) arising from interactions of the axion with gravity spoil the axion solution by misaligning the axion potential minimum. These contributions must therefore be sufficiently suppressed, leading to the so-called axion quality problem, which can be addressed by one of several mechanisms in the literature (see for example Refs.~\cite{Witten:1984dg,Randall:1992ut,Choi:2003wr,Lillard:2018fdt,Gavela:2018paw,Hook:2019qoh,Contino:2021ayn,Cox:2023dou}).
 
However, even under the assumption that shift-violating operators in the axion EFT are sufficiently suppressed, another aspect of the axion quality problem occurs within the Standard Model Effective Field Theory (SMEFT) where higher-dimensional CP-violating terms induced at a scale \CPV can shift $\bar{\theta}$ and misalign the axion potential~\cite{Dine:1986bg,Dine:2022mjw,Bedi:2022qrd}. 
These new sources of CP violation depend on the specific UV completion. If QCD is not modified in the UV completion then the effects on the axion potential and the neutron EDM constrain \CPV and these effects decouple as \CPV$\rightarrow \infty$.
Alternatively, there has been renewed interest in the old idea to modify QCD at a UV scale, $\Lambda_{\rm SI}$ that can increase the axion mass while still solving the strong CP problem~\cite{Dimopoulos:1979pp,Holdom:1982ex,Holdom:1985vx,Dine:1986bg,Flynn:1987rs,Rubakov:1997vp, Fukuda:2015ana,Gherghetta:2016fhp,Agrawal:2017ksf,Csaki:2019vte,Gherghetta:2020ofz}. In this case, new CP-violating sources can also be enhanced by small (UV) instantons whose effects are no longer suppressed due to the assumed larger QCD coupling at the scale $\Lambda_{\rm SI}$. The leading contributions to $\bar\theta$ then scale as $\Lambda^2_{\rm SI}/\CPVm^2$~\cite{Bedi:2022qrd} which do not necessarily decouple (i.e. when $\Lambda_{\rm SI},\CPVm\to \infty$ with a finite ratio $\Lambda_{\rm SI}/\CPVm$) and give rise to important constraints on CP violation in certain UV scenarios. For a sufficiently small (although large) QCD gauge coupling
these effects can be computed by performing a one-instanton calculation which provides the dominant contribution to the action. However, when the QCD gauge coupling becomes non-perturbative, the dilute instanton gas approximation breaks down and non-perturbative methods must be used.

The effects of CP-violation arising from higher-dimension operators in an instanton background, including a four-quark SMEFT operator were computed in Ref.~\cite{Kitano:2021fdl, Demirtas:2021gsq, Bedi:2022qrd} and estimated using an instanton naive dimensional analysis (NDA) in Ref.~\cite{Csaki:2023ziz}. However, different CP-violating UV scenarios can give rise to many other operators~\cite{deBlas:2017xtg,Guedes:2023azv} and therefore previous estimates of the contributions to $\bar{\theta}$ should be generalized for the complete list of SMEFT operators.
These new contributions to $\bar{\theta}$ must be independent of the flavor basis, and hence can be written in terms of flavor-invariant quantities constructed from the Wilson coefficients. Flavor-invariant quantities allow for an estimation of the physical consequences of the Wilson coefficients -- especially when used together with other NDA techniques~\cite{Csaki:2023ziz} -- prior to any explicit computation.
Indeed, knowledge of CP-violating invariants has previously been used for physical 
estimates~\cite{ELLIS1979141,Khriplovich:1993pf,Farrar:1993hn,Smith:2017dtz,Denton:2019yiw,Yu:2022ttm}.
This is particularly relevant in instanton calculations where the computation can become quite cumbersome. 
Besides providing an order parameter to estimate CP-violating physical effects, CP-odd invariants also provide selection rules on the contribution from a particular Wilson coefficient, such as the number of Yukawa coupling insertions or loop factor suppressions. For example, we will show that semi-leptonic operators generate a $\overline{\theta}$ but are suppressed by one extra loop-order and an extra lepton Yukawa factor compared with four-quark operators. Furthermore, knowing the number of up, down and lepton Yukawa couplings needed to construct the invariants, we will be able to classify the leading contributions for arbitrary Wilson coefficients. The use of flavor invariants also helps to explain the size of the instanton effects in different flavor scenarios. Moreover, most computations in the literature have focused on the limit of completely diagonal Yukawa matrices but, with the help of flavor invariants, the extension to general flavorful matrices becomes simpler.

Due to the CP-nature of $\bar{\theta}$, the induced $\bar{\theta}$ in the presence of different SMEFT operators can be parameterized via the CP-violating invariants introduced in Refs.~\cite{Bonnefoy:2021tbt,Bonnefoy:2023bzx}, where a basis of CP-odd flavor invariants was proposed at leading order in the corresponding Wilson coefficient. The invariants introduced were all constructed as the imaginary part of a trace of flavorful matrices. However, as will be made clear throughout this work, we find that the instanton calculations can be more directly captured by a different basis, built out of determinant-like structures, which were first pointed out in Appendix F of Ref.~\cite{Bonnefoy:2021tbt}. 
Indeed, we find that the instanton computations of the enhanced $\bar{\theta}$ directly result in quantities proportional to invariants in our new basis, instead of a complicated combination of CP-even and -odd invariants if the results were projected onto the trace-like basis. We will show that these determinant-like structures naturally arise in the path integral calculation of $\bar{\theta}$, after performing the Grassmann integration over fermion zero modes. While any CP-odd quantity can clearly be projected into both bases, the invariants built in this paper allow for an immediate estimation of the CP-violating effects in the SMEFT, in the presence of instanton backgrounds.

While the new flavor invariants introduced in this paper improve the estimate of $\bar{\theta}$ induced in the SMEFT, we will also perform the detailed computation of $\bar{\theta}$ in the presence of small instantons, using the one-instanton (or dilute instanton gas) approximation, to explicitly show how the new invariants appear and more naturally describe physical processes that arise from instanton computations. In particular, we generalize previous results on the insertion of the CP-violating four-quark operator $\cO^{(1)}_{\rm quqd}$ and calculate the effects of the semi-leptonic, $\cO^{(1)}_{\rm lequ}$ and gluon dipole, $\cO_{\rm dG}$ operators. These results can then be used to translate the stringent upper bound on $\bar{\theta}$ into limits on the scale of new physics generating these operators. For the leading order operators, such as the four-quark and gluon dipole operators, we find that $\CPVm\gtrsim 10^{6}\,\Lambda_{\rm SI}$, assuming a minimally flavor violating (MFV) scenario for the SMEFT couplings. Under these same flavor assumptions, the loop suppressed contributions, arising from $\cO^{(1)}_{\rm lequ}$, lead to the weaker constraint $\CPVm\gtrsim 10^ {4}\,\Lambda_{\rm SI}$. The bounds become more stringent if there is no flavor structure in the Wilson coefficients,  such as the anarchic flavor scenario. In this case, assuming all the Wilson coefficients are order one, we obtain $\CPVm\gtrsim 10^{11}\,\Lambda_{\rm SI}$ for the four-quark operator $\cO^{(1)}_{\rm quqd}$, $\CPVm\gtrsim 10^{8}\,\Lambda_{\rm SI}$ for the gluon dipole operator $\cO_{\rm dG}$ and $\CPVm\gtrsim 10^{7}\,\Lambda_{\rm SI}$ for the semi-leptonic operator $\cO^{(1)}_{\rm lequ}$.
As a non-trivial check, we also verify the expected independence of the renormalization scheme as well as the cancellation of divergences in the renormalized effective theory that arise from loop integrals. To arrive at this cancellation, one has to include the appropriate counterterms which lead to the renormalization group equations (RGEs) of the SMEFT.

The paper is organized as follows. In Section~\ref{sec:Invariants}, we introduce the determinant-like invariants, showing that they form a basis of CP-violating invariants, and then build such a basis for certain SMEFT operators. The instanton calculations are presented in Section~\ref{sec:Instantons} where we show explicitly that the determinant-like invariants appear directly from the topological susceptibility computation; furthermore, we also explore why the determinant structures appear in this sort of calculation and the information it contains prior to any actual computation. In Section~\ref{sec:Pheno}, we compute the enhanced $\bar{\theta}$ and obtain new bounds on the ratio between the small-instanton scale $\Lambda_{\rm SI}$ and the CP-violating scale \CPV. We conclude and suggest future directions for this work in Section~\ref{sec:Conclusions}. The Appendices review background information and contain further details of the calculation. In Appendix~\ref{app:SMEFT}, we introduce our conventions and the relevant SMEFT operators. A complete basis of determinant-like invariants is given in Appendix~\ref{app:invs}, while in Appendix~\ref{app:basics} we briefly review the basics of instanton calculations that are relevant for this paper. Finally, in Appendix~\ref{sec:SolveInt} we perform the integrals over collective coordinates used throughout the main text; in particular, we verify the cancellation of divergences with the appropriate counterterms, which is a new result obtained from an instanton calculation.

\section{Flavor invariants featuring \texorpdfstring{$\thetaQCD$}{thetaQCD}}\label{sec:Invariants}
Perturbative CP violation in the SM induced by the CKM phase is an intricate collective effect that can only be properly described by a combination of Lagrangian parameters. The most effective way to capture this effect in the SM is by using the
Jarlskog invariant $J_4$, defined as~\cite{Jarlskog:1985ht,Jarlskog:1985cw,Bernabeu:1986fc}
\begin{equation}
    J_4 = \Im\(\Tr\[X_{\rm u},X_{\rm d}\]^3\) \,,
    \label{eq:Jarlinv}
\end{equation}
where $X_{\rm u,d}^{\vphantom{\dagger}} \equiv Y^{\vphantom{\dagger}}_{\rm u,d}Y_{\rm u,d}^\dagger$. 
The quantity $J_4$ captures, in a flavor basis-invariant way (c.f. Table~\ref{tab:FlavorTrafo}), the single physical phase that appears in the renormalizable part of the SM Lagrangian and hence is the order parameter of CP violation in the SM. Here, we have used the fact that $U(3)^5\equiv U(3)_{\rm Q}\times U(3)_{\rm u} \times U(3)_{\rm d} \times U(3)_{\rm L} \times U(3)_{\rm e}$ is the largest possible flavor group allowed by the SM fermion kinetic terms and is only broken by the SM Yukawa couplings and global anomalies. The Lagrangian can be formally made invariant under this symmetry by promoting the Yukawa couplings to spurions transforming under $U(3)^5$ as given in the Table~\ref{tab:FlavorTrafo}.

\begin{table}[t]
        \centering
        \begin{tabular}{c|c|c|c|c|c}
             & $U(3)_{\rm Q}$ & $U(3)_{\rm u}$ & $U(3)_{\rm d}$ & $U(3)_{\rm L}$ & $U(3)_{\rm e}$ \\ \hline
            $e^{i \theta_{\text{QCD}}}$ & $\mathbf{1}_{+6}$ & $\mathbf{1}_{-3}$ & $\mathbf{1}_{-3}$ & $\mathbf{1}_0$ & $\mathbf{1}_0$ \\[0.1cm]
            $Y_{\rm u}$ & $\mathbf{3}_{+1}$ & $\mathbf{\bar{3}}_{-1}$ & $\mathbf{1}_0$ & $\mathbf{1}_0$ & $\mathbf{1}_0$ \\[0.1cm]
            $Y_{\rm d}$ & $\mathbf{3}_{+1}$ & $\mathbf{1}_0$ & $\mathbf{\bar{3}}_{-1}$ & $\mathbf{1}_0$ & $\mathbf{1}_0$ \\[0.1cm]
            $Y_{\rm e}$ & $\mathbf{1}_0$ & $\mathbf{1}_0$ & $\mathbf{1}_0$ & $\mathbf{3}_{+1}$ & $\mathbf{\bar{3}}_{-1}$ \\[0.1cm]

        \end{tabular}
        \caption{Flavor transformation properties of $\theta_{\rm QCD}$ and the Yukawa coupling matrices $Y_{\rm u,d,e}$. The subscripts of the $SU(3)$ representations denote the charge under the $U(1)$ part of the flavor symmetry.}
        \label{tab:FlavorTrafo}
\end{table}

Following the same logic, Refs.~\cite{Bonnefoy:2021tbt,Bonnefoy:2023bzx} constructed the flavor invariants which capture new CP-violating phases present at lowest order in the SMEFT parametrization of UV physics. In Appendix \ref{app:SMEFT}, we briefly introduce the SMEFT and the relevant operators which will be considered throughout our work.
The construction of these invariants specifically adopted the philosophy of the Jarlskog invariant in Eq.~\eqref{eq:Jarlinv} in the sense that they are built out of traces of flavorful matrices. Similarly, we can construct an invariant for the QCD theta angle, $ \theta_{\rm QCD} $. Using the charges introduced in Table~\ref{tab:FlavorTrafo}, the flavor invariant given by\footnote{ \label{foot:Ktheta}For instance, the correction to the axion mass in the SM is proportional to $K_{\theta} = \Re[e^{-i\thetaQCD}\det(Y_{\rm u} Y_{\rm d})] \propto \cos\bar\theta$, leading to the well-known cosine potential~\cite{Marsh:2015xka,DiLuzio:2020wdo}, whereas the linear term of the axion potential generated via non-perturbative QCD effects in the SM is proportional to $J_\theta \propto \sin\bar\theta$. }
\begin{equation}
    J_\theta = \Im [e^{-i\thetaQCD}\det(Y_{\rm u} Y_{\rm d})]\,,
\end{equation}
captures the \emph{non}-perturbative source of CP violation in the SM Lagrangian, specifically $\bar\theta=\theta_{\rm QCD}-{\rm arg}\,{\rm det}[Y_{\rm u} Y_{\rm d}]$. 
Indeed, contributions to the $\theta$ potential in the presence of an instanton background gives the following dependence~\cite{tHooft:1976snw,Flynn:1987rs}:
\begin{equation}
     V\(\thetaQCD,Y_{\rm u},Y_{\rm d}\) \propto e^{-i \thetaQCD} \prod_{i=1}^3 \hat{y}_{{\rm u},i} \, \hat{y}_{{\rm d},i} \,,
\end{equation}
where $\hat{y}_{{\rm u},i}, \, \hat{y}_{{\rm d},i}$ are the Yukawa matrix eigenvalues and $i$ labels the quark flavors. Later, the eigenvalues will sometimes be referred to by their particle name, i.e. for instance $\hat{y}_{{\rm u},3} = y_{\rm t}$. This result does not appear to be flavor invariant. However, the result can be reproduced from $J_\theta$ or $K_\theta = \Re[e^{-i\thetaQCD}\det(Y_{\rm u} Y_{\rm d})]$ (depending on the CP parity of the contribution to $V$) that is calculated by expanding the invariants in the limit of diagonal SM Yukawa matrices. This suggests that the flavor invariants can appear directly in the instanton calculations provided general flavorful couplings are used in the computation.

In this paper, our goal is to consider the contribution of SMEFT operators to the topological susceptibility in the presence of (small) instantons. We begin by considering the effect of the four-fermion operator $\mathcal{O}_{\rm quqd}^{(1)} = \bar{Q}u\bar{Q}d$ in the Lagrangian, ${\cal L}\supset C_{{\rm quqd}}^{(1)} \cO_{\rm quqd}^{(1)}/\CPVm^2$. The effect of inserting this effective operator in an instanton background to the topological susceptibility, $\chi_{\rm quqd}$, or equivalently the potential, was calculated in Ref.~\cite{Bedi:2022qrd} (see figure \ref{fig:Cquqd}). In the limit of diagonal SM Yukawa couplings,\footnote{The CKM matrix is assumed to be unity here, which is possible in the SM below the $W$-boson mass, since all effects of the CKM matrix can be put into effective operators.} we obtain a contribution proportional to
\begin{equation}
\label{eq:quqdref}
     V\(\thetaQCD,Y_{\rm u},Y_{\rm d}, C_{{\rm quqd}}^{(1)}\) \propto e^{-i \thetaQCD}  
     \frac{c_{ij} }{\hat{y}_{{\rm u},i} \hat{y}_{{\rm d},j}}\prod_{k=1}^{3} \hat{y}_{{\rm u},k} \hat{y}_{{\rm d},k}\,,
\end{equation}
where $c_{ij}$ captures the contribution from the two possible flavor structures $c_{ij}= C_{{\rm quqd},iijj}^{(1)}$ or $c_{ij}= C_{{\rm quqd},ijji}^{(1)}$ and $k$ labels the six entries of the diagonal Yukawa matrices. The proportionality factor in Eq.~\eqref{eq:quqdref} depends on the details of the instanton calculation and these factors will be derived in Section~\ref{sec:Instantons}.

It is not immediately apparent that Eq.~\eqref{eq:quqdref} is related in any way to the trace-like invariants introduced in Ref.~\cite{Bonnefoy:2021tbt}. As mentioned in the Introduction, determinant-like invariants are much better suited to describe instanton calculations. Respecting the charge assignments introduced in 
Table~\ref{tab:FlavorTrafo}, we can build the following simplest leading order (in the EFT power counting) invariant
\begin{equation} \label{eq:quqdinv}
   \mathcal{I}(C_{\rm quqd}^{(1,8)})= \Im \[e^{-i \thetaQCD} \epsilon^{ABC}\epsilon^{abc}\epsilon^{DEF}\epsilon^{def} Y_{{\rm u},Aa}^{\phantom{(}}Y_{{\rm u},Bb}^{\phantom{(}} C_{{\rm quqd},CcDd}^{(1,8)}Y_{{\rm d},Ee}^{\phantom{(}}Y_{{\rm d},Ff}^{\phantom{(}} \]\,,
\end{equation}
which contains the Wilson coefficient $C_{{\rm quqd},ijkl}^{(1,8)}$ and we sum over repeated indices.
Note that this exact invariant had already been proposed in Appendix F of Ref.~\cite{Bonnefoy:2021tbt}. In the limit of diagonal Yukawa couplings and vanishing $\thetaQCD$ (assumed in  Ref.~\cite{Bedi:2022qrd}), Eq.~\eqref{eq:quqdinv} can be expanded as
\begin{equation}\label{eq:cquqd_Inv_diag}
    \mathcal{I}(C_{\rm quqd}^{(1,8)}) =   4 \left(\prod_{k=1}^{3} \hat{y}_{{\rm u},k}\hat{y}_{{\rm d},k}\right) \sum_{i,j=1}^{3} \frac{\mathrm{Im}[C_{\rm quqd}^{(1,8)}]_{iijj}}{\hat{y}_{{\rm u},i} \hat{y}_{{\rm d},j}} \,,
\end{equation}
which matches the result obtained in Eq.~\eqref{eq:quqdref} and explains the flavor structure appearing in the instanton contribution. Similarly, the $C_{{\rm quqd},ijji}^{(1,8)}$ flavor structure arises if we consider a second invariant where the indices $C$ and $D$ are interchanged.

At this point several questions arise regarding these determinant-like invariants: why do they seem to more naturally appear in instanton calculations? How can their knowledge help in these computations? Can we construct similar determinant-like invariants for all effective operators and how do they appear in instanton calculations? Furthermore, one should also connect with the previously constructed trace-like basis of invariants: can a complete basis be built out of the determinant-like invariants and how do they relate with the previously constructed basis? While the former questions will be addressed in detail in Section~\ref{sec:Instantons}, the remainder of Section~\ref{sec:Invariants} will answer the latter questions regarding the construction of a determinant-like basis suitable for instanton calculations.

\subsection{A basis of determinant-like flavor invariants} \label{sec:basis}

In this section, we will discuss how one can, in principle, construct a complete basis of flavor invariants for all SMEFT operators that are suitable for instanton calculations featuring $\thetaQCD$. By complete, we mean that the basis captures all CP-violating effects from UV phases in the Wilson coefficients of SMEFT operators.
Hence, we do not include \textit{opportunistic} effects where the interference of the real part of a Wilson coefficient with the CKM phase induces CP violation, as previously considered in Ref.~\cite{Bonnefoy:2023bzx}. We also want to emphasize that we only work at leading order in the EFT, i.e., all flavor invariants will be linear in the Wilson coefficients. Higher-order terms are negligible and we have estimated their effects in Appendix~\ref{app:Four-fermion}.

To test our set for completeness we will use the transfer matrix method introduced in Ref.~\cite{Bonnefoy:2021tbt}. The linearity of the invariants in the Wilson coefficients implies that there is a linear map between the flavor invariants and the entries of the Wilson coefficients $C^{(6)}$
\begin{equation}
    \cI_a^{\vphantom{(}}\left(C^{(6)}\right) = \cT_{ai}^{\vphantom{(}} \, \vec{C}_i^{(6)} \,,
\end{equation}
where we have defined the transfer matrix $\cT$ and the vector of Wilson coefficients 
$\vec{C}_i^{(6)} = \(\(\Re\, C^{(6)}\)_1,\(\Re\, C^{(6)}\)_2,\dots,\(\Im\, C^{(6)}\)_1,\(\Im\, C^{(6)}\)_2,\dots\)$ 
which is a list of the real and imaginary parts of the entries of the Wilson coefficients. The index $a$ ranges from 1 to the total number of independent flavor invariants built out of $C^{(6)}$ and the index $i$ ranges from 1 to the total number of parameters in the theory which cannot be removed by field redefinitions and that can appear in an interference amplitude with the SM.
The transfer matrix has a block-diagonal form $\cT = \( \cT^R \ \cT^I \)$ where the block $\cT^R$ will be ignored because it captures the interference of the real part of the Wilson coefficients with the phases of the SM. To check if the set captures all sufficient and necessary conditions of CP violation at the leading order in the EFT, we simply check if the block $\cT^I$ of the transfer matrix has full rank, i.e., if the rank equals the number of phases which can interfere with the SM (see Ref.~\cite{Bonnefoy:2021tbt}, where the maximal ranks are given for all operators in the Warsaw basis). Note also that, in this paper, we do not consider Yukawa matrices with any special values, e.g. degenerate masses, zero masses or texture zeros in the CKM matrix, that enlarge the flavor symmetry of the SM left unbroken by the Yukawa couplings.

\paragraph{$\mathbf{\cO_{\rm uH}}$ operator:} As a first example for building a complete basis with determinant-like invariants, we consider the higher-dimensional Yukawa interactions of the up-type quarks, $\cO_{\rm uH} = |H|^2 \bar{Q} \wtilde{H} u$. This requires constructing an object that simultaneously removes the $U(1)$ transformations of $e^{-i\thetaQCD}$ (which appears in instanton calculations) and is invariant under the remaining non-Abelian part of the flavor symmetry, while at the same time being linear in the Wilson coefficients. Following the previous discussion, the simplest flavor invariant object that fulfills all these requirements is
\begin{equation}
    \Im\[e^{-i \thetaQCD} 
\epsilon^{IJK}\epsilon^{ijk} Y_{{\rm u},Ii} Y_{{\rm u},Jj} C_{{\rm uH},Kk} \det Y_{\rm d}\] \,,
\label{eq:flavorinv}
\end{equation}
where the rephasings of the Yukawa couplings and the Wilson coefficients precisely cancel those of $e^{-i\thetaQCD}$ and the determinant-like structure of the Levi-Civita symbols allows the construction of $SU(3)$-invariant structures. 
Starting from the form in Eq.~\eqref{eq:flavorinv}, we can now systematically construct flavor invariants that can capture all phases in the Wilson coefficient $C_{\rm uH}$ by using the matrices $X_{\rm u,d}^{\vphantom{\dagger}} = Y^{\vphantom{\dagger}}_{\rm u,d} Y_{\rm u,d}^\dagger$, transforming in the adjoint of $SU(3)_{\rm Q}$, to project out different entries of the Wilson coefficients. 

With the help of the transfer matrix method, one can check that a set of flavor invariants which captures all the sources of CP violation for the operator $\cO_{\rm uH}$, for $J_4 = J_\theta = 0$, is
\begin{equation}
\begin{split}
    & \cI_{0000}(C_{\rm uH}), \, \cI_{1000}(C_{\rm uH}), \, \cI_{0100}(C_{\rm uH}), \, \cI_{1100}(C_{\rm uH}), \, \cI_{0110}(C_{\rm uH}), \\
    & \qquad\quad \cI_{2200}(C_{\rm uH}), \, \cI_{0220}(C_{\rm uH}), \, \cI_{1220}(C_{\rm uH}), \, \cI_{0122}(C_{\rm uH}) \,,
\end{split} \label{eq:CuH_list_of_inv}
\end{equation}
where we have defined 
\begin{equation} \label{eq:CuHinv}
    \cI_{abcd}(C_{\rm uH}) \equiv \Im \[e^{-i \thetaQCD} 
\epsilon^{IJK}\epsilon^{ijk} Y_{{\rm u},Ii}^{\phantom{d}} Y_{{\rm u},Jj}^{\phantom{d}} \left(X_{\rm u}^a X_{\rm d}^b X_{\rm u}^c X_{\rm d}^d C_{\rm uH} \right)_{Kk} \det Y_{\rm d} \] \,.
\end{equation}

\paragraph{$\mathbf{\cO_{\rm quqd}^{(1,8)}}$ operators:} After discussing this simple complete example for an operator that only contains 9~CPV phases, we next return to the four-fermion operator $\cO_{\rm quqd}^{(1)}$, and its $SU(3)$ adjoint form $\cO_{\rm quqd}^{(8)}$, that appeared in the previous section. A complete invariant basis can also be built for this operator by defining the following two structures
\begin{equation} \label{eq:ABquqd}
    \begin{split}
        \cA_{a_2,b_2,c_2,d_2}^{a_1,b_1,c_1,d_1}(C_{\rm quqd}^{(1,8)}) & = \Im \[ e^{-i\thetaQCD} \epsilon^{ABC} \epsilon^{abc} \epsilon^{DEF} \epsilon^{def}Y^{\phantom{\dagger}}_{{\rm u},Aa} Y^{\phantom{\dagger}}_{{\rm u},Bb} \( X_{\rm u}^{a_1} X_{\rm d}^{b_1} X_{\rm u}^{c_1} X_{\rm d}^{d_1} \)_{C}^{\hphantom{C}C^\prime} \right. \\
        & \qquad\quad \left. \times \,C_{{\rm quqd},C^\prime c D^\prime d}^{(1,8)} \( X_{\rm u}^{a_2} X_{\rm d}^{b_2} X_{\rm u}^{c_2} X_{\rm d}^{d_2} \)_{D}^{\hphantom{D}D^\prime} Y^{\phantom{\dagger}}_{{\rm d},Ee} Y^{\phantom{\dagger}}_{{\rm d},Ff} \] \,, \\
        \cB_{a_2,b_2,c_2,d_2}^{a_1,b_1,c_1,d_1}(C_{\rm quqd}^{(1,8)}) & = \Im \[ e^{-i\thetaQCD} \epsilon^{ABC} \epsilon^{abc} \epsilon^{DEF} \epsilon^{def} Y^{\phantom{\dagger}}_{{\rm u},Aa} Y^{\phantom{\dagger}}_{{\rm u},Bb} \( X_{\rm u}^{a_1} X_{\rm d}^{b_1} X_{\rm u}^{c_1} X_{\rm d}^{d_1} \)_{D}^{\hphantom{D}C^\prime} \right. \\
        & \qquad\quad \left. \times \,C_{{\rm quqd},C^\prime c D^\prime d}^{(1,8)} \( X_{\rm u}^{a_2} X_{\rm d}^{b_2} X_{\rm u}^{c_2} X_{\rm d}^{d_2} \)_{C}^{\hphantom{C}D^\prime} Y^{\phantom{\dagger}}_{{\rm d},Ee} Y^{\phantom{\dagger}}_{{\rm d},Ff} \] \,. 
    \end{split}
\end{equation}
Here, the index assignment $\cA_{0000}^{0000}(C_{\rm quqd}^{(1,8)})$ corresponds to the invariant in Eq.~\eqref{eq:quqdinv} and $\cB_{0000}^{0000}(C_{\rm quqd}^{(1,8)})$ corresponds to the second invariant mentioned in the last section, where the indices C and D are interchanged. The operator $\cO_{\rm quqd}^{(1,8)}$ has 81 phases that can interfere with the dimension-4 terms of the SM. We
list a full set of 81 invariants that capture all these phases in non-perturbative calculations in Appendix~\ref{app:MoreInvs}. 

\paragraph{$\mathbf{\cO_{\rm lequ}^{(1,3)}}$ operators:} One can also build determinant-like invariants for (semi-)leptonic operators. For instance, the invariants capturing the 27 CP-odd phases of the Wilson coefficients $C_{\rm lequ}^{(1,3)}$ of the semi-leptonic operator $\cO_{\rm lequ}^{(1)} = \left(\bar{L} e\right) \left(\bar{Q} u\right)$  and its $SU(2)$ adjoint form $\cO_{\rm lequ}^{(3)}$ are 
\begin{equation} \label{eq:Ilequ}
  \hspace{-2mm}  \cI_{abcd}^f(C_{\rm lequ}^{(1,3)}) \equiv \Im \[e^{-i \thetaQCD} \epsilon^{IJK}\epsilon^{ijk} Y^{\phantom{\dagger}}_{{\rm u},Ii} Y^{\phantom{\dagger}}_{{\rm u},Jj}
\(X_{\rm u}^a X_{\rm d}^b X_{\rm u}^c X_{\rm d}^d\)_{K}^{\hphantom{K}L} \(Y_{\rm e}^\dagger X_{\rm e}^f\)^{mN} C_{{\rm lequ},NmLk}^{(1,3)} \det Y^{\phantom{\dagger}}_{\rm d} \].
\end{equation}
Here again the index assignments for the insertion of $X_{\rm u,d,e}$ are the same as those in the trace invariants of Ref.~\cite{Bonnefoy:2021tbt} for the same operators, which are also mentioned in 
Section~\ref{app:MoreInvs}.

\paragraph{$\mathbf{\cO_{\rm Hq}^{(1,3)}}$ operators:} As a last example, one can also build invariants for SMEFT operators that are not charged under the $U(1)$ rephasings of the flavor group. For instance, for the phases in the Wilson coefficients $C_{\rm Hq}^{(1,3)}$ of the operator $\cO_{\rm Hq}^{(1)} = \left( H^\dagger i \overleftrightarrow{D}_{\hspace{-2pt}\mu} H \right) \left( \bar{Q} \gamma^{\mu} Q \right)$ and its $SU(2)$ adjoint form $\cO_{\rm Hq}^{(3)}$, one can write down the following invariants
\begin{equation}\label{eq:CHqinv}
    \cI_{1100}(C_{\rm Hq}^{(1,3)}), \, \cI_{2200}(C_{\rm Hq}^{(1,3)}), \, \cI_{1122}(C_{\rm Hq}^{(1,3)}) \,,
\end{equation}
where 
\begin{equation}
\cI_{abcd}(C_{\rm Hq}^{(1,3)}) \equiv \Im \[e^{-i\thetaQCD} \epsilon^{IJK} \epsilon^{ijk} Y_{{\rm u},Ii}^{\phantom{d}} Y_{{\rm u},Jj}^{\phantom{d}}
\left(X_{\rm u}^a X_{\rm d}^b X_{\rm u}^c X_{\rm d}^d\, C_{\rm Hq}^{(1,3)}\, Y^{\phantom{\dagger}}_{\rm u} \right)_{Kk} \det Y^{\phantom{\dagger}}_{\rm d} \].
\end{equation}
Following this procedure, a complete set of flavor invariants capturing all CP-violating effects at leading order in the SMEFT expansion can be built for all operators in the Warsaw basis~\cite{Grzadkowski:2010es}. We present a few more complete examples in Appendix~\ref{app:MoreInvs}.

Let us once again emphasize that these new invariants are redundant with respect to the trace-like invariants introduced in Ref.~\cite{Bonnefoy:2021tbt}, as those were already a complete basis of invariants which fully characterize the CP-violating phases of the theory. Therefore, the determinant-like invariants \emph{must} be redundant in regards to the trace-like invariants. For example, the invariants in Eq.~\eqref{eq:CHqinv} can be rewritten as
\begin{align} 
   \cI_{abcd}(C_{\rm Hq}^{(1,3)}) =& \Im\left[\left(e^{-i\thetaQCD} \det \left(Y_{\rm u} Y_{\rm d}\right) \right)\epsilon^{IJK} \epsilon_{IJL} 
\left(X_{\rm u}^a X_{\rm d}^b X_{\rm u}^c X_{\rm d}^d C_{\rm Hq}^{(1,3)}\right)_{K}^{\hphantom{K}L} \right] \nonumber\\
=& 2\,\left( J_{\theta}\,R_{abcd}(C_{\rm Hq}^{(1,3)})+ K_{\theta}\,L_{abcd}(C_{\rm Hq}^{(1,3)})\right)~, \label{eq:relCHq}
\end{align}
where $R_{abcd}(C)=\Re\,\left[\Tr\left(X_{\rm u}^a X_{\rm d}^b X_{\rm u}^c X_{\rm d}^d C \right)\right]$ and $L_{abcd}(C)=\Im\,\left[\Tr\left(X_{\rm u}^a X_{\rm d}^b X_{\rm u}^c X_{\rm d}^d C \right)\right]$, as defined in Ref.~\cite{Bonnefoy:2021tbt}.
There are also similar relations for other operators such as
\begin{equation} 
\label{eq:CuHRelToTrace}
     \cI_{abcd}(C_{\rm uH})=2\,\left( J_{\theta}\,R_{(a-1)bcd}(C_{\rm uH}Y_{\rm u}^\dagger) + K_{\theta}\,L_{(a-1)bcd}(C_{\rm uH}Y_{\rm u}^\dagger)\right)~,
\end{equation}
\begin{equation} 
\label{eq:LequRelToTrace}
       \cI_{abcd}^f(C_{\rm lequ}^{(1,3)}) =2 \left( J_{\theta}\, \Im\, A^f_{(a-1)bcd}(C_{\rm lequ}^{(1,3)}) + K_{\theta}\,\Re\, A^f_{(a-1)bcd}(C_{\rm lequ}^{(1,3)} )\right)~,
\end{equation}
where $ A^f_{abcd}(C_{\rm lequ}^{(1,3)} )=
X_{{\rm e}, {ji}}^f\(X_{\rm u}^a X_{\rm d}^b X_{\rm u}^c X_{\rm d}^d\)_{lk} Y_{{\rm e}, {{mj} }}^\dagger Y^{\vphantom{\dagger}}_{{\rm u}, {nl}}
  C_{{\rm lequ}, imkn}^{(1,3)}.$
This procedure allows us to map all determinant-like invariants directly to the trace invariants of Ref.~\cite{Bonnefoy:2021tbt} for all operators up to the invariants of the form $\cI_{0bcd}(C_{\rm uH})$, where inverse Yukawa couplings appear in the trace invariants (c.f. $\cI_{0000}(C_{\rm uH})$ in Eq.~\eqref{eq:CuHRelToTrace}). We will show in Appendix~\ref{app:RelInvYuk} how these latter invariants can also be mapped to the old basis, or alternatively a different basis from the one given in Ref.~\cite{Bonnefoy:2021tbt} would be required, {including invariants with inverse Yukawa couplings}. It also becomes apparent that $ \cI_{abcd}$ captures both the CP-violation due to the phases introduced by SMEFT operators and that due to the interference between these SMEFT operators and the SM strong CP phase.

In the next section we will explore why (and when) the determinant-like invariants are better suited to describe CP violation.

\section{The interplay of Topological susceptibilities and Flavor invariants}
\label{sec:Instantons}

As mentioned earlier, the determinant-like invariants capture the interference between the strong CP phase $\bar{\theta}$ and the Wilson coefficients of various SMEFT operators. In particular, we focus on CP-violating SMEFT operators
\begin{align}
    \cL \supset \dfrac{C_{_{\mathcal{O}}}^{ij\cdots}}{\CPVm^{D-4}}\mathcal{O}^{D,\,ij\cdots}~,
\end{align}
where $D$ is the mass dimension of the EFT operator $\cal O$ and $i,j,\dots$ are the flavor indices. 
Since both the parameters $\bar{\theta}$ and $C_{_{\mathcal{O}}}$ are odd under $CP$, the vacuum energy is given by~\cite{Bigi:1991rh}
\begin{align}
    V(\bar{\theta}, C_{_{\mathcal{O}}})=\frac{1}{2}\(u_{\theta}\,\bar{\theta}\,^2+2u_{\theta\cO}\,\bar{\theta}\, C_{_\cO}+u_{\cO}\, C_{_\cO}^2\)~,\label{eq:theta_potential}
\end{align}
where $u_\theta,u_{\theta\,\cO},u_{\cO}$ can be computed in terms of SM and SMEFT operators.
We can arrive at Eq.~\eqref{eq:theta_potential} by computing the terms quadratic in the $\theta$ term and the SMEFT operator $\cal O$ in the effective action, which implies
\begin{align}
    u_{\theta}\sim \left\langle \(G\tilde{G}\)^2\right\rangle,\qquad u_{\theta\cO}\sim \frac{1}{\CPVm^{D-4}}
    \left\langle G\tilde{G}\,\cO\right\rangle,\qquad u_{\cO}\sim \frac{1}{\CPVm^{2D-8}}
    \left\langle  \cO^2\right\rangle~.
\end{align}
As such, the $ u_{\cO}$ term can be neglected compared to the other two terms. Note that Eq.~\eqref{eq:theta_potential} introduces a linear term in $\bar{\theta}$, implying that the minimum of the energy is shifted. In the PQ mechanism, i.e., when
$\bar{\theta}$ is promoted to $ a/f_a$, a dynamical degree of freedom, the effective value of $\bar{\theta}\equiv\langle a/f_a\rangle$ is determined by Eq.~\eqref{eq:theta_potential}. To be concrete, the potential can be re-written in terms of the axion field $a$
\begin{align}
    V(a) = \chi_{_{\mathcal{O}}}(0) \dfrac{a}{f_a} + \dfrac{1}{2}\chi (0) \left( \dfrac{a}{f_a} \right)^2 ~,
    \label{eq;axionpot}
\end{align}
where we have introduced $\chi (0)$ and $\chi_{_{\mathcal{O}}}(0)$ to replace $u_{\theta}$ and $u_{\theta\,\cO}$, respectively, which can be defined as~\cite{Witten:1979vv,Shifman:1979if,Pospelov:1997uv,Pospelov:2005pr}
\begin{align}
\chi (0) &= -i\lim_{k\rightarrow 0}\int d^4x \, e^{ikx} 
    \left\langle 0\left| T \left\{ \dfrac{g^2}{32\pi^2} G \tilde{G}(x) \,,\, \dfrac{g^2}{32\pi^2} G \tilde{G}(0) \right\} 
    \right| 0 \right\rangle
    \,,
    \label{Definition: QCD-Topo-suscep}
\end{align}
known as the QCD topological susceptibility and 
\begin{align}
    \chi_{_{\mathcal{O}}}(0) &= -i\lim_{k\rightarrow 0}\int d^4x \, e^{ikx} 
    \left\langle 0\left| T \left\{ \dfrac{g^2}{32\pi^2} G^a_{\,\mu\nu} \tilde{G}_a^{\,\mu\nu}(x) \,,\, \dfrac{C_{_{\mathcal{O}}}^{ij\cdots}}{\CPVm^{D-4}}\mathcal{O}^{D,\,ij\cdots}(0) \right\} \right| 0 \right\rangle
    \,.
    \label{eq:suscepdef}
    \end{align} 
The shift in the axion potential of Eq.~\eqref{eq;axionpot} is then given by
\begin{align}
    \theta_{\text{ind}} \equiv -\dfrac{
    \chi_{_{\mathcal{O}}}(0)
    }{\chi (0)}
    \,.
    \label{eq:thetaind}
\end{align}
Experimental bounds on the neutron EDM lead to the constraint $\theta_{\text{ind}}\lesssim 10^{-10}$, which can then be used to obtain limits on any UV parameters contained in $\chi_{_{\mathcal{O}}}(0)$.

Usually, models of axions or axion-like particles (ALPs) are constructed with a $U(1)$ Peccei--Quinn symmetry in mind, which dictates the ALP couplings to the SM particles -- either directly or in an EFT after integrating out the heavy modes from the theory. If one allows for some explicit breaking of the $U(1)$ symmetry,\footnote{Even if this explicit breaking is not introduced by hand it will be generated by quantum gravity effects~\cite{Georgi:1981pu,Holman:1992us,Kamionkowski:1992mf, Barr:1992qq, Ghigna:1992iv}.} responsible for the Nambu--Goldstone boson nature of the ALP, an axion potential can be generated in ordinary perturbation theory. The interactions of the ALP with the SM particles, including those breaking the shift symmetry, can be captured in an EFT in a relatively model-independent way. In this case, the axion potential can be determined by calculating the Coleman--Weinberg potential in the ALP EFT including operators that break the shift symmetry of the ALP explicitly. The tadpole term of the resulting potential should be proportional to the invariants presented in Ref.~\cite{Bonnefoy:2022rik} that capture all sources of shift symmetry breaking in the effective theory (see also Ref.~\cite{Bonnefoy:2022vop}).

In this work, we focus on the dimension-six SMEFT operators in the Warsaw basis~\cite{Grzadkowski:2010es} -- see also Appendix~\ref{app:SMEFT} for definitions and conventions. 
We will show in Section~\ref{sec:WhyDetInvs} how the determinant-like CP-odd invariants introduced in Section~\ref{sec:Invariants} arise in the vacuum-to-vacuum amplitude in  Eq.~\eqref{eq:suscepdef} in the presence of the one-(anti)instanton background. To illustrate how the invariants for different types of SMEFT operators appear in the instanton calculations, we will focus on the operators $\mathcal{O}_{\rm quqd}^{(1)},\, \mathcal{O}^{(1)}_{\rm lequ},\, \mathcal{O}_{\rm dG}$. The full instanton computation to estimate the contribution of the SMEFT operators to $\chi_{_{\mathcal{O}}}(0)$ requires the substitution of the zero mode profile of fermions and gauge fields, which are reviewed in Appendix~\ref{sec: Instanton: review}, as well as the evaluation of loop and collective coordinates integrals. These final steps will be carried out in detail in Appendix~\ref{sec:SolveInt}, where the calculated expressions will then be used in 
Section~\ref{sec:Pheno} to obtain phenomenological bounds on the CP-odd invariants.

\subsection{Topological Susceptibilities}
In quantum field theory, much of the essential information (e.g. the S-matrix elements, power spectrum) can be accessed by evaluating correlation functions of operators. In this section, we schematically evaluate the two-point correlation functions which are relevant for our calculations in the presence of a one-(anti)-instanton background by using the path integral formalism together with the result derived by 't~Hooft~\cite{tHooft:1976snw} (see Eq.~\eqref{Definition: 'tHooft-result} in Appendix~\ref{app:basics}.).

As an example to illustrate how to evaluate the correlator defined in Eq.~\eqref{eq:suscepdef}, we consider a generic dimension-six operator $\mathcal{O}[\varphi_{\rm I}, \varphi]$, where $\varphi_{\rm I}$ are fields with instanton solutions (e.g. gluon and quark fields), and $\varphi$ denotes the other fields unrelated to instanton dynamics (e.g. Higgs or lepton fields). The susceptibility associated with $\cal O$ is given by
\begin{align}
    \chi_{_{\mathcal{O}}}(0) &= -i\lim_{k\rightarrow 0}\int d^4x \, e^{ikx} 
    \left\langle 0\left| T \left\{ \dfrac{g^2}{32\pi^2} G \tilde{G}(x) \,,\, \dfrac{C_{_{\mathcal{O}}}}{\CPVm^2}  \mathcal{O}[\varphi_{\rm I}, \varphi] (0) \right\} 
    \right| 0 \right\rangle\,,
    \nonumber \\
    &= e^{-i\thetaQCD} \int d^4x_0 \int \dfrac{d\rho}{\rho^5} d_N(\rho) \int \prod_{f=1}^{N_f}\big( \rho \, d\xi^{(0)}_f d\bar{\xi}^{(0)}_f \big)
	\nonumber \\    
    &\times \int \mathcal{D}\varphi \, e^{-S_0[\varphi] -S_{\rm int}[\varphi_{\rm I}, \varphi]} 
\int d^4x\dfrac{g^2}{32\pi^2}G\tilde{G}(x) \times \dfrac{C_{_{\mathcal{O}}}}{\CPVm^2} \mathcal{O}[ \varphi_{\rm I}, \varphi] (0) \bigg|_{\text{1-(a.-)inst.}}
	\,,
	\label{Mixed-correlator: example}
\end{align}
where $S_0[\varphi]$, $S_{\rm int}[\varphi_{\rm I},\varphi]$ describe the action containing the $\varphi$ kinetic terms and the interactions between $\varphi_{\rm I}$ and $\varphi$, respectively. The key points to evaluate Eq.~\eqref{Mixed-correlator: example} depend on the different treatments of the fields $\varphi_{\rm I},~ \varphi$ and their corresponding path integrals. The essential steps are summarized as follows:
\begin{itemize}
    \item \textit{Fields with instanton solutions $\varphi_{\rm I}$.} The field $\varphi_{\rm I}$ is expanded in eigenmodes, where only the zero modes of $\varphi_{\rm I}$ are replaced by the instanton solutions. In particular, the fermion fields are expanded as
    \begin{align}
        \psi_f(x) = \sum_k \xi_f^{(k)}\psi^{(k)}
        \,;\quad
        \bar{\psi}_f(x) = \sum_k \bar{\xi}_f^{(k)}\bar{\psi}^{(k)}
        \,,
    \end{align}
    where $\xi_f^{(k)}$, $\bar{\xi}_f^{(k)}$ are Grassmann variables, $f$ is a fermion flavor index and the explicit form of $\psi^{(0)}$ is given in Eq.~\eqref{Inst-fermion: zero-modes}. Importantly, the non-zero modes of $\varphi_{\rm I}$ are integrated out and the path integral over the zero modes is interpreted as an integration over collective coordinates (see Appendix~\ref{app:basics} for further details). Thus, we can directly replace the path integral of $\varphi_{\rm I}$ using 't~Hooft's result~\cite{tHooft:1976snw}
    \begin{align}
    \int\mathcal{D}\varphi_{\rm I} \, e^{-S_E[\varphi_{\rm I}]} \rightarrow e^{-i\thetaQCD} \int d^4x_0 \int \dfrac{d\rho}{\rho^5} d_N(\rho) \int \prod_{f=1}^{N_f}\big( \rho \, d\xi_f^{(0)} d\bar{\xi}_f^{(0)} \big)
    	\,,
     \label{tHooftsresult}
    \end{align}
    where $d\xi^{(0)},\, d\bar{\xi}^{(0)}$ are Grassmann integration measures associated with the fermion zero modes and $d_N(\rho)$ is the instanton density in the $SU(N)$ theory (see Eq.~\eqref{Definition: inst-density}) with $\rho$ denoting the instanton size.
    
    \item \textit{Fields without instanton solutions $\varphi$.} The remaining fields, $\varphi$, 
    are integrated over without performing the eigenmode expansion. This procedure can be diagrammatically seen as closing the external legs $\varphi_{\rm I}$ (e.g. quark fields) which are coupled to the instanton vertex by using the fields $\varphi$ (e.g. Higgs fields), e.g. see Fig.~\ref{fig:Cquqd}. The crucial step is to expand the interaction terms of $e^{-S_{\rm int}[\varphi_{\rm I},\varphi]}$, obtaining the contributions of the $\varphi_{\rm I}$ zero modes.

    \item The remaining steps require substituting the zero mode profiles of $\varphi_{\rm I}$ (given by Eqs.~\eqref{Definition: BPST instantons},
    ~\eqref{Inst-fermion: zero-modes}) and evaluating the remaining loop integrals induced by $\varphi$ and collective coordinate integrals. Most of these calculations are carried out in Appendix~\ref{sec:SolveInt}. Finally, the integral over instanton size, $\rho$, is performed in Section~\ref{sec:Pheno}, where some UV scenarios responsible for the instanton dynamics are specified.
\end{itemize}
Another relevant correlation function is the QCD topological susceptibility defined in Eq.~\eqref{Definition: QCD-Topo-suscep}. This two-point correlation function 
has been computed in the literature, assuming $\chi(0)$ only receives contributions from the Standard Model, and within the perturbative regime and one-instanton approximation is given by~\cite{Callan:1977gz,Flynn:1987rs}
\begin{align} 
   \chi(0) = -3! \(2 K_{\theta}\) \, i \int \dfrac{d\rho}{\rho^5} d_N(\rho)  \dfrac{1}{(6\pi^2)^3}
   \,,
   \label{eq:QCD_susc_ferm}
\end{align}
where $K_\theta = \Re\[e^{-i\thetaQCD} \det\(Y_{\rm u} Y_{\rm d}\)\]$ as introduced in the footnote~\ref{foot:Ktheta}.

\subsection{Relevance of determinant-like flavor invariants}
\label{sec:WhyDetInvs}

The appearance of determinant-like structures in instanton computations, which are more directly parameterized by the invariants introduced in Section~\ref{sec:basis}, is related to the technicalities introduced in the previous section which are explored in further detail here.

The main point relates to the treatment of the fermionic contributions in the instanton background. The fermion fields are expanded in their eigenmodes, such that one can then isolate the zero modes as in Eq.~\eqref{Inst-fermion: eigenmodes}. The Grassmann integration measures $d\xi^{(0)}_f$ will effectively project out the zero modes of the fermions, since $\int d\xi^{(0)}_f \xi^{(0)}_f = 1$. In the full path integral calculation, the following Grassmann integral relations are useful
\begin{equation} \label{eq:grassmanid}
    \begin{split}
    \int d^3\xi_1 d^3\xi_2 ~ e^{\xi_1 A \xi_2} &= \det A \,,\\
    \int d^3\xi_1 d^3\xi_2 ~ e^{\xi_1 A \xi_2} \xi_1 B \xi_2 &= \frac{1}{2} \epsilon^{i_1i_2i_3} \epsilon^{j_1j_2j_3} A_{i_1j_1}A_{i_2j_2} B_{i_3j_3} \,,\\
    \int d^3\xi_1 d^3\xi_2 d^3\xi_3 d^3\xi_4 ~ e^{\xi_1 A \xi_2 + \xi_3 B \xi_4} \xi_1 C \xi_2 &= \frac{1}{2} \epsilon^{i_1i_2i_3} \epsilon^{j_1j_2j_3} A_{i_1j_1}A_{i_2j_2} C_{i_3j_3} \det B \,, \\
    \int d^3\xi_1 d^3\xi_2 d^3\xi_3 d^3\xi_4 ~ e^{\xi_1 A \xi_2 + \xi_3 B \xi_4} \xi_1 C \xi_2 \, \xi_3 D \xi_4 &= \frac{1}{4} \epsilon^{i_1i_2i_3} \epsilon^{j_1j_2j_3} A_{i_1j_1}A_{i_2j_2} C_{i_3j_3} \\ 
    &~~~\times \epsilon^{k_1k_2k_3}\epsilon^{l_1l_2l_3} B_{k_1l_1}B_{k_2l_2} D_{k_3l_3} \,,
    \end{split}
\end{equation}
where $\xi_{1,\dots,4}$ are three-dimensional Grassmann variables and $A,B,C,D$ are $3\times3$ matrices.
These identities are at the origin of the appearance of flavorful objects contracted with Levi-Civita symbols in the calculation, which we describe as determinant-like.

As such, in computations where the fermion zero modes in an effective operator are integrated over, the determinant-like invariants introduced in Section~\ref{sec:basis} are better suited at describing CP-violation. This happens not only because the final result is more easily connected to them, but also because they capture the full dependency of the final result in terms of the flavorful couplings of the theory. In other words, the result obtained will be proportional to a determinant-like invariant times instanton-related quantities; all the dependence on the rest of the theory (in this case Yukawa couplings) is captured by the invariant. This would not be the case if we were considering the trace-like basis of invariants. While the results could be projected into this basis, this would occur as  complicated combinations of the invariants and with coefficients which include other SM flavor invariants as we explicitly show in Appendix~\ref{app:MoreInvs}.

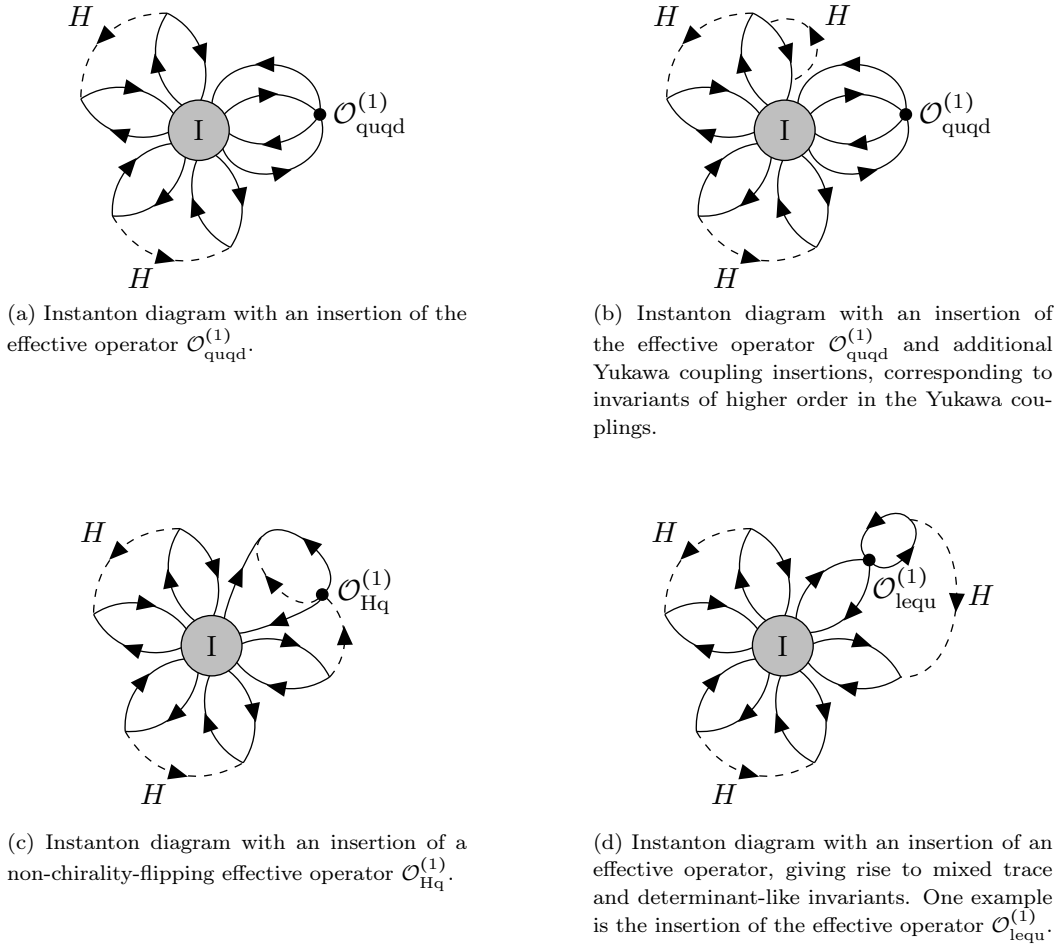
\begin{figure}[ht]
    \centering
    \begin{minipage}{\textwidth}\centering
    \begin{subfigure}[t]{0.4\textwidth}
    	\centering
    	\begin{tikzpicture}[scale=0.450]
    \begin{feynhand}
    \setlength{\feynhandblobsize}{8mm}
 \vertex  (h1) at (-1.2941/1.4,4.82963/1.4);  \vertex  (h2) at (-4.82963/1.4,1.2941/1.4); \vertex  (h3) at (-3.53553/1.4,-3.53553/1.4); \vertex  (h4) at (1.2941/1.4,-4.82963/1.4);  \vertex  [dot] (f4) at (4.95722/1.4,0.652631/1.4) {};
    \vertex (v1) at (1.29904/2,.75/2);
     \vertex (v2) at (.75/2,1.29904/2);
       \vertex (v3) at (0,1.5/2);
        \vertex (v4) at (-0.75/2,1.29904/2);
        \vertex (v5) at (-1.29904/2,0.75/2);
     \vertex (v6) at (-1.5/2,0);
     \vertex (v7) at (-1.29904/2,-0.75/2);
      \vertex (v8) at (-0.75/2,-1.29904/2);
       \vertex (v9) at (0,-1.5/2);
        \vertex (v10) at (.75/2,-1.29904/2);
        \vertex (v11) at (1.29904/2,-.75/2);
          \vertex (v12) at (1.5/2,0);  
    \propag[fer] (h1) to [quarter left](v3);
     \propag[fer] (v4) to [quarter left](h1);
      \propag[fer] (h2) to [quarter left ](v5);
     \propag[fer] (v6) to [quarter left](h2);
      \propag[fer] (h3) to [quarter left ](v7);
     \propag[fer] (v8) to [quarter left](h3);
      \propag[fer] (h4) to [quarter left ](v9);
     \propag[fer] (v10) to [quarter left](h4);
      \propag[fer] (v1) to [quarter left](f4);
     \propag[fer] (v11) to [half right, looseness=1.4](f4); 
     \propag[fer] (f4) to [half right, looseness=1.4](v2);
    
     \propag[fer] (f4) to  [quarter left](v12);
     \propag[chasca](h1) to [quarter right, looseness=1.1, edge label'=$H$](h2);
      \propag[chasca](h3) to  [quarter right, looseness=1.1, edge label'=$H$](h4);
       \vertex[grayblob] (tv) at (0,0) {};
       \node at (0,0) {{I}};
       \node at (6/1.2,0.65/1.4) {$\mathcal{O}_{\rm quqd}^{(1)}$}; 
    \end{feynhand}
    \end{tikzpicture}\vspace{-5pt}
    	\caption{Instanton diagram with an insertion of the effective operator $\mathcal{O}_{\rm quqd}^{(1)}$.}
    	\label{fig:Cquqd}
    \end{subfigure} \hspace{40pt}
    \begin{subfigure}[t]{0.4\textwidth}
    	\centering
    	\begin{tikzpicture}[scale=0.450]
    \begin{feynhand}
    \setlength{\feynhandblobsize}{8mm}
 \vertex  (h11) at (-0.7 /1.4,4.4 /1.4); 
 \vertex  (h12) at (0.3,2.1 /1.4); 
 
 \vertex  (h1) at (-1.2941/1.4,4.82963/1.4);  \vertex  (h2) at (-4.82963/1.4,1.2941/1.4); \vertex  (h3) at (-3.53553/1.4,-3.53553/1.4); \vertex  (h4) at (1.2941/1.4,-4.82963/1.4);  \vertex  [dot] (f4) at (4.95722/1.4,0.652631/1.4) {};
    \vertex (v1) at (1.29904/2,.75/2);
     \vertex (v2) at (.75/2,1.29904/2);
       \vertex (v3) at (0,1.5/2);
        \vertex (v4) at (-0.75/2,1.29904/2);
        \vertex (v5) at (-1.29904/2,0.75/2);
     \vertex (v6) at (-1.5/2,0);
     \vertex (v7) at (-1.29904/2,-0.75/2);
      \vertex (v8) at (-0.75/2,-1.29904/2);
       \vertex (v9) at (0,-1.5/2);
        \vertex (v10) at (.75/2,-1.29904/2);
        \vertex (v11) at (1.29904/2,-.75/2);
          \vertex (v12) at (1.5/2,0);  
    \propag[fer] (h1) to [quarter left](v3);
     \propag[fer] (v4) to [quarter left](h1);
      \propag[fer] (h2) to [quarter left ](v5);
     \propag[fer] (v6) to [quarter left](h2);
      \propag[fer] (h3) to [quarter left ](v7);
     \propag[fer] (v8) to [quarter left](h3);
      \propag[fer] (h4) to [quarter left ](v9);
     \propag[fer] (v10) to [quarter left](h4);
      \propag[fer] (v1) to [quarter left](f4);
     \propag[fer] (v11) to [half right, looseness=1.4](f4); 
     \propag[fer] (f4) to [half right, looseness=1.4](v2);
    
     \propag[fer] (f4) to  [quarter left](v12);
     
      \propag[chasca](h12) to [half right, looseness=2, edge label'=$H$](h11);
      
     \propag[chasca](h1) to [quarter right, looseness=1.1, edge label'=$H$](h2);
      \propag[chasca](h3) to  [quarter right, looseness=1.1, edge label'=$H$](h4);
       \vertex[grayblob] (tv) at (0,0) {};
       \node at (0,0) {{I}};
       \node at (6/1.2,0.65/1.4) {$\mathcal{O}_{\rm quqd}^{(1)}$}; 
    \end{feynhand}
    \end{tikzpicture}\vspace{-5pt}
    	\caption{Instanton diagram with an insertion of the effective operator $\mathcal{O}_{\rm quqd}^{(1)}$ and additional Yukawa coupling insertions, corresponding to invariants of higher order in the Yukawa couplings.}
    	\label{fig:CquqdHiOrd}
    \end{subfigure}
    \end{minipage} \\ \vspace{20pt}
    
    \begin{minipage}{\textwidth}\centering
    \begin{subfigure}[t]{0.4\textwidth}
    	\centering
    	\begin{tikzpicture}[scale=0.450]
     \begin{feynhand}
    \setlength{\feynhandblobsize}{8mm}
 \vertex  (h1) at (-1.2941/1.4,4.82963/1.4); 
 \vertex  (h2) at (-4.82963/1.4,1.2941/1.4); 
 \vertex  (h3) at (-3.53553/1.4,-3.53553/1.4); 
 \vertex  (h4) at (1.2941/1.4,-4.82963/1.4);  
  
 \vertex  (h5) at (4.82963/1.4,-1.2941/1.4); 
 \vertex  (h6) at (2.11/1.4,4.53/1.4); 
 
 \vertex  [dot] (hq) at (4.53/1.4,2.11/1.4) {}; 

 \vertex   (hqaux) at (8.69/1.4,2.32/1.4) ;
    \vertex (v1) at (1.29904/2,.75/2);
     \vertex (v2) at (.75/2,1.29904/2);
       \vertex (v3) at (0,1.5/2);
        \vertex (v4) at (-0.75/2,1.29904/2);
        \vertex (v5) at (-1.29904/2,0.75/2);
     \vertex (v6) at (-1.5/2,0);
     \vertex (v7) at (-1.29904/2,-0.75/2);
      \vertex (v8) at (-0.75/2,-1.29904/2);
       \vertex (v9) at (0,-1.5/2);
        \vertex (v10) at (.75/2,-1.29904/2);
        \vertex (v11) at (1.29904/2,-.75/2);
          \vertex (v12) at (1.5/2,0);  
    \propag[fer] (h1) to [quarter left](v3);
     \propag[fer] (v4) to [quarter left](h1);
      \propag[fer] (h2) to [quarter left ](v5);
     \propag[fer] (v6) to [quarter left](h2);
      \propag[fer] (h3) to [quarter left ](v7);
     \propag[fer] (v8) to [quarter left](h3);
      \propag[fer] (h4) to [quarter left ](v9);
     \propag[fer] (v10) to [quarter left](h4);

      \propag[fer] (hq) to [quarter left, looseness=.3 ](v1);
     \propag[fer] (v2) to [quarter left, looseness=.3](h6);
      \propag[fer] (h5) to [quarter left ](v11);
     \propag[fer] (v12) to [quarter left](h5);
     
     \propag[chasca](h1) to [quarter right, looseness=1.1, edge label'=$H$](h2);
      \propag[chasca](h3) to  [quarter right, looseness=1.1, edge label'=$H$](h4);

     \propag[chasca](h5) to [quarter right ](hq);
       \propag[chasca](hq) to  [half left  ](h6);
       \propag[fer](hq) to  [half right  ](h6);  
       \vertex[grayblob] (tv) at (0,0) {};
       \node at (0,0) {{I}};
       \node at (4.5,1.6) {$\mathcal{O}_{\rm Hq}^{(1)}$}; 
      
    \end{feynhand}
    \end{tikzpicture}
    	\caption{Instanton diagram with an insertion of a non-chirality-flipping effective operator $\mathcal{O}_{\rm Hq}^{(1)}$.}
    	\label{fig:CHq}
    \end{subfigure} \hspace{40pt}
    \begin{subfigure}[t]{0.4\textwidth}
    	\centering
    	\begin{tikzpicture}[scale=0.450]
    \begin{feynhand}
    \setlength{\feynhandblobsize}{8mm}
 
 \vertex  (h1) at (-1.2941/1.4,4.82963/1.4); 
 \vertex  (h2) at (-4.82963/1.4,1.2941/1.4); 
 \vertex  (h3) at (-3.53553/1.4,-3.53553/1.4); 
 \vertex  (h4) at (1.2941/1.4,-4.82963/1.4);  
  
 \vertex  (h5) at (4.82963/1.4,-1.2941/1.4); 
 \vertex  (h6) at (5.2/1.4,5.2/1.4); 
 
 \vertex  [dot] (leq) at (3.53553/1.4,3.53553/1.4) {};
  \node at (5/1.4,2.3/1.4) {$\mathcal{O}_{\rm lequ}^{(1)}$}; 
 \vertex   (hqaux) at (8.69/1.4,2.32/1.4) ;
    \vertex (v1) at (1.29904/2,.75/2);
     \vertex (v2) at (.75/2,1.29904/2);
       \vertex (v3) at (0,1.5/2);
        \vertex (v4) at (-0.75/2,1.29904/2);
        \vertex (v5) at (-1.29904/2,0.75/2);
     \vertex (v6) at (-1.5/2,0);
     \vertex (v7) at (-1.29904/2,-0.75/2);
      \vertex (v8) at (-0.75/2,-1.29904/2);
       \vertex (v9) at (0,-1.5/2);
        \vertex (v10) at (.75/2,-1.29904/2);
        \vertex (v11) at (1.29904/2,-.75/2);
          \vertex (v12) at (1.5/2,0);  
    \propag[fer] (h1) to [quarter left](v3);
     \propag[fer] (v4) to [quarter left](h1);
      \propag[fer] (h2) to [quarter left ](v5);
     \propag[fer] (v6) to [quarter left](h2);
      \propag[fer] (h3) to [quarter left ](v7);
     \propag[fer] (v8) to [quarter left](h3);
      \propag[fer] (h4) to [quarter left ](v9);
     \propag[fer] (v10) to [quarter left](h4);

      \propag[fer] (leq) to [quarter left ](v1);
     \propag[fer] (v2) to [quarter left](leq);
      \propag[fer] (h5) to [quarter left ](v11);
     \propag[fer] (v12) to [quarter left](h5);

     \propag[chasca](h6) to [half left, looseness=1.1](h5);
     \node at (5.8,1.5) {$H$};
     \propag[chasca](h1) to [quarter right, looseness=1.1, edge label'=$H$](h2);
      \propag[chasca](h3) to  [quarter right, looseness=1.1, edge label'=$H$](h4);
      
       \propag[fer](leq) to  [half right, looseness=1.2 ](h6);  \propag[fer](h6) to  [half right, looseness=1.2 ](leq);
       
       \vertex[grayblob] (tv) at (0,0) {};
       \node at (0,0) {{I}};
    \end{feynhand}
    \end{tikzpicture}
    	\caption{Instanton diagram with an insertion of an effective operator, giving rise to mixed trace and determinant-like invariants. One example is the insertion of the effective operator $\mathcal{O}_{\rm lequ}^{(1)}$.}
    	\label{fig:Clequ}
    \end{subfigure}
    \end{minipage} \\ \vspace{20pt}
    
    \caption{Examples of instanton diagrams corresponding to invariants discussed in the text. Here, the gray blob depicts the instanton background that the fermions (solid lines) are coupled to. The fermion lines are closed via Yukawa interactions with the Higgs (dashed lines).}
    \label{fig:FlowerDiagramsSMEFT}
\end{figure}
Furthermore, a direct relation between diagrammatic contributions and invariants seems to be clear. Consider the example of the calculation performed in Ref.~\cite{Bedi:2022qrd}, where the topological susceptibility from an insertion of the effective operator $\mathcal{O}_{\rm quqd}^{(1)}$ was studied. Diagrammatically this process can be understood as that of Fig.~\ref{fig:Cquqd}; one can observe in the diagram that since all fermion legs of the effective operator are directly connected to the instanton background, this corresponds to their zero modes being projected out in the path integral calculations. Therefore, as previously shown, the resulting contribution will follow a determinant-like structure on all indices of the Wilson coefficient. Indeed, as we will prove explicitly in the next section, this diagram gives a contribution proportional to the introduced invariants $\cA_{0000}^{0000}(C_{\rm quqd}^{(1)})$ and $\cB_{0000}^{0000}(C_{\rm quqd}^{(1)})$.

Another interesting contribution which illustrates the previous points is the contribution from the insertion of the operator $\mathcal{O}_{\rm lequ}^{(1)}$. The corresponding diagram is that of Fig.~\ref{fig:Clequ} and only the quarks emerging from the effective operator have zero modes, as the leptons are assumed not to be charged under the symmetry group responsible for the instanton dynamics. Indeed, looking at the constructed invariants $I_{abcd}^f(C_{\rm lequ}^{(1)})$, we see exactly that only the quark indices are contracted with the anti-symmetric $\epsilon$-structure (determinant-like) whereas the lepton indices are contracted in a trace-like manner over a matrix product with a lepton Yukawa coupling.

A final illustrative example is that of rephasing invariant operators such as $\mathcal{O}_{\rm Hq}^{(1)}$. In this case, even at the lowest order in Yukawa couplings, one cannot build an invariant where both quark indices are directly contracted with an $\epsilon$-structure; at most one index is contracted, as shown in the invariant $\cI_{abcd}(C_{\rm Hq}^{(1)})$. This means that diagrammatically only one fermionic propagator is directly connected to the instanton background, that is, only one zero mode is projected out from the effective operator. This case is illustrated in Fig.~\ref{fig:CHq}.

We have so far argued that the result of instanton calculations are proportional to the determinant-like invariants (and no extra flavor structures). Next, we will show how these patterns arise explicitly.

\subsection{Four-quark operator}
\label{sec:4Fermi_op}
Diagrammatically speaking, having effective operators in the theory allows for a different way to contract the open fermion legs coupled to the instanton vertex apart from utilizing mass terms or Yukawa 
couplings~\cite{Csaki:2019vte}. Let us start by considering the operator $\mathcal{O}_{\rm quqd}^{(1)}$, which can give rise to the instanton diagram in Fig.~\ref{fig:Cquqd}. Since the SM $SU(2)$ gauge group is unrelated to the instanton dynamics, the $SU(2)$ quark structure can be treated in the same way as a flavor index. As such, zero modes in the instanton background will not depend on the $SU(2)$ index. The topological susceptibility, Eq.~\eqref{eq:suscepdef}, induced by the four-fermion operator can be calculated as\footnote{Note, that all computations are done in Euclidean space by Wick-rotating the time coordinate everywhere in the calculations.}
\begin{equation}
\begin{split}
    & \chi_{\rm quqd}^{(1)}(0)^{1-\text{inst.}} = -i \lim _{k \rightarrow 0} \int d^4 x\, e^{i k x}\left\langle 0\left|T\left\{\frac{1}{32 \pi^2} G \wtilde{G}(x), \frac{C^{(1)}_{\rm quqd}}{\CPVm^2} \mathcal{O}^{(1)}_{\rm quqd}(0)\right\}\right| 0\right\rangle\,, \\
    & = e^{-i\thetaQCD} \int d^4x_0 \int \frac{d\rho}{\rho^5} d_N(\rho) \int \mathcal{D}H\mathcal{D}H^\dagger \, e^{- S_0[H,H^\dagger]} \int \prod_{f=1}^3 \left( \rho^2 \, d\xi_{u_f}^{(0)}d\xi_{d_f}^{(0)}d^2\bar{\xi}_{Q_f}^{(0)}\right) \\
    & \times  e^{\int d^4x (\bar{Q} Y_{\rm u} \wtilde{H} u + \bar{Q} Y_{\rm d} H d + \hc)(x)} \frac{1}{32 \pi^2} \int d^4x\, G\wtilde{G}(x)\left( \frac{C^{(1)}_{\rm quqd}}{\CPVm^2}\bar{Q} u\bar{Q} d(0) + \hc \right) \,,
\end{split}\label{eq:4quark_top_susc1}
\end{equation}
where $d^2{\bar\xi}_{Q_f^{\vphantom{1}}}^{(0)}\equiv d{\bar\xi}_{Q_f^1}^{(0)} \, d{\bar\xi}_{Q_f^2}^{(0)}$ for the two components of the $SU(2)$ quark doublet. The fermions are expanded in their eigenmodes (c.f. Eq.~\eqref{Inst-fermion: eigenmodes}) and only those terms containing fields with zero modes in the instanton background have been kept. As can be seen in Eq.~\eqref{Inst-fermion: zero-modes} this will be $u,d,Q^\dagger$ in the instanton background and the conjugates will contribute to the anti-instanton scenario.

The next step is to expand the exponential of the interacting action of the fermions and Higgs, such that precisely enough fermion fields appear in the Grassmann integral to obtain a non-vanishing result. We will also make the $SU(2)$ indices of all $SU(2)$ doublets explicit in the following calculations by giving all $SU(2)$ doublets upper case indices. We find
\begin{align}
    & \chi_{\rm quqd}^{(1)}(0)^{1-\text{inst.}} = e^{-i\thetaQCD} \int d^4x_0 \int \frac{d\rho}{\rho^5} d_N(\rho) \int \mathcal{D}H\mathcal{D}H^\dagger \, e^{- S_0[H,H^\dagger]} \prod_{f=1}^3 \left( \rho^2 \, d\xi_{u_f}^{(0)}d\xi_{d_f}^{(0)}d^2\bar{\xi}_{Q_f}^{(0)}\right) \nonumber \\
    & \times \int d^4x_1 d^4x_2 d^4x_3 d^4x_4 \, \frac{1}{4!}  \left[ \sum_{\substack{\text{perm. over}\\ \text{fermion fields}}} \bar{\xi}_{Q_{i_1}^I}^{(0)}(\bar{\psi}^{(0)} Y_{{\rm u},i_1j_1} \wtilde{H}^I P_R \psi^{(0)})(x_1) \xi_{u_{j_1}}^{(0)} \right. \nonumber  \\
    &\left.\times \bar{\xi}_{Q_{i_2}^J}^{(0)}(\bar{\psi}^{(0)} Y_{{\rm u},i_2j_2} \wtilde{H}^J P_R \psi^{(0)})(x_2) \xi_{u_{j_2}}^{(0)} \quad \bar{\xi}_{Q_{k_1}^K}^{(0)}(\bar{\psi}^{(0)} Y_{{\rm d},k_1l_1} H^K P_R \psi^{(0)})(x_3) \xi_{d_{l_1}}^{(0)} \right. \\
    & \left. \times \bar{\xi}_{Q_{k_2}^L}^{(0)}(\bar{\psi}^{(0)} Y_{{\rm d},k_2l_2} H^L P_R \psi^{(0)})(x_4) \xi_{d_{l_2}}^{(0)} \, \int d^4x \frac{G\wtilde{G}(x)}{32 \pi^2} \left( \frac{C_{{\rm quqd},mnop}^{(1)}}{\CPVm^2} \bar{\xi}_{Q_{m}^M}^{(0)} (\bar{\psi}^{(0)} P_R \psi^{(0)} ) \xi_{u_{n}}^{(0)} \, \epsilon_{MN}\right. \right. \nonumber  \\
    &  \times \bar{\xi}_{Q_{o}^N}^{(0)} (\bar{\psi}^{(0)} P_R \psi^{(0)} ) \xi_{d_{p}}^{(0)} \bigg)\,(0) \, \bigg]\,, \nonumber 
\end{align}
where the indices $m,M$ of $\xi_{Q_{m}^M}^{(0)}$ denote the $SU(2)$ and flavor indices, respectively, of the zero mode Grassmann vector $\xi_Q$, which in this case is six-dimensional. After integrating over all Grassmann variables of the zero modes and considering all the permutations over the fermion fields, we find that the flavor invariants constructed in Section~\ref{sec:Invariants} appear explicitly
\begin{equation}
\begin{split}
    & \chi_{\rm quqd}^{(1)}(0)^{1-\text{inst.}} = \frac{1}{4\CPVm^2}\[ e^{-i\thetaQCD} \epsilon^{i_1i_2m} \epsilon^{j_1j_2n} Y_{{\rm u},i_1j_1} Y_{{\rm u},i_2j_2} C_{{\rm quqd},mnop}^{(1)}\epsilon^{k_1k_2o}\epsilon^{l_1l_2p} Y_{{\rm d},k_1l_1} Y_{{\rm d},k_2l_2} \right. \\
    & \left. + e^{-i\thetaQCD} \epsilon^{i_1i_2m} \epsilon^{j_1j_2n} Y_{{\rm u},i_1j_1} Y_{{\rm u},i_2j_2} C_{{\rm quqd},onmp}^{(1)} \epsilon^{k_1k_2o}\epsilon^{l_1l_2p} Y_{{\rm d},k_1l_1} Y_{{\rm d},k_2l_2} \] \, \int d^4x_0 \int \frac{d\rho}{\rho^5} d_N(\rho) \rho^6 \\
    & \times \underbrace{ \int \mathcal{D}H \mathcal{D}H^\dagger \, e^{- S_0[H,H^\dagger]}\left[ \int d^4x_1 d^4x_2 (\bar{\psi}^{(0)} H_I^\dagger \epsilon^{IJ} P_R \psi^{(0)})(x_1) \, (\bar{\psi}^{(0)} \epsilon_{JK} H^K P_R \psi^{(0)})(x_2) \right]^2}_{= \, 2! \, \left[ \int d^4x_1d^4x_2 (\bar{\psi}^{(0)} P_R \psi^{(0)})(x_1) \Delta_H(x_1-x_2) \epsilon_{IJ}\epsilon^{JI} (\bar{\psi}^{(0)} P_R \psi^{(0)})(x_2) \right]^{2} \equiv \, 2! \, \mathcal{I}^2} \\
    & \times \left(\epsilon_{MN}\epsilon^{MN} \bar{\psi}^{(0)} P_R \psi^{(0)} \, \bar{\psi}^{(0)} P_R \psi^{(0)} \right)(0) \int d^4x \frac{G\wtilde{G}(x)}{32 \pi^2}\,.
    \label{result-4fermions: upto-inv-functional-int}
\end{split}
\end{equation}
The factor of $1/4$ at the beginning of Eq.~\eqref{result-4fermions: upto-inv-functional-int} appears because the integral over the fermion zero modes is expressed in terms of the Levi-Civita symbols (see also Eq.~\eqref{eq:grassmanid}). The last step is to integrate over the Higgs field in the Euclidean path integral; using the definition of the Higgs propagator in position space
\begin{equation} \label{Definition: Higgs-propagator}
    \int \cD H \cD H^\dagger e^{- S_0[H,H^\dagger]} H_I^{\vphantom{\dagger}}(x_1) H_J^\dagger(x_2) = \Delta_H(x_1-x_2) \, \delta_{IJ} \,,
\end{equation} 
we are left with the integral
\begin{equation}
    \mathcal{I} = 2 \int d^4x_1d^4x_2 (\bar{\psi}^{(0)} P_R \psi^{(0)})(x_1) \Delta_H(x_1-x_2) (\bar{\psi}^{(0)} P_R \psi^{(0)})(x_2) \,,
\end{equation}
multiplying the invariant that we set out to find. After some simplifications, we finally arrive at
\begin{equation}
\begin{split}
    & \chi_{\rm quqd}^{(1)}(0)^{1-\text{inst.}} = \frac{A_{\rm quqd}^{(1)} + B_{\rm quqd}^{(1)}}{\CPVm^2} \int d^4x_0 \int \frac{d\rho}{\rho^5} d_N(\rho) \rho^6 \mathcal{I}^2 \left(\bar{\psi}^{(0)} P_R \psi^{(0)}\bar{\psi}^{(0)} P_R \psi^{(0)} \right)(0) \,,
\end{split}
\end{equation}
where we have defined
\begin{equation}
    \begin{split}
        A_{\rm quqd}^{(1)} & = e^{-i\thetaQCD} \epsilon^{i_1i_2m} \epsilon^{j_1j_2n} Y_{{\rm u},i_1j_1} Y_{{\rm u},i_2j_2} C_{{\rm quqd},mnop}^{(1)} \epsilon^{k_1k_2o}\epsilon^{l_1l_2p} Y_{{\rm d},k_1l_1} Y_{{\rm d},k_2l_2} \,, \\
        B_{\rm quqd}^{(1)} & = e^{-i\thetaQCD} \epsilon^{i_1i_2m} \epsilon^{j_1j_2n} Y_{{\rm u},i_1j_1} Y_{{\rm u},i_2j_2} C_{{\rm quqd},onmp}^{(1)} \epsilon^{k_1k_2o}\epsilon^{l_1l_2p} Y_{{\rm d},k_1l_1} Y_{{\rm d},k_2l_2} \,.
    \end{split}
    \label{eq:ABdefn}
\end{equation}

The same calculation can be performed with the anti-instanton solution. In this case, the non-vanishing contributions will arise from the Hermitian conjugate terms in the calculation. Furthermore, the winding number $\int d^4x \,G \wtilde{G}(x)$ will flip its sign, which also induces a sign flip in the exponential of $\thetaQCD$. Therefore, the full one-instanton and anti-instanton contribution to the topological susceptibility induced by a four-fermion operator reads
\begin{equation}
\begin{split}
    & \chi_{\rm quqd}^{(1)}(0) = \chi_{\rm quqd}^{(1)}(0)\big|_{1-\text{inst.}} + \chi_{\rm quqd}^{(1)}(0)\big|_{1-\text{a.-inst.}}  \\
    &=  \frac{1}{\CPVm^2}\(A_{\rm quqd}^{(1)} + B_{\rm quqd}^{(1)}\) \int d^4x_0 \int \frac{d\rho}{\rho^5} d_N(\rho) \rho^6 \mathcal{I}^2 \left(\bar{\psi}^{(0)} P_R \psi^{(0)} \, \bar{\psi}^{(0)} P_R \psi^{(0)} \right)(0)\bigg|_{\text{1-inst.}}
    \\ 
    & - \frac{1}{\CPVm^2}\(A_{\rm quqd}^{(1)} + B_{\rm quqd}^{(1)}\)^* \int d^4x_0 \int \frac{d\rho}{\rho^5} d_N(\rho) \rho^6 \mathcal{I}^2 \left(\bar{\psi}^{(0)} P_L \psi^{(0)} \, \bar{\psi}^{(0)} P_L \psi^{(0)} \right)(0)\bigg|_{\text{1-a.-inst.}}
    \,.
\end{split}\label{eq:4Fermi_op_AB_inv}
\end{equation}
Substituting the explicit form of the fermion zero modes from Eq.~\eqref{Inst-fermion: zero-modes} gives
\begin{equation}
    \bar{\psi}_i^{(0)} P_{R} \psi^{(0)}_i \bar{\psi}_j^{(0)} P_{R} \psi^{(0)}_j \bigg|_{\text{1-inst.}} = \dfrac{4\rho^4}{\pi^4} \dfrac{1}{(x_0^2+\rho^2)^6} = \bar{\psi}_i^{(0)} P_{L} \psi^{(0)}_i \bar{\psi}_j^{(0)} P_{L} \psi^{(0)}_j \bigg|_{\text{1-a.-inst.}}\,,
\end{equation}
which in turn leads to the result
\begin{align}
   \chi_{\rm quqd}^{(1)}(0) &= \frac{2i}{\CPVm^2}\, \Im\(A_{\rm quqd}^{(1)} + B_{\rm quqd}^{(1)}\) \int d^4x_0 \int \frac{d\rho}{\rho^5} d_N(\rho) \rho^6 \mathcal{I}^2 \left[\dfrac{4\rho^4}{\pi^4} \dfrac{1}{(x_0^2+\rho^2)^6} \right]\,.
   \label{result-4fermions: upto-inv}
\end{align}
As expected, the final result depends explicitly on the determinant-like invariants introduced in Eq.~\eqref{eq:ABquqd}, since using Eq.~\eqref{eq:ABdefn} we find
\begin{equation}
    \Im (A_{\rm quqd}^{(1)}) = \cA_{0000}^{0000}\(C_{\rm quqd}^{(1)}\) \, , \quad \Im (B_{\rm quqd}^{(1)}) = \cB_{0000}^{0000}\(C_{\rm quqd}^{(1)}\) \, .
\end{equation}

It is instructive to compare our results with the NDA estimates introduced in Ref.~\cite{Csaki:2023ziz}, which states that the loop factor suppression, $(4\pi)^{-\alpha}$, can be predicted by
\begin{equation}
\label{eq:NDA}
    \alpha = z - 2v + 2 p\,,
\end{equation}
where $z$ is the number of fermion zero modes, $v$ the number of vertices and $p$ the number of propagators in the instanton calculation. From Fig.~\ref{fig:Cquqd}, we verify that for an insertion of $\cO^{(1)}_{\rm quqd}$ we have $\alpha=12-10+4=6$. After appropriately substituting the fermion zero modes, our result in Eq.~\eqref{Appendix: final-result-4F} gives a suppression of $1/(45\pi^6)$. Clearly, the power of $\pi$ matches the NDA prediction, but the numerical factor is smaller than $4^6$ obtained from NDA. This difference arises from the fact that we have assumed an unbroken $SU(2)$ group and there are also combinatoric factors. Taking these effects into account and summing both the instanton and anti-instanton configurations, the estimation from NDA would predict a suppression factor $1/(256\pi^6)$ in Eq.~\eqref{Appendix: final-result-4F}, which is within one order of magnitude compared to the full calculation.

\subsection{Semileptonic four-fermion operator}
\label{sec:semileptonic}
In Section~\ref{sec:basis}, we showed that invariants featuring $\thetaQCD$ can also be constructed for the semileptonic operator $\cO_{\rm lequ}^{(1)}$, so it is important to verify whether they arise in instanton calculations. As the leptons are not charged under the gauge group generating the instantons, they are not coupled to the instanton vertex directly and should, hence, be treated perturbatively like the Higgs field in the last section. The invariants of a semi-leptonic operator can therefore only enter by treating the leptons perturbatively on top of the instanton background giving the special functional form to the quark zero modes (c.f. Eq.~\eqref{Inst-fermion: zero-modes}), as we will now show. The topological susceptibility will be calculated with an insertion of the operator $\cO_{\rm lequ}^{(1)}$, where the leptons will be kept in the path integral. As before, we will split off the zero modes and integrate over the non-zero modes of the quark fields. This gives 
\begin{equation}
\begin{split}
    & \chi_{\rm lequ}^{(1)}(0)^{1-\text{inst.}} = -i \lim _{k \rightarrow 0} \int d^4 x e^{i k x}\left\langle 0\left|T\left\{\frac{1}{32 \pi^2} G \wtilde{G}(x), \frac{C^{(1)}_{\rm lequ}}{\CPVm^2} \mathcal{O}^{(1)}_{\rm lequ}(0)\right\}\right| 0\right\rangle\,, \\
    & = e^{-i\thetaQCD} \int d^4x_0 \int \frac{d\rho}{\rho^5} d_N(\rho) \int \mathcal{D}H\mathcal{D}H^\dagger \mathcal{D}L\mathcal{D}\bar{L} \mathcal{D}e\mathcal{D}\bar{e} \, e^{- S_0[H,H^\dagger]} \, e^{- S_0[L, \bar{L}]} \, e^{- S_0[e,\bar{e}]} \\
    & \times\int \prod_{f=1}^3 \left( \rho^2 \, d\xi_{u_f}^{(0)}d\xi_{d_f}^{(0)}d^2\bar{\xi}_{Q_f}^{(0)}\right) e^{\int d^4x (\bar{Q} Y_{\rm u} \wtilde{H} u + \bar{Q} Y_{\rm e} H d + \bar{L} Y_{\rm e} H e + \hc)(x)} \\
    & \times \frac{1}{32 \pi^2} \int d^4x\, G\wtilde{G}(x) \left( \frac{C^{(1)}_{\rm lequ}}{\CPVm^2} \bar{L} e \bar{Q} u(0) + \hc \right)\,.
\end{split}
\end{equation}
As previously, we will now expand the exponential of the action containing the fermion and Higgs field. We expand the exponential over the quark Yukawa couplings in the zero modes as before and neglect the quark non-zero modes. Then, as is usually done in perturbation theory, we expand the exponential of the lepton Yukawa interaction order by order in the small Yukawa coupling, since expanding the exponential to first order will be sufficient to obtain a non-vanishing result.
\begin{equation}
\begin{split}
    & \chi_{\rm lequ}^{(1)}(0)^{1-\text{inst.}} = e^{-i\thetaQCD} \int d^4x_0 \int \frac{d\rho}{\rho^5} d_N(\rho) \int \mathcal{D}H\mathcal{D}H^\dagger \mathcal{D}L\mathcal{D}\bar{L} \mathcal{D}e\mathcal{D}\bar{e} \, e^{- S_0[H,H^\dagger]} \\
    & \times \, e^{-S_0[L, \bar{L}]} \, e^{-S_0[e,\bar{e}]} \prod_{f=1}^3 \left( \rho^2 \, d\xi_{u_f}^{(0)}d\xi_{d_f}^{(0)}d^2\bar{\xi}_{Q_f}^{(0)}\right) \int d^4x_1 d^4x_2 d^4x_3 d^4x_4 d^4x_5  \\
    & \times \frac{1}{6!}\left[ \sum_{\substack{\text{perm. over}\\ \text{fermion fields}}} \bar{\xi}_{Q_{i_1}^I}^{(0)}(\bar{\psi}^{(0)} Y_{{\rm u},i_1j_1} \wtilde{H}^I P_R \psi^{(0)})(x_1) \xi_{u_{j_1}}^{(0)} \quad \bar{\xi}_{Q_{i_2}^J}^{(0)}(\bar{\psi}^{(0)} Y_{{\rm u},i_2j_2} \wtilde{H}^J P_R \psi^{(0)})(x_2) \xi_{u_{j_2}}^{(0)} \right. \\
    & \left. \times ~\bar{\xi}_{Q_{k_1}^K}^{(0)}(\bar{\psi}^{(0)} Y_{{\rm d},k_1l_1} H^K P_R \psi^{(0)})(x_3) \xi_{d_{l_1}}^{(0)} \quad \bar{\xi}_{Q_{k_2}^L}^{(0)}(\bar{\psi}^{(0)} Y_{{\rm d},k_2l_2} H^L P_R \psi^{(0)})(x_4) \xi_{d_{l_2}}^{(0)}\right. \\
    & \left. \times ~\bar{\xi}_{Q_{k_3}^M}^{(0)}(\bar{\psi}^{(0)} Y_{{\rm d},k_3l_3} H^M P_R \psi^{(0)})(x_5) \xi_{d_{l_3}}^{(0)} \, \int d^4x \frac{G\wtilde{G}(x)}{32 \pi^2} 
    \left( \frac{C_{{\rm lequ},mnop}^{(1)}}{\CPVm^2} \( \bar{L}_m^N e_n \) \, \epsilon_{NO} \right. \right. \\
    & \times ~\bar{\xi}_{Q_{o}^O}^{(0)} (\bar{\psi}^{(0)} P_R \psi^{(0)} ) \xi_{u_{p}}^{(0)} \bigg)\,(0)\, \bigg] \, \int d^4x_6 \, \big(\bar{e}_q Y_{{\rm e},qr}^\dagger H^{\dagger,P} L_r^P \big) \, (x_6)\,.
\end{split}
\end{equation}
Note that in comparison to the computation of $\cO_{\rm quqd}^{(1)}$, an extra down Yukawa coupling is needed to complete the zero mode expansion because $\cO_{\rm lequ}^{(1)}$ has no down-quark bilinear. In the next step, we will perform the integration in the path integral over the Higgs and lepton fields, as well as the zero mode integrals. 
Using
\begin{equation}
    \int \mathcal{D}\psi \mathcal{D}\bar{\psi} \, e^{-S_0[\psi,\bar{\psi}]} \, 
    \psi_I(x_1) \bar{\psi}_J(x_2) \equiv \Delta_F(x_1-x_2) \, \delta_{IJ}
    \,,
\end{equation}
the lepton fields are contracted to form a loop.\footnote{The indices $I,J$ represent all internal indices, like the flavor and $SU(2)$ gauge group indices.} The resulting expression reads
\begin{equation}
\begin{split}
    & \chi_{\rm lequ}^{(1)}(0)^{1-\text{inst.}} = \frac{1}{2\CPVm^2} \underbrace{e^{-i\thetaQCD} \epsilon^{i_1i_2m} \epsilon^{j_1j_2n} Y_{{\rm u},i_1j_1}^{\phantom{(}} Y_{{\rm u},i_2j_2}^{\phantom{(}} C_{{\rm lequ},opmn}^{(1)}Y_{\vphantom{l}{\rm e},po}^\dagger \det Y_{\rm d}}_{\equiv\,I_{\rm lequ}^{(1)}} \\
    & \times 3! \, \int d^4x_0 \int \dfrac{d\rho}{\rho^5} d_N(\rho) \rho^6 \, \mathcal{I}^2 \, \int d^4x_5 d^4x_6 \( \bar{\psi}^{(0)} P_R \psi^{(0)} \)(x_5) \Delta_H(x_5-x_6) \\
    & \times \tr \Big( P_R \Delta_F(x_6-0) P_L \Delta_F(0-x_6) \Big)  \( \bar{\psi}^{(0)} P_R \psi^{(0)} \)(0) \, \epsilon_{OP}\epsilon^{PO} \int d^4x \frac{G\wtilde{G}(x)}{32 \pi^2}\,.
    \label{result-lequ: upto-inv}
\end{split}
\end{equation}
As before, we add the anti-instanton contribution to obtain the full result, which leads to the complex conjugate invariants appearing with the opposite sign in the final result. Thus, the final result 
is proportional to the invariant
\begin{equation}
    \begin{split}
        \Im\(I_{\rm lequ}^{(1)}\) & = \Im \[ e^{-i\thetaQCD} \epsilon^{i_1i_2m} \epsilon^{j_1j_2n} Y_{\vphantom{l}{\rm u},i_1j_1}^{\phantom{(}} Y_{\vphantom{l}{\rm u},i_2j_2}^{\phantom{(}} C_{{\rm lequ},opmn}^{(1)} Y_{\vphantom{l}{\rm e},po}^\dagger \det Y_{\rm d} \]
        = \cI_{0000}^0\(C_{\rm lequ}^{(1)}\) \,,
    \end{split}
\end{equation}
that we have defined in Section~\ref{sec:basis} multiplied by a complicated integral. In Appendix~\ref{app:Semilept}, the integrals in Eq.~\eqref{result-lequ: upto-inv} are evaluated; in particular the integral over the leptonic loop is divergent. We have explicitly verified that when considering the appropriate renormalized effective field theory, the counterterms cancel this divergence, as expected. Furthermore, the NDA estimation of Ref.~\cite{Csaki:2023ziz} also works in this case: following Eq.~\eqref{eq:NDA}, we expect a suppression of $(4\pi)^{-8}$ which matches the $\pi$ suppression of the result obtained in Eq.~\eqref{eq:susc_lequ_finite}, where the numerical factor $\simeq 1/(450\pi^{8})$. For this factor, if we take into account the combinatoric factors, the unbroken $SU(2)$ and the sum over instanton and anti-instanton configurations, the NDA estimation of $4^{-8}$ becomes approximately half of the full result.

This analysis can be repeated for all other operators in the SMEFT following the same procedure. We present calculations for the insertion of the gluon dipole operator $\cO_{\rm dG}$ in Appendix~\ref{app:GluonDip}, that will also be considered in a phenomenological study in Section~\ref{sec:Pheno}. For some SMEFT operators, their leading contribution might not arise from projecting the zero modes out of all the fermion legs. Indeed, considering non-zero modes of the quarks in the effective operators is also needed to obtain the invariants with more powers of Yukawa couplings introduced in Section~\ref{sec:basis}. We next discuss the calculations in these cases.

\subsection{Higher-order invariants and selection rules}
\label{Higher-order inv.}

In our calculations we have only considered contributions at leading order with the least possible power of the Yukawa couplings. However, one could compute higher-loop diagrams with more powers of Yukawa couplings, which would be captured by higher-order invariants, such as the diagram depicted in Fig.~\ref{fig:CquqdHiOrd}. The explicit calculation of higher-loop diagrams works differently than what was performed in Sections~\ref{sec:4Fermi_op} and~\ref{sec:semileptonic}, because some interactions mix the zero and non-zero modes of the fermions charged under the instanton group. Explicitly performing this calculation is beyond the scope of this paper but we will comment on how these calculations work in principle. Instead of just including the quark zero modes in the calculations, one would have to include the non-zero mode interactions in the action as well. These should be treated perturbatively as was done for the leptons in Section~\ref{sec:semileptonic}. As a consequence, one can no longer simply integrate over the free part of the action containing just the non-zero modes to remove them from the path integral as done in Eq.~\eqref{tHooftsresult}, since their interactions with the zero modes necessarily appear in the action as well. Hence, we would have to reevaluate ’t~Hooft's result for the non-zero mode integration (i.e., remove the factor of $e^{0.292N_f}$ from the instanton density in Eq.~\eqref{Definition: inst-factor-CN}) and treat the non-zero modes of the colored fields as perturbations around the instanton background.  

In summary, including extra powers of Yukawas from the interacting part of the action forces us to consider terms mixing zero- and non-zero modes of the quarks. The contraction of the flavor indices of the non-zero mode fermions in the Yukawa interactions is done by Kronecker deltas as a result of flavorful propagators introduced in the perturbative calculation (c.f. the calculation with the semi-leptonic operator in Section~\ref{sec:semileptonic}). This results in the higher-order invariants introduced in Section \ref{sec:basis}, where the extra Yukawas are contracted in a matrix product. The indices corresponding to non-zero modes remain contracted in a determinant-like manner, i.e. via the $\epsilon$ symbol.

In addition, flavor invariants offer an explanation as to why operators invariant under the flavor-diagonal $U(1)$ quark rephasings (shown in Table~\ref{tab:FlavorTrafo}) cannot enter through zero mode contributions in instanton calculations. One example for such an operator is $\cO_{\rm Hq}^{(1)}$, which is invariant under $Q \to e^{i \alpha_Q} Q$, where all flavors are rephased with the same parameter. Because the operator is rephasing invariant, other flavorful objects besides the Wilson coefficient are needed to counteract the rephasing of $e^{-i \thetaQCD}$ that necessarily appears in instanton calculations. Due to the linearity of the flavor invariants in the Wilson coefficient, this object can only be constructed by SM Yukawa couplings. There are two options to construct a flavor invariant given these constraints. The object counteracting the rephasing of $e^{-i \thetaQCD}$ can either be $\det Y_{\rm u} Y_{\rm d}$ with the Wilson coefficient appearing in a trace invariant or a determinant-like invariant where, even at lowest order, the Wilson coefficient multiplies one of the Yukawa couplings (c.f. Eq.~\eqref{eq:CHqinv}).\footnote{These two types of invariants are equivalent as we show explicitly in Eq.~\eqref{eq:relCHq} for the operator $\cO_{\rm Hq}^{(1)}$ and all arguments presented here work for both forms.} As we have discussed previously, both traces and matrix products can only appear through propagators in perturbative calculations of the non-zero modes of quarks around the instanton background. Hence, the flavor invariants imply a selection rule on all operators that are invariant under rephasings to only contribute in instanton calculations when the non-zero modes of its fermions are considered.

As a more general statement, flavor invariants can be used to understand how the contribution from any SMEFT operator will contract with the flavor structure of the theory without explicitly doing the path integral computations, as we have anticipated in Section~\ref{sec:WhyDetInvs}. Furthermore, knowing how the non-vanishing contributions scale and connect with an instanton diagram, allows to correctly account for the loop factors coming from the zero modes and from the rest of the perturbative calculation; invariants can therefore allow for a more refined NDA estimate of the instanton effects in the spirit of what was done in~Ref.~\cite{Csaki:2023ziz}. 

\section{Constraints on dimension-six CP-violating operators}\label{sec:Pheno}

The results derived in the previous sections can now be used to place bounds on the scale $\CPVm$ associated with dimension-six CP violating operators. We will assume that QCD is modified at a scale $\Lambda_{\rm SI}$ where the one-instanton approximation remains valid and small instantons induce a shift in $\bar\theta$ proportional to $\Lambda^2_{\rm SI}/\CPVm^2$. Using the determinant-like invariants arising from the one-instanton calculation, we will then obtain limits on the ratio $\Lambda_{\rm SI}/\CPVm$, under the assumption that this induced $\bar\theta$ in the small instanton background saturates the experimental bound from the neutron EDM, $\bar\theta \lesssim 10^{-10}$~\cite{Abel:2020pzs}.

The effects from small (UV) instantons are enhanced, provided the gauge coupling becomes larger in the UV. In this limit, the one instanton calculation remains consistent and the invariants derived in Section~\ref{sec:Invariants} can be directly used. As such, we can reliably compute the effects of various dimension-six operators and compare them to the topological susceptibility $\chi (0) $ shown in Eq.~\eqref{eq:QCD_susc_ferm}. There are two UV models that modify QCD UV dynamics to enhance the topological susceptibility for which the instanton computation is relevant and these will be briefly reviewed below:

\paragraph{Product group models}
A natural way to explicitly compute the instanton integrals discussed in Section~\ref{sec:Instantons} is to Higgs an enlarged gauge group containing QCD color. This modifies the instanton measure $d_N(\rho)$ by an exponential factor 
 \begin{align}
       d_N(\rho) \to   d_N(\rho) \, e^{-2\pi^2\rho^2 \sum | \langle\sigma\rangle|^2}
       ~,
       \label{eq:sclar_inst_measure}
\end{align} 
where the sum extends over all the scalars $\sigma$ that Higgs the gauge group. 
This provides a cutoff $ \sim 1/| \langle\sigma\rangle| $ for the instanton integrals.
In particular, we consider the product group model introduced in Refs.~\cite{Agrawal:2017ksf, Csaki:2019vte}, where the gauge group $SU(3)_1 \times SU(3)_2 \times\dots \times SU(3)_k$ is Higgsed to the diagonal $SU(3)_c$ via bifundamental scalars $\sigma$. The gauge couplings of the individual gauge groups at the UV scale can be chosen to be larger than the QCD coupling while still in the perturbative regime so that small instantons are enhanced and instanton computations are applicable.

\paragraph{5D Instantons}
Another model where small instantons provide a significant contribution to the topological susceptibility was studied in Ref.~\cite{Gherghetta:2020keg}. This involves uplifting the BPST instanton presented in Eq.~\eqref{Definition: BPST instantons} to a compact extra dimension of size $R$, which modifies the running of the effective gauge coupling in Eq.~\eqref{eq:RunningCoupling} above the compactification scale $1/R$. The effective action then becomes
\begin{equation}
    S_{\rm eff}\simeq   \dfrac{8\pi^2}{g^2(1/R)}-\frac{R}{\rho}+b_0\ln\frac{R}{\rho}~,
    \label{eff_action}
\end{equation}
where $b_0$ is the $\beta$-function coefficient of the zero modes and the linear term $R/\rho$ is 
due to additional contributions in Eq.~\eqref{E-action: full-action} from the Kaluza-Klein modes. As such, the instanton measure becomes modified by an amount
\begin{align}
       d_N(\rho) \to   d_N(\rho) \, e^{R/\rho}
       ~.
       \label{eq:5D_inst_measure}
\end{align}The dilute instanton gas approximation can then be used to compute the topological susceptibility in this model by imposing the 5D perturbativity condition
\begin{align}
\Lambda_{\rm SI}R\lesssim \frac{24\pi^2}{g^2}~, 
\end{align}
where $\Lambda_{\rm SI}$ is identified with the cutoff scale of the 5D gauge theory.

\subsection{Bounds from induced $\bar\theta$ }
As discussed in Section~\ref{sec:Instantons}, new sources of CP violation in the SMEFT can induce a shift in $\bar{\theta}$, which leads to observable effects such as the neutron EDM. In principle, all the invariants discussed in Section \ref{sec:basis} and Appendix \ref{app:MoreInvs} will give contributions to $\bar{\theta}$.
However, due to the different flavor structures, there are only a few invariants that contribute to leading order. 

In the following, we consider three different flavor scenarios to study their impact on the bounds obtained from the induced $\bar{\theta}$-angle. We will briefly introduce them here.
\begin{enumerate}
    \item The simplest is the anarchic flavor scenario, in which all Wilson coefficients have an $\cO(1)$ value. Compared to the SM, this will in particular lead to large flavor-changing interactions.

    \item A slightly more restrictive flavor assumption is the MFV~\cite{Buras:2000dm,DAmbrosio:2002vsn,Isidori:2012ts} scenario. As we have noted earlier, in the SM the only breaking of the $U(3)^5$ flavor symmetry of the fermion kinetic term is due to the SM Yukawa couplings. Taking the Yukawa couplings to be spurions under this symmetry (c.f. Table~\ref{tab:FlavorTrafo}), makes the Lagrangian formally invariant under this approximate symmetry. In MFV we assume that the non-renormalizable operators of the SMEFT follow the same symmetry scheme. Thus, all SMEFT Wilson coefficients are polynomials in the Yukawa couplings dictated by the spurious transformations of the Wilson coefficients under the flavor group.

    \item Lastly, we consider a Froggatt--Nielsen (FN) scenario~\cite{Froggatt:1978nt} that offers an explanation for the size of the SM lepton and quark masses as well as the parameters in the CKM matrix. In this scenario, the SM fields are extended by a complex scalar field $\phi$ which is a singlet under the SM gauge group. The new scalar field has charge $-1$ under a global $U(1)$ symmetry. Constructing a Lagrangian invariant under the SM gauge group and the newly postulated $U(1)$ yields
\begin{equation}
\label{eq:FN}
    \cL = - \left( \frac{\phi^\star}{\Lambda_{\rm FN}} \right)^{q_{Q_i}+q_{u_j}} C_{ij}^u \bar{Q}_i^{\vphantom{u}} \tilde{H} u_j^{\vphantom{u}} - \left( \frac{\phi^\star}{\Lambda_{\rm FN}} \right)^{q_{Q_i}+q_{d_j}} C_{ij}^d \bar{Q}_i^{\vphantom{d}} H d_j^{\vphantom{d}} - \left( \frac{\phi^\star}{\Lambda_{\rm FN}} \right)^{q_{L_i}+q_{e_j}} C_{ij}^e \bar{L}_i^{\vphantom{e}} H e_j^{\vphantom{e}}\,,
\end{equation}
where the FN charges of the left-handed fermions $Q$, $u^\dagger$, $d^\dagger$, $L$, $e^\dagger$ are denoted as $q_Q$, $q_u$, $q_d$, $q_L$, $q_e$, respectively, $q_H=0$, $\Lambda_{\rm FN}$ is the effective scale where the Froggatt--Nielsen scenario is UV completed and the coefficients $C_{ij}^{u,d,e}$  are $\cO(1)$ complex numbers. Eventually, the global $U(1)$ symmetry is broken by the VEV of the complex scalar, which yields hierarchical Yukawa couplings as powers of $\lambda = \frac{\langle \phi \rangle}{\Lambda_{\rm FN}} \sim 0.2$ dictated by the FN charges. One set of charge assignments that can reproduce the SM Yukawa couplings to large accuracy is
\begin{equation}
    q_Q=\{3,2,0\},\quad q_u=\{5,2,0\},\quad q_d=\{4,3,3\}~,\label{eq:FN_quark_charges}
\end{equation}
for the quarks and
 \begin{equation}
     q_L=\{9,5,3\},\quad q_e=\{0,0,0\} ~,\label{eq:FN_lepton_charges}
 \end{equation}
for the leptons. This construction can be extended to the effective operators of the SMEFT~\cite{Bordone:2019uzc}, resulting in hierarchical entries for the Wilson coefficients.

\end{enumerate}
We begin by identifying the leading order invariants amongst those given in Section~\ref{sec:Invariants}. This can be easily achieved by studying the  FN scaling of the invariant with the least number of Yukawa matrices for each operator. Consider the topological susceptibility of QCD, $\chi(0)\propto K_\theta$, e.g., which scales as $ \propto \lambda^{27}.$ This compares with the SMEFT invariants in Section~\ref{sec:basis} which scale as
\begin{align}
    \cI_{0000}(C_{\rm uH}),\, \cA_{0000}^{0000}(C_{\rm quqd}^{(1,8)}),\,\cB_{0000}^{0000}(C_{\rm quqd}^{(1,8)})\propto\lambda^{27}~,
    \label{eq:invscal1}
\end{align}
\begin{align}
   \cI_{1100}(C_{\rm Hq}^{(1,3)}),\,  \cI_{0000}^0(C_{\rm lequ}^{(1,3)}) \propto \lambda^{33} ~.
   \label{eq:invscal2}
\end{align}
This scaling helps to determine which invariants are important and hence phenomenologically the most interesting. For instance, Eqs.~\eqref{eq:invscal1} and \eqref{eq:invscal2} indicate that the operators ${\cal O}_{\rm u H}$ and ${\cal O}_{\rm quqd}^{(1,8)}$ lead to larger effects compared to  ${\cal O}_{\rm H q}^{(1,3)}$ or ${\cal O}_{\rm lequ}^{(1,3)}$, i.e. if the Wilson coefficients are assumed to be of the same order (up to the appropriate power of the FN parameter $\lambda$), the contribution of the operator ${\cal O}_{\rm quqd}^{(1,8)}$ (or ${\cal O}_{\rm u H}$) to $\theta_{\rm ind}$ dominates over that of ${\cal O}_{\rm lequ}^{(1,3)}$ (or  ${\cal O}_{\rm H q}^{(1,3)}$). This can also be understood from Fig.~\ref{fig:Cquqd} and Fig.~\ref{fig:Clequ} -- the latter figure contains additional loops and Yukawa couplings, compared to the former figure 
and the leading order contribution from $\chi(0)$. 

Below, we study in more detail how these invariants contribute to the shift in the axion potential minimum, $\theta_{\rm ind}$, for two leading-order operators -- $\cO_{\rm quqd}^{(1)}$ as well as the dipole operator  $\cO_{\rm dG}$, and the sub-leading semi-leptonic operator $\cO_{\rm lequ}^{(1)}$.

For this analysis, the MFV (at leading order) and FN flavor scenarios result in the same scaling for the Wilson coefficients, which occurs because we are only considering the contribution from chirality-flipping operators to one observable. For instance, the scaling of $\cO^{(1)}_{\rm quqd}$ is 
 \begin{equation}
     C^{(1)}_{\mathrm{quqd},ijkl } \, \overset{\rm MFV}{\sim} c_1 Y_{\mathrm{u},ij} Y_{\mathrm{d},kl} + \cO(Y^3_{\rm u,d})\,, \qquad C^{(1)}_{\mathrm{quqd},ijkl } \, \overset{\rm FN}{\sim} c_{1,ijkl} \lambda^{q_{Q_i}+q_{u_j}+q_{Q_k}+q_{d_l}}\,,   
 \end{equation}
where $c_1$ are $\cO(1)$ coefficients. Since by the FN construction, Eq.~\eqref{eq:FN}, $Y_{\rm u}\sim \lambda^{q_{Q}+q_{u}}$ and $Y_{\rm d}\sim \lambda^{q_{Q}+q_{d}}$, we explicitly see the same scaling in both scenarios. Therefore, we will only present constraints on $\CPVm$ (for a given $\Lambda_{\rm SI}$) under anarchic and MFV scenarios for the considered operators in the following.

\subsubsection{Four-quark operators}
For the four-quark operator $\cO_{\rm quqd}^{(1)}$, the topological susceptibility is computed in Appendix~\ref{app:Four-fermion} and the result is given in Eq.~\eqref{Appendix: final-result-4F}.
Performing the integral over $\rho$ in the  product group model $ SU(3)^k\to SU(3)_c$ and assuming $| \langle\sigma\rangle|=\Lambda_{\rm SI}$, we obtain
\begin{align}
    \theta_{\rm ind}= \dfrac{16\pi^2}{5(b_0-6)K_\theta}
    \left(\cA_{0000}^{0000}\(C_{\rm quqd}^{(1)}\)+\cB_{0000}^{0000}\(C_{\rm quqd}^{(1)}\)\right) \dfrac{\Lambda_{\rm SI}^2}{\CPVm^2}~,
    \label{eq:4q_theta}
\end{align}
where $b_0=13/2$ for $SU(3)_1$ and $b_0=21/2$ for $SU(3)_k$\,. For $SU(3)_2,\,\dots\, SU(3)_{k-1}$\,, $b_0=10$ and we get an additional factor of $2$ on the RHS of Eq~\eqref{eq:4q_theta}. 
In the case of 5D instantons, we obtain  
\begin{align}
    \theta_{\rm ind}= \dfrac{2}{5 K_\theta}\left(\cA_{0000}^{0000}\(C_{\rm quqd}^{(1)}\)+\cB_{0000}^{0000}\(C_{\rm quqd}^{(1)}\)\right) \dfrac{\Lambda_{\rm SI}^2}{\CPVm^2} \label{eq:4q_theta5D}
    ~.
\end{align}
Note that as a consequence of the Higgsed theory, the product group model has a smooth cutoff on the instanton size $\rho$, which leads to a  mild dependence on the $\beta$ function coefficient, $b_0$ in Eq.~\eqref{eq:4q_theta}.

In contrast, Eq.~\eqref{eq:5D_inst_measure} implies that the integral over $\rho$ is dominated by instantons of size $\rho\sim 1/\Lambda_{\rm SI}$. Therefore, all the susceptibilities for the 5D model only depend on $\Lambda_{\rm SI}$, up to an overall factor. This factor cancels when taking the ratio of susceptibilities, implying that $ \theta_{\rm ind}$ is independent of the $\beta$ function coefficient, $b_0$.

 \begin{figure}[h!]
    \centering
    \begin{subfigure}[t]{0.9\textwidth}
    	\centering
    	\includegraphics[width=\textwidth]{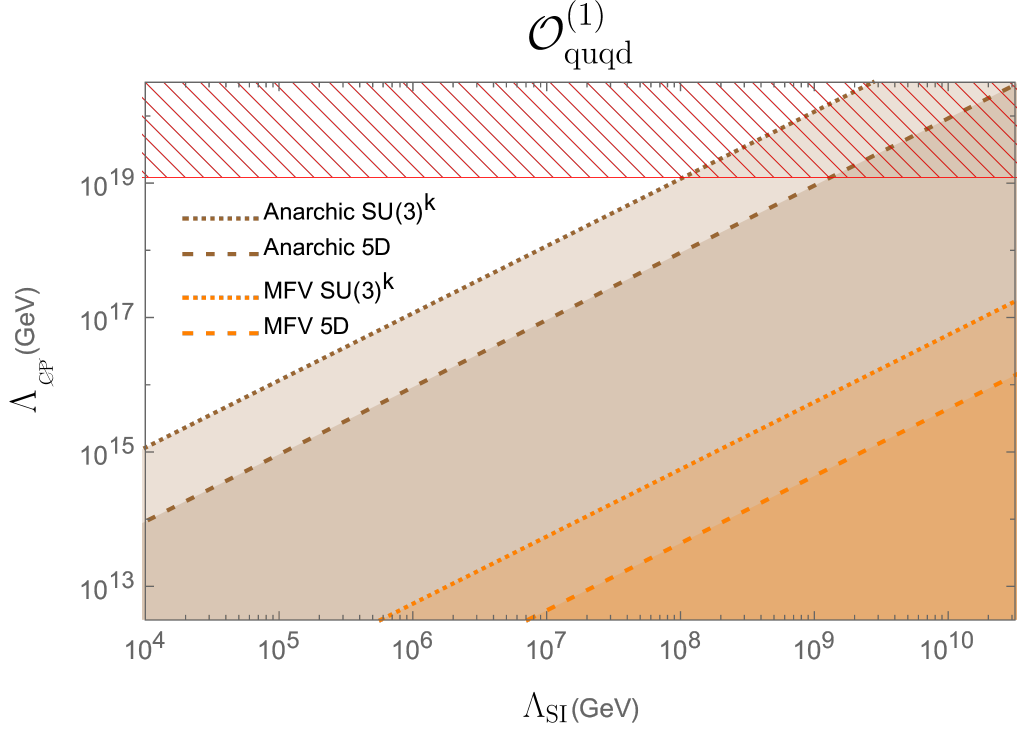}
    \end{subfigure} 
     
    \caption{Limits on the UV scale $ \CPVm $ of the four-quark operator $\cO_{\rm quqd}^{(1)}$ in different flavor scenarios as a function of the small instanton scale $\Lambda_{\rm SI}$. The shaded regions are excluded by the non-observation of the neutron EDM. The striped red region, which corresponds to scales above the Planck mass, is plotted for reference.} 
      \label{fig:LimitsOn4FermiModel}
\end{figure}

The constraints arising from Eqs.~\eqref{eq:4q_theta} and \eqref{eq:4q_theta5D} are shown in Fig.~\ref{fig:LimitsOn4FermiModel}, where for the product group model, we use $b_0=13/2$ since it gives the most stringent constraints. Note that the same value of $b_0$ will be used in constraining the semi-leptonic and gluon dipole operators. In the anarchic flavor scenario, we find that $\CPVm\gtrsim 10^{10}(10^{11})\,\Lambda_{\rm SI}$ for the 5D (product group) model. On the other hand, the MFV scenario provides a much weaker constraint -- $\CPVm\gtrsim 5\times 10^{5}(10^{6})\,\Lambda_{\rm SI}$ for the 5D (product group) model, which differs exactly by a factor of $\sim\sqrt{y{\rm _u}\,y_{\rm d}}$, as indicated by Eq~\eqref{eq:cquqd_Inv_diag}.
This matches the MFV scenario considered in Ref.~\cite{Bedi:2022qrd}, up to an overall factor due to the Higgs doublet structure. In addition, the invariants help us to easily incorporate off-diagonal Yukawa couplings which changes the previous estimate of $\theta_{\rm ind}$ in Ref.~\cite{Bedi:2022qrd} by $\sim 6\%$.

\subsubsection{Semi-leptonic operator}
\label{sec:semileptop}

For the four-fermion operator $\cO_{\rm lequ}^{(1)} $, the susceptibility is given by Eq.~\eqref{eq:susc_lequ_finite}. In the case of the product group model, performing the integral over $\rho$ gives
\begin{align} 
 \theta_{\rm ind}  &= \dfrac{\cI^0_{0000}(C_{\rm lequ}^{(1)})}{(b_0-6) K_\theta}\left[\dfrac{11}{25}+\frac{6}{5}\(\log\(\frac{\CPVm}{\Lambda_{\rm SI}} \) +\gamma_{\rm E}-\log4\pi \)+\frac{3}{5}\Psi\left(\frac{b_0}{2}-3\right) \right]\frac{\Lambda_{\rm SI}^2}{\CPVm^2}\,,
\label{eq:lequ_theta_prod}
\end{align}
where $\Psi(z) $ is the digamma function, while in the 5D instanton model we obtain
\begin{align}
 \theta_{\rm ind}  &= \dfrac{\cI^0_{0000}(C_{\rm lequ}^{(1)})}{8 \pi^2K_\theta}\,\left[\frac{11}{25}+\frac{6}{5}\(\log\(\frac{\CPVm}{\Lambda_{\rm SI}} \) +\gamma_{\rm E}-\log2 \) \right] \frac{\Lambda_{\rm SI}^2}{\CPVm^2}\,.
 \label{eq:lequ_theta_5D}
\end{align}\begin{figure}[h!]
    \centering
    
  \begin{subfigure}[t]{0.9\textwidth}
    	\centering
    	\includegraphics[width=\textwidth]{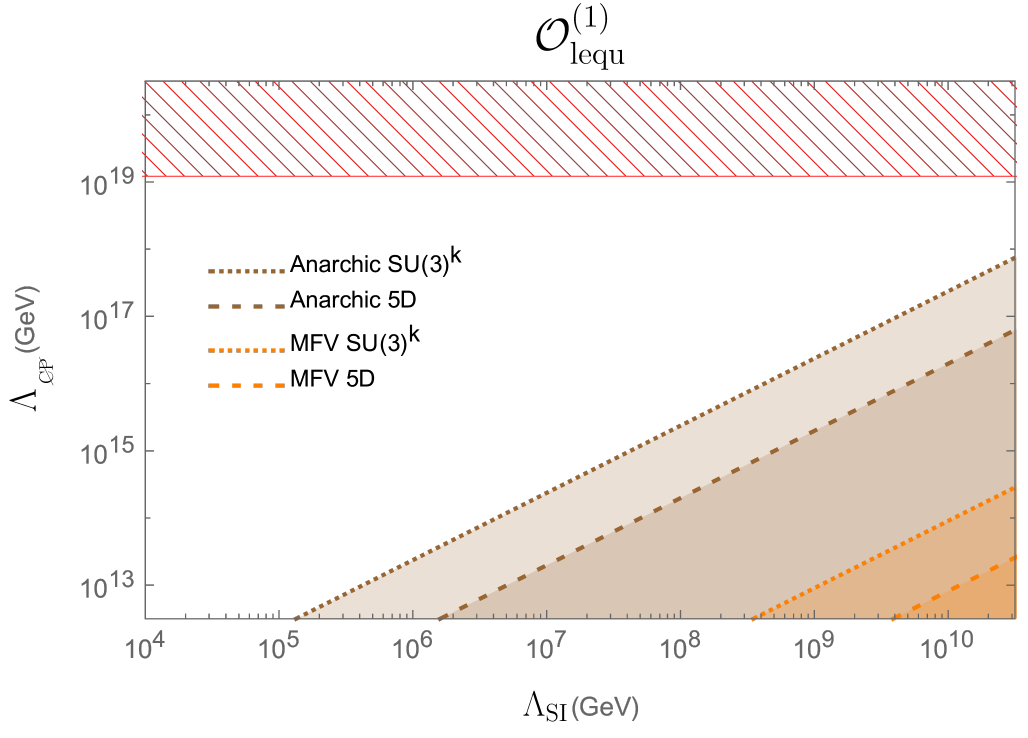}
    \end{subfigure} 
    \caption{Limits on the UV scale $ \CPVm $ of the semi-leptonic operator $\cO_{\rm lequ}^{(1)}$ in different flavor scenarios as a function of the small instanton scale $\Lambda_{\rm SI}$. The shaded regions are excluded by the non-observation of the neutron EDM. The striped red region, which corresponds to scales above the Planck mass, is plotted for reference.} 
      \label{fig:LimitsOnlequModel}
\end{figure}
We present the constraints arising from Eqs.~\eqref{eq:lequ_theta_prod} and \eqref{eq:lequ_theta_5D} in Figure~\ref{fig:LimitsOnlequModel}. As expected, the constraints in this case are much weaker than the four quark operators. For the anarchic and MFV cases, we obtain $\CPVm\gtrsim 10^{6}(10^{7})\,\Lambda_{\rm SI}$ and $\CPVm\gtrsim 5\times 10^{3}(10^{4})\,\Lambda_{\rm SI}$, respectively, for the 5D (product group) model. From Eq.~\eqref{eq:Ilequ}, we see that the largest term in $\cI^0_{0000}(C_{\rm lequ}^{(1)})$ is approximately $\sim K_\theta \, y_\tau/y_{\rm u}$ for the anarchic case ($C_{\rm lequ}^{(1)}\sim \cO(1)$), and  $\sim K_\theta \, y_\tau^2$ for the MFV scenario ($C_{{\rm lequ}, NmLk}^{(1)}\sim Y_{{\rm e}, Nm}Y_{{\rm u}, Lk}$). This results in a difference by a factor of $\sim\sqrt{y_{{\rm u}}\,y_\tau}$ in the two flavor scenarios. In comparison to the result for the four-quark operator, the difference can again be understood in terms of a loop factor and different Yukawa couplings entering the invariants $\cA_{0000}^{0000}(C_{\rm quqd}^{(1)})$, $\cB_{0000}^{0000}(C_{\rm quqd}^{(1)})$ and $\cI^0_{0000}(C_{\rm lequ}^{(1)})$ -- for MFV, there is a relative factor of $\sim \sqrt{y_\tau^2/16\pi^2}\equiv \sqrt{ \lambda^6/16\pi^2}$ whereas for the anarchic case the factor is $\sim \sqrt{y_\tau y_{\rm d}/16\pi^2}$.

\subsubsection{Gluon dipole operator}
\label{sec:gluondipop}
Next, we consider the gluon dipole operator $\cO_{\rm dG} = (\bar Q \sigma^{\mu\nu} T^A d) H\, G_{\mu\nu}^A$. This operator contributes to the topological susceptibility at the same order as $\cO_{\rm quqd}^{(1,8)}$, and has the same functional form in terms of instanton parameters. The flavor structure of this operator is similar to $\cO_{\rm uH}$ presented in Eq.~\eqref{eq:CuHinv}, and the leading order invariant is given by 
\begin{equation}\label{eq:dG_Inv}
    \cI_{0000}(C_{\rm dG}) \equiv \Im \[e^{-i \thetaQCD} 
\epsilon^{IJK}\epsilon^{ijk} Y_{{\rm d},Ii} Y_{{\rm d},Jj} C_{{\rm dG},\,Kk} \det Y_{\rm u} \] \,,
\end{equation}
The computation of the susceptibility for this operator is given in App.~\ref {app:GluonDip} and the result is presented in Eq.~\eqref{eq:cEDM_op}. In this case, the product group model gives the result 
\begin{align}\label{eq:dG_theta_prod}
    \theta_{\rm ind}= \dfrac{\cI_{0000}\(C_{\rm dG} \)}{K_\theta} \dfrac{144\pi^2}{5(b_0-6)}\dfrac{\Lambda_{\rm SI}^2}{\CPVm^2}~,
\end{align} 
where $b_0$ and $|\langle\sigma\rangle|$ are similarly defined as in Eq.~\eqref{eq:4q_theta}. In the case of the 5D instanton model we obtain
\begin{align}\label{eq:dG_theta_5D}
    \theta_{\rm ind}= \dfrac{\cI_{0000}\(C_{\rm dG} \)}{K_\theta} \dfrac{18}{5}
    \dfrac{\Lambda_{\rm SI}^2}{\CPVm^2}
    ~.
\end{align} 

\begin{figure}[h!]
    \centering

    \begin{subfigure}[t]{0.9\textwidth}
    	\centering
    	\includegraphics[width=\textwidth]{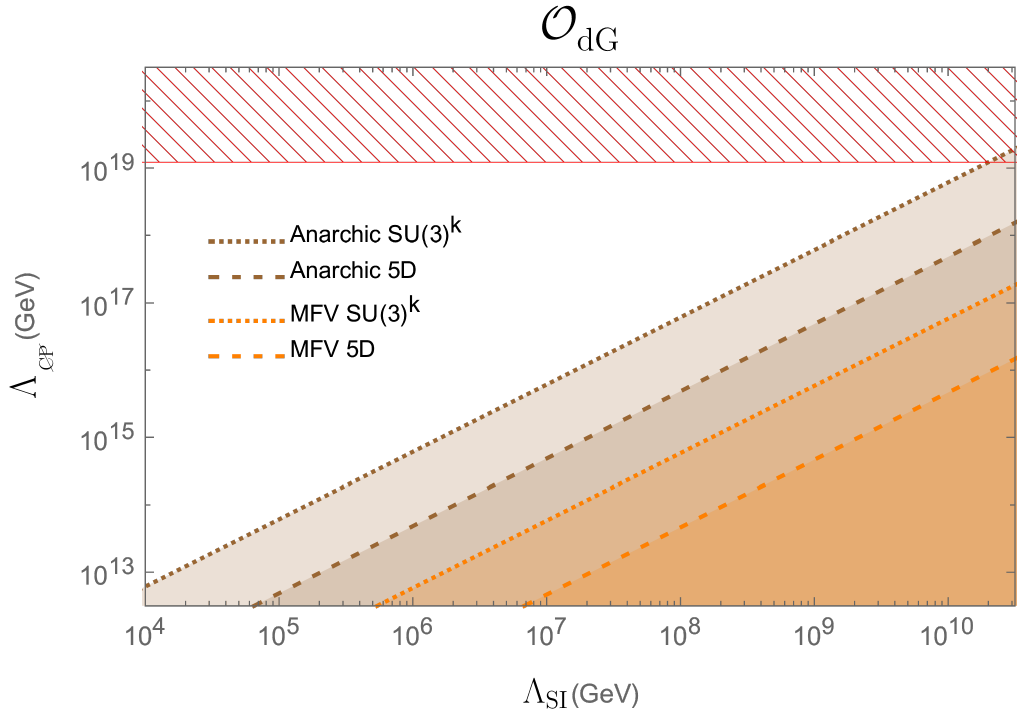}
    \end{subfigure} 
    \caption{Limits on the UV scale $ \CPVm $ of the gluon dipole operator $\cO_{\rm dG}$ in different flavor scenarios as a function of the small instanton scale $\Lambda_{\rm SI}$. The shaded regions are excluded by the non-observation of the neutron EDM. The striped red region, which corresponds to scales above the Planck mass, is plotted for reference.} 
      \label{fig:LimitsOnDipoleModel}
\end{figure}
The constraints coming from $ \theta_{\rm ind}$ in Eqs.~\eqref {eq:dG_theta_prod} and \eqref{eq:dG_theta_5D} are presented in Figure~\ref{fig:LimitsOnDipoleModel}. In the anarchic scenario, the constraint is approximately, $\CPVm\gtrsim 5\times 10^{7}(10^{8})\,\Lambda_{\rm SI}$ for the 5D (product group) model. 
It is worth noting that the constraints arising from both the leading order operators $\cO_{\rm quqd}^{(1)}$ and $\cO_{\rm dG}$ are similar in the MFV scenario, while those from $\cO_{\rm lequ}^{(1)}$ are the weakest among the three, as expected. The anarchic and MFV flavor scenarios are defined by $C_{\rm dG}\sim 1$ and $C_{\rm dG}\sim Y_{\rm d}$, respectively, resulting in bounds differing by a factor of $\sim \sqrt{y_{\rm d}}$ (see Eq.~\eqref{eq:dG_Inv}). This is much less pronounced compared to $\cO_{\rm quqd}^{(1)} $ and $\cO_{\rm lequ}^{(1)} $ which have multiple Yukawas.

Finally, note that for simplicity we have assumed the small-instanton induced $\overline{\theta}$ provides the entire contribution to the neutron EDM. However, in principle, there can be direct contributions to the neutron EDM from the SMEFT operators which should also be taken into account. These contributions from parameters other than $\overline{\theta}$ have been considered in Refs.~\cite{Pospelov:2005pr,Dekens:2013zca,Cirigliano:2016nyn,Alioli:2017ces,deVries:2018mgf,Kley:2021yhn}.

\section{Conclusion}
\label{sec:Conclusions}

The enhanced effect of small (UV) instantons due to new high-energy dynamics that modifies QCD at a UV scale $\Lambda_{\rm SI}$ can be used to increase the QCD axion mass while still solving the strong CP problem~\cite{Holdom:1982ex,Holdom:1985vx,Dine:1986bg,Flynn:1987rs,Agrawal:2017ksf,Csaki:2019vte,Gherghetta:2020keg}. However, in the presence of higher-dimension CP-odd operators at a scale $\CPVm$, an extra contribution to $\bar\theta$ can also be induced from these small instantons which can misalign the axion potential and give rise to a large neutron EDM, leading to stringent constraints on the ratio $\Lambda_{\rm SI}/\CPVm$~\cite{Bedi:2022qrd}. In this paper, we have further explored these contributions for generic UV CP-violating scenarios parameterized by SMEFT operators. In order to estimate the induced shift $\theta_{\rm ind}$, the calculations have been performed in the one-(anti-)-instanton (or dilute instanton gas) approximation, valid when the instanton dynamics can be treated as perturbative. This assumes that the QCD coupling near the scale $\Lambda_{\rm SI}$ is large or semi-perturbative (in order to amplify the effect of small instantons). The calculation is not applicable for non-perturbatively large QCD couplings, where non-perturbative methods must be used.

Given that the instanton calculations involve complicated integrals and are usually considered as estimations~\cite{Csaki:2023ziz}, a more accurate estimation can be obtained by including the effect of flavorful couplings in the theory. However, since physical observable are flavor basis independent, the flavorful couplings should be arranged into  rephasing flavor invariants. In particular, the topological susceptibility from small instantons can be described in terms of SMEFT CP-odd invariants introduced in Ref.~\cite{Bonnefoy:2021tbt}. 
However, the basis of trace invariants presented in Ref.~\cite{Bonnefoy:2021tbt} is not well-suited to characterize the results from the instanton calculation of topological susceptibilities. The results, when projected into the basis of trace invariants, yield complicated linear combinations of the invariants with coefficients that may contain inverse powers of Yukawa couplings, therefore making the task of estimating physical effects with these invariants impractical. Instead, in this work we have proposed a new basis of CP-odd SMEFT invariants, built from determinant-like structures which are much better suited to describe instanton computations. We have explicitly shown and argued that the instanton calculations give results that are directly proportional to elements of our new basis with no extra powers of flavorful couplings.
The flavor invariants derived in this paper therefore allow a more refined estimation of the effects of including CP-violating new physics in small-instanton calculations, complementing the instanton NDA estimates in Ref.~\cite{Csaki:2023ziz}.

Furthermore, the new invariants directly imply selection rules on which kind of operators can appear at the leading order in the instanton calculation since they determine the number of Yukawa couplings and loop factors. We have also shown that rephasing invariant operators cannot contribute only via fermion zero modes which usually give the dominant contribution. For example, we show in the case of the semi-leptonic operator contribution how invariants encapsulate the expected lepton Yukawa dependence and extra loop suppression. Performing the computations in a flavor-invariant fashion also allows us to easily test different flavor assumptions for the SMEFT Wilson coefficients.
For instance, we find that for the four-quark operator $\cO_{\rm quqd}^{(1)}$, the leading order invariant in the MFV scenario 
contains an extra product of the up and down Yukawa couplings (or just the down Yukawa coupling for the gluon dipole operator $\cO_{\rm dG}$) compared to the anarchic scenario.
 Using the experimental bound on $\overline{\theta}$, we obtain constraints on the scale of the higher-dimension CP-violating operators by assuming that the contribution to $\overline{\theta}$ is entirely due to the calculated effect of the small instantons. We also show that for the leading order operators $\cO_{\rm quqd}^{(1)}$ and $\cO_{\rm dG}$ the invariants are approximately $\sim \sin\bar{\theta}$ and result in similar bounds $\CPVm\gtrsim 10^ {6}\,\Lambda_{\rm SI}$ for the MFV scenario.
 However, for the anarchic scenario, the limits are operator dependent and become much more stringent-- for $\cO_{\rm quqd}^{(1)}$, we obtain $\CPVm\gtrsim  10^{11}\,\Lambda_{\rm SI}$ while for $\cO_{\rm dG}$, the bound becomes $\CPVm\gtrsim   10^8\,\Lambda_{\rm SI}$.

The cancellation of divergences appearing in the instanton loop integrals can be used as a non-trivial cross-check of our calculations. The divergences in the correlation functions are canceled by including the counterterms of the SMEFT in a Green's basis. This cancellation has been explicitly shown for the semi-leptonic operator $\cO_{\rm lequ}^{(1)}$ to obtain a divergence-free result that is then used for a phenomenological study.  

Our work can be extended in several directions. The most immediate one is to perform similar calculations by including all relevant SMEFT operators systematically, explicitly verifying the appearance of the constructed invariants. In addition, considering higher orders in the Yukawa couplings would also prove interesting. Another possible extension of our work is to consider the effect of higher-dimensional effective operators (e.g. considering the double insertions of dimension-six SMEFT operators, or a single insertion of the dimension-eight SMEFT operators). Since these operators could allow for different topologies than those of Fig.~\ref{fig:FlowerDiagramsSMEFT}, their effect might not be in general trivially extrapolated from our results. While we estimated  that a higher-dimensional effect would be suppressed, as expected, in the case of the double insertion of $\cO^{(1)}_{\mathrm{quqd}}$ in Appendix \ref{app:Four-fermion}, a more systematic study of higher-dimensional operators could be relevant.

Furthermore, we have solely focused on contributions to the linear term in the axion potential in this work. However, similar computations could be performed to estimate the small instanton effects on the axion mass term using CP-even SMEFT invariants. Having more reliable estimates of observables to probe SMEFT operators can help to better direct experimental searches.

\section*{Acknowledgments}

We thank Quentin Bonnefoy, Pablo Quílez, Jasper Roosmale Nepveu and Maximilian Ruhdorfer for useful discussions.
The work of R.B. and T.G. is supported in part by the Department of Energy under Grant No.~DE-SC0011842 at the University of Minnesota. 
The work of C.G., G.G., J.K. and P.N.H.V. is supported by the Deutsche Forschungsgemeinschaft under Germany’s Excellence Strategy EXC 2121 “Quantum Universe” -- 390833306, as well as by the grant 491245950. This project also has received funding from the European Union’s Horizon Europe research and innovation programme under the Marie Skłodowska-Curie Staff Exchange grant agreement No 101086085 - ASYMMETRY. T.G. and C.G. acknowledge the Aspen Center for Physics, which is supported by National Science Foundation grant PHY-2210452, where this work was initiated.

\appendix
\section{SMEFT conventions}
\label{app:SMEFT}
In this work, we consider the SMEFT up to dimension-six terms and assume dimensionless Wilson coefficients $C_{{\rm a},\,i_1...i_n}$ defined as
\begin{equation}
    \mathcal{L}^{\mathrm{SMEFT}} = \mathcal{L}^{\mathrm{SM}} + \frac{1}{\CPVm^2}\sum_{\rm a} C_{{\rm a},\,{i_1...i_n}} \mathcal{O}_{\rm a}^{i_1...i_n}\,,
\end{equation}
where \CPV is the cutoff scale associated with the CP-violating operator, $\rm a$ labels the type of dimension-six operator and $i_1...i_n$ correspond to the $n$ flavor indices of fermionic operators. Above the electroweak breaking scale, the SM Lagrangian is given by 
\begin{align}
    \mathcal{L}^{\rm SM} &= -\dfrac{1}{4}G_{\mu\nu}^AG^{A,\mu\nu} -\dfrac{1}{4}W_{\mu\nu}^IW^{I,\mu\nu} -\dfrac{1}{4}B_{\mu\nu}B^{\mu\nu}
    + \thetaQCD\dfrac{g^2}{32\pi^2}G^A_{\mu\nu}\tilde{G}^{A,\,\mu\nu}
    \nonumber\\
     & + i\big(\bar{Q}\slashed{D}Q + \bar{u}\slashed{D}u + \bar{d}\slashed{D}d + \bar{L}\slashed{D}L + \bar{e}\slashed{D}e \big) - \big( \bar{Q}Y_{\rm u}\tilde{H}u + \bar{Q}Y_{\rm d} Hd + \bar{L}Y_{\rm e}He +\text{h.c.} \big)
    \nonumber\\
    & + (D^{\mu}H)^{\dagger}(D_{\mu}H) + m_H^2(H^{\dagger}H) - \dfrac{\lambda}{2}(H^{\dagger}H)^2 
    \,,
\end{align}
where $Y_{\rm u,d,e}$ are the Yukawa coupling matrices and $\tilde{H}^i = \epsilon^{ij} H_j^*$. Our sign convention for covariant derivatives is mostly positive, e.g. the covariant derivative acting on a field $\phi$ reads
\begin{equation}
    D_{\mu}\phi = \big( \partial_{\mu} + ig T_C^A G^A_{\mu} + ig_2 T^I W^I_{\mu} + ig_1\mathbf{Y}_{\phi}B_{\mu} \big)\phi
    \,,
\end{equation}
where $g,~g_2,~g_1$ are dimensionless gauge coupling constants and $T_C^A,~T^I$ are the $SU(3)$ and $SU(2)$ generators in the representations of $\phi$ respectively; $\mathbf{Y}_{\phi}$ stands for the hypercharge of $\phi$.   
Here, it is convenient to define the covariant derivatives acting to the field and its Hermitian conjugate,
\begin{equation}
    H^\dag i\overleftrightarrow{D}_{\mu} H \equiv H^\dag (iD_{\mu}H) - (iD_{\mu} H^\dag)H
    \,,\quad
    H^\dag i\overleftrightarrow{D}^I_{\mu} H \equiv H^\dag \tau^I (iD_{\mu}H) - (iD_{\mu} H^\dag)\tau^I H
    \,,
\end{equation}
where $\tau^I$ are the Pauli matrices.
We adopt the Warsaw basis~\cite{Grzadkowski:2010es} conventions for the definitions of the effective operators. The fermionic operators are given for completeness in Tables~\ref{tab:bilinearlist} and \ref{tab:4fermionlist}. We only consider fermionic operators and neglect fully bosonic ones in our analysis because only the former have zero modes projected out resulting in determinant-like structures, as explored in detail in Section~\ref{sec:Instantons}.
\begin{table}[ht!]
\centering
\scalebox{0.6}{
	\renewcommand{\arraystretch}{1.5}
	\resizebox{1\columnwidth}{!}{%
	\begin{tabular}{p{0.9cm}|c|c|c}
		Label & Operator & \# phases & \# primary phases \\ \hline
	
    $\mathcal{O}_{\rm eH}$ & $(H^\dag H)(\bar L_i e_j H)+\text{h.c.}$ & 9 & 3 \\
    $\mathcal{O}_{\rm uH}$ & $(H^\dag H)(\bar Q_i u_j \widetilde H )+\text{h.c.}$ & 9 & 9 \\
    $\mathcal{O}_{\rm dH}$ & $(H^\dag H)(\bar Q_i d_j H)+\text{h.c.}$ & 9 & 9 \\ \hline
    $\mathcal{O}_{\rm eW}$ & $(\bar L_i \sigma^{\mu\nu} e_j) \tau^I H W_{\mu\nu}^I+\text{h.c.}$ & 9 & 3 \\
    $\mathcal{O}_{\rm eB}$ & $(\bar L_i \sigma^{\mu\nu} e_j) H B_{\mu\nu}+\text{h.c.}$ & 9 & 3 \\
    $\mathcal{O}_{\rm uG}$ & $(\bar Q_i \sigma^{\mu\nu} T^A u_j) \widetilde H \, G_{\mu\nu}^A+\text{h.c.}$ & 9 & 9 \\
    $\mathcal{O}_{\rm uW}$ & $(\bar Q_i \sigma^{\mu\nu} u_j) \tau^I \widetilde H \, W_{\mu\nu}^I+\text{h.c.}$ & 9 & 9 \\
    $\mathcal{O}_{\rm uB}$ & $(\bar Q_i \sigma^{\mu\nu} u_j) \widetilde H \, B_{\mu\nu}+\text{h.c.}$ & 9 & 9 \\
    $\mathcal{O}_{\rm dG}$ & $(\bar Q_i \sigma^{\mu\nu} T^A d_j) H\, G_{\mu\nu}^A+\text{h.c.}$ & 9 & 9 \\ 
    $\mathcal{O}_{\rm dW}$ & $(\bar Q_i \sigma^{\mu\nu} d_j) \tau^I H\, W_{\mu\nu}^I+\text{h.c.}$ & 9 & 9 \\ 
    $\mathcal{O}_{\rm dB}$ & $(\bar Q_i \sigma^{\mu\nu} d_j) H\, B_{\mu\nu}+\text{h.c.}$ & 9 & 9 \\ \hline
    $\mathcal{O}_{\rm Hl}^{(1)}$ & $(H^\dag i\overleftrightarrow{D}_\mu H)(\bar L_i \gamma^\mu L_j)$ & 3 & 0 \\ 
    $\mathcal{O}_{\rm Hl}^{(3)}$      & $(H^\dag i\overleftrightarrow{D}^I_\mu H)(\bar L_i \tau^I \gamma^\mu L_j)$ & 3 & 0 \\
    $\mathcal{O}_{\rm H e}$ & $(H^\dag i\overleftrightarrow{D}_\mu H)(\bar e_i \gamma^\mu e_j)$ & 3 & 0 \\
    $\mathcal{O}_{\rm Hq}^{(1)}$ & $(H^\dag i\overleftrightarrow{D}_\mu H)(\bar Q_i \gamma^\mu Q_j)$ & 3 & 3 \\ 
    $\mathcal{O}_{\rm Hq}^{(3)}$ & $(H^\dag i\overleftrightarrow{D}^I_\mu H)(\bar Q_i \tau^I \gamma^\mu Q_j)$&3 & 3 \\ 
    $\mathcal{O}_{\rm H u}$ & $(H^\dag i\overleftrightarrow{D}_\mu H)(\bar u_i \gamma^\mu u_j)$ & 3 & 3 \\
    $\mathcal{O}_{\rm H d}$ & $(H^\dag i\overleftrightarrow{D}_\mu H)(\bar d_i \gamma^\mu d_j)$ & 3 & 3 \\
    $\mathcal{O}_{\rm H u d}$ & $(\widetilde H ^\dag i D_\mu H)(\bar u_i \gamma^\mu d_j)+\text{h.c.}$ & 9 & 9 \\
\end{tabular}%
}
}
\caption{\label{tab:bilinearlist}
Table of bilinear fermionic operators in the Warsaw basis~\cite{Grzadkowski:2010es}. For each operator, we also indicate the number of phases and the number of primary phases~\cite{Bonnefoy:2021tbt}, i.e. the number of flavor-invariant CP-odd quantities capturing the interference with the SM. The lower-case indices $i,j,k,l$ denote flavor indices, while the uppercase indices $I$ and $A$ denote the indices of the adjoint representation of the $SU(2)_L$ and $SU(3)_c$ gauge groups, respectively.}
\end{table}

\begin{table}[ht!]
\centering
\scalebox{0.6}{
	\renewcommand{\arraystretch}{1.5}
	\resizebox{1\columnwidth}{!}{%
	\begin{tabular}{p{0.9cm}|c|c|c}
		Label & Operator & \# phases & \# primary phases \\ \hline
    $\cO_{\rm ll}$ & $(\bar L_i \gamma_\mu L_j)(\bar L_k \gamma^\mu L_l)$ & 18 & 0 \\
    $\cO_{\rm qq}^{(1)}$  & $(\bar Q_i \gamma_\mu Q_j)(\bar Q_k \gamma^\mu Q_l)$ & 18 & 18 \\
    $\cO_{\rm qq}^{(3)}$  & $(\bar Q_i \gamma_\mu \tau^I Q_j)(\bar Q_k \gamma^\mu \tau^I Q_l)$ & 18 & 18 \\
    $\cO_{\rm lq}^{(1)}$ & $(\bar L_i \gamma_\mu L_j)(\bar Q_k \gamma^\mu Q_l)$ & 36 & 9 \\
    $\cO_{\rm lq}^{(3)}$ & $(\bar L_i \gamma_\mu \tau^I L_j)(\bar Q_k \gamma^\mu \tau^I Q_l)$ & 36 & 9 \\ \hline
    $\cO_{\rm ee}$ & $(\bar e_i \gamma_\mu e_j)(\bar e_k \gamma^\mu e_l)$ & 15 & 0 \\
    $\cO_{\rm uu}$ & $(\bar u_i \gamma_\mu u_j)(\bar u_k \gamma^\mu u_l)$ & 18 & 18 \\
    $\cO_{\rm dd}$ & $(\bar d_i \gamma_\mu d_j)(\bar d_k \gamma^\mu d_l)$ & 18 & 18 \\
    $\cO_{\rm eu}$ & $(\bar e_i \gamma_\mu e_j)(\bar u_k \gamma^\mu u_l)$& 36 & 9 \\
    $\cO_{\rm ed}$ & $(\bar e_i \gamma_\mu e_j)(\bar d_k\gamma^\mu d_l)$ & 36 & 9 \\
    $\cO_{\rm ud}^{(1)}$ & $(\bar u_i \gamma_\mu u_j)(\bar d_k \gamma^\mu d_l)$ & 36 & 36 \\
    $\cO_{\rm ud}^{(8)}$ & $(\bar u_i \gamma_\mu T^A u_j)(\bar d_k \gamma^\mu T^A d_l)$ & 36 & 36 \\ \hline
    $\cO_{\rm le}$ & $(\bar L_i \gamma_\mu L_j)(\bar e_k \gamma^\mu e_l)$ & 36 & 3 \\
    $\cO_{\rm lu}$ & $(\bar L_i \gamma_\mu L_j)(\bar u_k \gamma^\mu u_l)$ & 36 & 9 \\
    $\cO_{\rm ld}$ & $(\bar L_i \gamma_\mu L_j)(\bar d_k \gamma^\mu d_l)$ & 36 & 9 \\
    $\cO_{\rm qe}$ & $(\bar Q_i \gamma_\mu Q_j)(\bar e_k \gamma^\mu e_l)$ & 36 & 9 \\
    $\cO_{\rm qu}^{(1)}$ & $(\bar Q_i \gamma_\mu Q_j)(\bar u_k \gamma^\mu u_l)$ & 36 & 36 \\
    $\cO_{\rm qu}^{(8)}$ & $(\bar Q_i \gamma_\mu T^A Q_j)(\bar u_k \gamma^\mu T^A u_l)$ & 36 & 36 \\
    $\cO_{\rm qd}^{(1)}$ & $(\bar Q_i \gamma_\mu Q_j)(\bar d_k \gamma^\mu d_l)$ & 36 & 36 \\
    $\cO_{\rm qd}^{(8)}$ & $(\bar Q_i \gamma_\mu T^A Q_j)(\bar d_k \gamma^\mu T^A d_l)$ & 36 & 36 \\\hline
    $\cO_{\rm ledq}$ & $(\bar L_i^a e_j)(\bar d_k Q_{la})+\text{h.c.}$ & 81 & 27 \\ \hline
    $\cO_{\rm quqd}^{(1)}$ & $(\bar Q_i^a u_j) (\bar Q_k^b d_l)+\text{h.c.}$ & 81 & 81 \\
    $\cO_{\rm quqd}^{(8)}$ & $(\bar Q_i^a T^A u_j) (\bar Q_k^b T^A d_l)+\text{h.c.}$ & 81 & 81 \\
    $\cO_{\rm lequ}^{(1)}$ & $(\bar L_i^a e_j)  (\bar Q_k^b u_l)+\text{h.c.}$ & 81 & 27\\
    $\cO_{\rm lequ}^{(3)}$ & $(\bar L_i^a \sigma_{\mu\nu} e_j)  (\bar Q_s^k \sigma^{\mu\nu} u_t)+\text{h.c.}$ & 81 & 27 \\
\end{tabular}%
}
}
\caption{\label{tab:4fermionlist}
Table of four-fermion operators in the Warsaw basis~\cite{Grzadkowski:2010es}. For each operator, we also indicate the number of phases and the number of primary phases~\cite{Bonnefoy:2021tbt}, i.e. the number of flavor-invariant CP-odd objects capturing the interference with the SM.
}
\end{table}

\section{Details on determinant-like invariants and their relation to trace invariants}
\label{app:invs}

\subsection{Complete set of flavor invariants featuring \texorpdfstring{$\thetaQCD$}{thetaQCD} for all SMEFT operators} \label{app:MoreInvs}

In this Appendix, we present a complete set of flavor invariants, featuring $\thetaQCD$ and are linear in the Wilson coefficient, for all dimension-six SMEFT operators, which were not shown in the main text due to their length. We begin with the operators that are used as examples in the main text. For the operator $\cO_{\rm quqd}^{(1,8)}$, a complete set can be built with the following 81 invariants
\begin{align}
        & \cA^{0000}_{0000}\(C_{\rm quqd}^{(1,8)}\),\, \cA^{0000}_{1000}\(C_{\rm quqd}^{(1,8)}\) ,\, \cA^{1000}_{0000}\(C_{\rm quqd}^{(1,8)}\) ,\,
		\cA^{1000}_{1000}\(C_{\rm quqd}^{(1,8)}\) ,\, \cA^{0000}_{0100}\(C_{\rm quqd}^{(1,8)}\) ,\,  \nonumber \\ 
        & \cA^{0100}_{0000}\(C_{\rm quqd}^{(1,8)}\) ,\, \cA^{0000}_{1100}\(C_{\rm quqd}^{(1,8)}\) ,\, \cA^{0000}_{0110}\(C_{\rm quqd}^{(1,8)}\) ,\, \cA^{0100}_{1000}\(C_{\rm quqd}^{(1,8)}\) ,\, \cA^{1000}_{0100}\(C_{\rm quqd}^{(1,8)}\) ,\,  \nonumber \\ 
        & \cA^{1100}_{0000}\(C_{\rm quqd}^{(1,8)}\) ,\, \cA^{0110}_{0000}\(C_{\rm quqd}^{(1,8)}\) ,\, \cA^{1000}_{1100}\(C_{\rm quqd}^{(1,8)}\) ,\, \cA^{1000}_{0110}\(C_{\rm quqd}^{(1,8)}\) ,\, \cA^{1100}_{1000}\(C_{\rm quqd}^{(1,8)}\) ,\,  \nonumber \\ 
        & \cA^{0100}_{0100}\(C_{\rm quqd}^{(1,8)}\) ,\, \cA^{0100}_{1100}\(C_{\rm quqd}^{(1,8)}\) ,\, \cA^{0100}_{0110}\(C_{\rm quqd}^{(1,8)}\) ,\, \cA^{0110}_{0100}\(C_{\rm quqd}^{(1,8)}\) ,\, \cA^{0000}_{2200}\(C_{\rm quqd}^{(1,8)}\) ,\,  \nonumber \\ 
        & \cA^{0000}_{0220}\(C_{\rm quqd}^{(1,8)}\) ,\, \cA^{0200}_{2000}\(C_{\rm quqd}^{(1,8)}\) ,\, \cA^{1100}_{1100}\(C_{\rm quqd}^{(1,8)}\) ,\, \cA^{1100}_{0110}\(C_{\rm quqd}^{(1,8)}\) ,\, \cA^{2000}_{0200}\(C_{\rm quqd}^{(1,8)}\) ,\,  \nonumber \\ 
        & \cA^{2100}_{0100}\(C_{\rm quqd}^{(1,8)}\) ,\, \cA^{0110}_{1100}\(C_{\rm quqd}^{(1,8)}\) ,\, \cA^{0110}_{0110}\(C_{\rm quqd}^{(1,8)}\) ,\, \cA^{0210}_{1000}\(C_{\rm quqd}^{(1,8)}\) ,\, \cA^{0000}_{1220}\(C_{\rm quqd}^{(1,8)}\) ,\,  \nonumber \\ 
        & \cA^{1200}_{2000}\(C_{\rm quqd}^{(1,8)}\) ,\, \cA^{0000}_{0122}\(C_{\rm quqd}^{(1,8)}\) ,\, \cA^{0100}_{1220}\(C_{\rm quqd}^{(1,8)}\) ,\, 	\cA^{1000}_{0122}\(C_{\rm quqd}^{(1,8)}\) ,\, \cA^{1100}_{2200}\(C_{\rm quqd}^{(1,8)}\) ,\,  \nonumber \\ 
        & \cA^{1100}_{0220}\(C_{\rm quqd}^{(1,8)}\) ,\, \cA^{1200}_{2100}\(C_{\rm quqd}^{(1,8)}\) ,\, \cA^{2100}_{1200}\(C_{\rm quqd}^{(1,8)}\) ,\, \cA^{2100}_{0210}\(C_{\rm quqd}^{(1,8)}\),\, \cA^{2200}_{0110}\(C_{\rm quqd}^{(1,8)}\) ,\, \nonumber \\ 
        & \cA^{0110}_{2200}\(C_{\rm quqd}^{(1,8)}\) ,\, \cA^{0110}_{0220}\(C_{\rm quqd}^{(1,8)}\) ,\, \cA^{0112}_{2000}\(C_{\rm quqd}^{(1,8)}\) ,\, \cA^{1100}_{1220}\(C_{\rm quqd}^{(1,8)}\),\,\cA^{2100}_{0112}\(C_{\rm quqd}^{(1,8)}\) ,\, \\ 
        & \cA^{1200}_{1220}\(C_{\rm quqd}^{(1,8)}\),\,\cA^{2200}_{2200}\(C_{\rm quqd}^{(1,8)}\),\,\cA^{0110}_{1122}\(C_{\rm quqd}^{(1,8)}\) ,\, \cA^{0122}_{2100}\(C_{\rm quqd}^{(1,8)}\),\,\cA^{0220}_{0220}\(C_{\rm quqd}^{(1,8)}\),\,  \nonumber \\ 
        & \cB^{0000}_{0000}\(C_{\rm quqd}^{(1,8)}\) ,\, \cB^{0000}_{0100}\(C_{\rm quqd}^{(1,8)}\),\, \cB^{0000}_{1000}\(C_{\rm quqd}^{(1,8)}\),\, \cB^{0000}_{1100}\(C_{\rm quqd}^{(1,8)}\) ,\, \cB^{0000}_{2200}\(C_{\rm quqd}^{(1,8)}\),\,  \nonumber \\ 
        & \cB^{0000}_{0110}\(C_{\rm quqd}^{(1,8)}\),\,\cB^{0000}_{0122}\(C_{\rm quqd}^{(1,8)}\) ,\, \cB^{0000}_{0220}\(C_{\rm quqd}^{(1,8)}\),\,\cB^{0100}_{0000}\(C_{\rm quqd}^{(1,8)}\),\, \cB^{0100}_{1000}\(C_{\rm quqd}^{(1,8)}\) ,\,  \nonumber \\ 
        & \cB^{0100}_{1100}\(C_{\rm quqd}^{(1,8)}\),\, \cB^{0100}_{2100}\(C_{\rm quqd}^{(1,8)}\),\,\cB^{0100}_{0120}\(C_{\rm quqd}^{(1,8)}\) ,\, \cB^{0100}_{1220}\(C_{\rm quqd}^{(1,8)}\),\,\cB^{0200}_{1120}\(C_{\rm quqd}^{(1,8)}\),\,  \nonumber \\ 
        & \cB^{1000}_{0000}\(C_{\rm quqd}^{(1,8)}\) ,\, \cB^{1000}_{0100}\(C_{\rm quqd}^{(1,8)}\),\,\cB^{1000}_{1200}\(C_{\rm quqd}^{(1,8)}\),\,\cB^{1000}_{0110}\(C_{\rm quqd}^{(1,8)}\) ,\, \cB^{1000}_{0122}\(C_{\rm quqd}^{(1,8)}\) ,\,  \nonumber \\ 
        & \cB^{1000}_{0210}\(C_{\rm quqd}^{(1,8)}\) ,\, \cB^{1100}_{0000}\(C_{\rm quqd}^{(1,8)}\) ,\, \cB^{1100}_{1100}\(C_{\rm quqd}^{(1,8)}\) ,\, \cB^{1100}_{2200}\(C_{\rm quqd}^{(1,8)}\) ,\, \cB^{1100}_{0110}\(C_{\rm quqd}^{(1,8)}\) ,\,  \nonumber \\ 
        & \cB^{1100}_{0220}\(C_{\rm quqd}^{(1,8)}\) ,\, \cB^{1100}_{1122}\(C_{\rm quqd}^{(1,8)}\) ,\, \cB^{1200}_{2100}\(C_{\rm quqd}^{(1,8)}\) ,\, \cB^{2100}_{0122}\(C_{\rm quqd}^{(1,8)}\) ,\, \cB^{2200}_{0000}\(C_{\rm quqd}^{(1,8)}\) ,\,  \nonumber \\ 
        & \cA^{2200}_{1122}\(C_{\rm quqd}^{(1,8)}\) ,\,  \nonumber
\end{align}
where the two structures $\cA(C_{\rm quqd}^{(1,8)})$ and $\cB(C_{\rm quqd}^{(1,8)})$ are defined in Eq.~\eqref{eq:ABquqd}.

We have also used the operator $\cO_{\rm lequ}^{(1,3)}$ in Section~\ref{sec:basis} for which the full list of 27 invariants reads
\begin{equation}
    \begin{split}
         & \cI^0_{0000}\(C_{\rm lequ}^{(1,3)}\),\, \cI^1_{0000}\(C_{\rm lequ}^{(1,3)}\),\, \cI^2_{0000}\(C_{\rm lequ}^{(1,3)}\),\, \cI^0_{1000}\(C_{\rm lequ}^{(1,3)}\),\, \cI^1_{1000}\(C_{\rm lequ}^{(1,3)}\),\\
         & \cI^2_{1000}\(C_{\rm lequ}^{(1,3)}\),\, \cI^0_{0100}\(C_{\rm lequ}^{(1,3)}\) ,\, \cI^1_{0100}\(C_{\rm lequ}^{(1,3)}\) ,\, \cI^2_{0100}\(C_{\rm lequ}^{(1,3)}\) ,\, \cI^0_{1100}\(C_{\rm lequ}^{(1,3)}\), \\
         & \cI^1_{1100}\(C_{\rm lequ}^{(1,3)}\) ,\, \cI^2_{1100}\(C_{\rm lequ}^{(1,3)}\) ,\, \cI^0_{0110}\(C_{\rm lequ}^{(1,3)}\) ,\, \cI^1_{0110}\(C_{\rm lequ}^{(1,3)}\) ,\, \cI^2_{0110}\(C_{\rm lequ}^{(1,3)}\),\\
         & \cI^0_{2200}\(C_{\rm lequ}^{(1,3)}\) ,\, \cI^1_{2200}\(C_{\rm lequ}^{(1,3)}\) ,\, \cI^2_{2200}\(C_{\rm lequ}^{(1,3)}\) ,\, \cI^0_{0220}\(C_{\rm lequ}^{(1,3)}\) ,\, \cI^1_{0220}\(C_{\rm lequ}^{(1,3)}\),\\
         & \cI^2_{0220}\(C_{\rm lequ}^{(1,3)}\) ,\, \cI^0_{1220}\(C_{\rm lequ}^{(1,3)}\) ,\, \cI^1_{1220}\(C_{\rm lequ}^{(1,3)}\) ,\, \cI^2_{1220}\(C_{\rm lequ}^{(1,3)}\) ,\, \cI^0_{0122}\(C_{\rm lequ}^{(1,3)}\),\\
         & \cI^1_{0122}\(C_{\rm lequ}^{(1,3)}\) ,\, \cI^2_{0122}\(C_{\rm lequ}^{(1,3)}\)
    \end{split}
\end{equation}
where $\cI(C_{\rm lequ}^{(1,3)})$ is defined in Eq.~\eqref{eq:Ilequ}.

We next continue with all other invariants in the Warsaw basis of the SMEFT. Only the form of the invariants will be given and we refer to Ref.~\cite{Bonnefoy:2021tbt} for the index assignments that are needed to obtain a complete set of invariants which will be exactly the same as those that are needed for the determinant-like invariants. Consider first the fermion bilinears where we have already defined the invariants for $\cO_{\rm uH}$ in Eq.~\eqref{eq:CuHinv} and the invariant for $\cO_{\rm dH}$ can be defined in a similar way. Furthermore, the dipole operators $\cO_{\rm uB},\cO_{\rm uW},\cO_{\rm uG}$ and $\cO_{\rm dB},\cO_{\rm dW},\cO_{\rm dG}$, fall into the same class of operators and the invariants have exactly the same form as those for $\cO_{\rm uH}$ and $\cO_{\rm dH}$. For $\cO_{\rm eH}$, the form of the invariants is
\begin{equation}
    \cI_a(C_{\rm eH}) \equiv \Im\[ e^{-i\thetaQCD} \Tr\( X_{\rm e}^a C_{\rm eH} \) \det \left(Y_{\rm u} Y_{\rm d}\right) \] \,,
\end{equation}
where the index assignments for $a$, for this operator and all other operators below, are the same as for the trace invariants in Ref.~\cite{Bonnefoy:2021tbt}. $\cO_{\rm eB}$ and $\cO_{\rm eW}$ fall again into the same class of operators.

For the bilinear current-current operators, the Hermitian leptonic operators $\cO_{\rm Hl}^{(1,3)}$ and $\cO_{\rm He}$ do not have any phases interfering with the SM at the leading order and consequently there exist no flavor invariants. For the operators containing quarks we have already defined the invariants for $\cO_{\rm Hq}^{(1,3)}$ in Eq.~\eqref{eq:CHqinv}. The phases introduced by the operators $\cO_{\rm Hu}$ and $\cO_{\rm Hd}$ can be obtained from the invariants of $\cO_{\rm Hq}^{(1,3)}$ by replacing $C_{\rm Hq}^{(1,3)} \to Y_{\rm u}^{\vphantom{\dagger}} C_{\rm Hu}^{\vphantom{\dagger}} Y_{\rm u}^\dagger, \, Y_{\rm d}^{\vphantom{\dagger}} C_{\rm Hd}^{\vphantom{\dagger}} Y_{\rm d}^\dagger$, while the phases in $\cO_{\rm Hud}$ can be obtained from the invariants of $\cO_{\rm uH}$ by replacing $C_{\rm uH}^{\vphantom{\dagger}} \to C_{\rm Hud}^{\vphantom{\dagger}} Y_{\rm d}^\dagger$.

Only the four-fermion operators remain to be treated. Again, for the purely leptonic ``Hermitian'' four-fermion operators $\cO_{\rm ll}$ and $\cO_{\rm ee}$, whose Wilson coefficients satisfy the identity $C_{ijkl}^* = C_{jilk}$, no CP-odd invariants arise at leading order in the Wilson coefficients. For $\cO_{\rm le}$, we can define the following invariants
\begin{equation}
    \cI_a(C_{\rm le}) \equiv \Im\[ e^{-i\thetaQCD} (X_{\rm e}^a)_{ij}^{\vphantom{\dagger}} Y_{{\rm e},kl}^\dagger C_{{\rm le},jlmi}^{\vphantom{\dagger}} Y_{{\rm e},mk}^{\vphantom{\dagger}} \det \left(Y_{\rm u} Y_{\rm d}\right) \] \,.
\end{equation}
The remaining semi-leptonic four-fermion operators can be divided into two classes. In the first class $\cO_{\rm qe}, \cO_{\rm ed}, \cO_{\rm eu}$, can be captured by the following invariant forms
\begin{equation}
\begin{split}
    & \cI_{abcd}^f(C_{\rm qe}) \equiv \Im\[ e^{-i\thetaQCD} (X_{\rm u}^a X_{\rm d}^b X_{\rm u}^c X_{\rm d}^d)_{ij}^{\vphantom{\dagger}} (X_{\rm e}^f Y_{\rm e}^{\vphantom{f}})_{kl}^{\vphantom{\dagger}}  C_{{\rm qe},jilm}^{\vphantom{\dagger}} Y_{{\rm e},mk}^\dagger \det \left(Y_{\rm u} Y_{\rm d}\right) \]\,, \\
    & \cI_{abcd}^f(C_{\rm ed}) \equiv \Im\[ e^{-i\thetaQCD} (X_{\rm e}^f Y_{\rm e}^{\vphantom{f}})_{ij}^{\vphantom{\dagger}} (X_{\rm u}^a X_{\rm d}^b X_{\rm u}^c X_{\rm d}^d Y_{\rm d})_{lm}^{\vphantom{\dagger}}  C_{{\rm ed},jkmn}^{\vphantom{\dagger}} Y_{{\rm e},ki}^\dagger Y_{{\rm d},nl}^\dagger \det \left(Y_{\rm u} Y_{\rm d}\right) \]\,,
\end{split}
\end{equation}
and the $\cO_{\rm eu}$ invariants  are obtained by $Y_{{\rm d},lm}^{\vphantom{\dagger}} C_{{\rm ed},jkmn}^{\vphantom{\dagger}} Y_{{\rm d},nl}^\dagger \to Y_{{\rm u},lm}^{\vphantom{\dagger}} C_{{\rm eu},jkmn}^{\vphantom{\dagger}} Y_{{\rm u},nl}^\dagger$.

\noindent The second class $\cO_{\rm lq}^{(1,3)}, \cO_{\rm ld}^{\vphantom{\dagger}}, \cO_{\rm lu}^{\vphantom{\dagger}}$ is captured by the invariants of the following form
\begin{equation}
\begin{split}
    & \cI_{abcd}^f(C_{\rm lq}^{(1,3)}) \equiv \Im\[ e^{-i\thetaQCD} (X_{\rm e}^f)_{ij}^{\vphantom{\dagger}} (X_{\rm u}^a X_{\rm d}^b X_{\rm u}^c X_{\rm d}^d)_{kl}^{\vphantom{\dagger}}  C_{{\rm lq},jilk}^{(1,3)} \det \left(Y_{\rm u} Y_{\rm d}\right) \]\,, \\
    & \cI_{abcd}^f(C_{\rm ld}) \equiv \Im\[ e^{-i\thetaQCD} (X_{\rm e}^f)_{ij}^{\vphantom{\dagger}} (X_{\rm u}^a X_{\rm d}^b X_{\rm u}^c X_{\rm d}^d Y_{\rm d})_{kl}^{\vphantom{\dagger}}  C_{{\rm ld},jilm}^{\vphantom{\dagger}}  Y_{{\rm d},mk}^\dagger \det \left(Y_{\rm u} Y_{\rm d}\right) \]\,,
\end{split}
\end{equation}
and the $\cO_{\rm lu}$ invariants  are obtained by $Y_{{\rm d},kl}^{\vphantom{\dagger}}  C_{{\rm ld},jilm}^{\vphantom{\dagger}}  Y_{{\rm d},mk}^\dagger \to Y_{{\rm u},kl}^{\vphantom{\dagger}} C_{{\rm lu},jilm}^{\vphantom{\dagger}} Y_{{\rm u},mk}^\dagger$. 

\noindent The phases introduced by the operator $\cO_{\rm ledq}$ are captured by the following invariant 
\begin{equation}
    \cI_{abcd}^f(C_{\rm ledq}) \equiv \Im\[ e^{-i\thetaQCD} (Y_{\rm e}^\dagger X_{\rm e}^f)_{ij}^{\vphantom{\dagger}} (Y_{\rm d} X_{\rm u}^a X_{\rm d}^b X_{\rm u}^c X_{\rm d}^d)_{kl}^{\vphantom{\dagger}}  C_{{\rm ledq},jilk}^{(1,3)} \det \left(Y_{\rm u} Y_{\rm d}\right) \]\,.
\end{equation}

The operators $\cO_{\rm qq}^{(1,3)}, \cO_{\rm uu}^{\vphantom{\dagger}}, \cO_{\rm dd}^{\vphantom{\dagger}}$ can all be described by invariants of the form
\begin{equation}
\begin{split}
    & \cA_{a_1b_1c_1d_1}^{a_2b_2c_2d_2}(C_{\rm qq}^{(1,3)}) \equiv \Im\[ e^{-i\thetaQCD} (X_{\rm u}^{a_1} X_{\rm d}^{b_1} X_{\rm u}^{c_1} X_{\rm d}^{d_1})_{ij}^{\vphantom{\dagger}} (X_{\rm u}^{a_2} X_{\rm d}^{b_2} X_{\rm u}^{c_2} X_{\rm d}^{d_2})_{kl}^{\vphantom{\dagger}}  C_{{\rm qq},jilk}^{(1,3)} \det \left(Y_{\rm u} Y_{\rm d}\right) \] \,, \\
    & \cB_{a_1b_1c_1d_1}^{a_2b_2c_2d_2}(C_{\rm qq}^{(1,3)}) \equiv \Im\[ e^{-i\thetaQCD} (X_{\rm u}^{a_1} X_{\rm d}^{b_1} X_{\rm u}^{c_1} X_{\rm d}^{d_1})_{ij}^{\vphantom{\dagger}} (X_{\rm u}^{a_2} X_{\rm d}^{b_2} X_{\rm u}^{c_2} X_{\rm d}^{d_2})_{kl}^{\vphantom{\dagger}}  C_{{\rm qq},lijk}^{(1,3)} \det \left(Y_{\rm u} Y_{\rm d}\right) \] \,,
\end{split}
\end{equation}
where the following replacements have to be made for the latter two operators $C_{{\rm qq},ijkl}^{(1,3)} \to Y_{{\rm u},im}^{\vphantom{\dagger}} Y_{{\rm u},nj}^\dagger C_{\vphantom{d}{\rm uu},mnop}^{\vphantom{\dagger}} Y_{{\rm u},ko}^{\vphantom{\dagger}} Y_{{\rm u},pl}^\dagger \,, Y_{{\rm d},im}^{\vphantom{\dagger}} Y_{{\rm d},nj}^\dagger C_{{\rm dd},mnop}^{\vphantom{\dagger}} Y_{{\rm d},ko}^{\vphantom{\dagger}} Y_{{\rm d},pl}^\dagger$.\footnote{Note, that we have chosen the same form for all invariants here while Ref.~\cite{Bonnefoy:2021tbt} chooses a form with fewer insertions of Yukawa couplings, whenever no insertions of $X_{u,d}$ are made in one of the bilinears of right-handed quarks. We chose to do this here to present the invariants in a more compact form. We will also do this for the following four-fermion operators containing only quark fields.}

\noindent For the operators $\cO_{\vphantom{d}\rm qu}^{(1,8)}, \cO_{\rm qd}^{(1,8)}$, a complete set of invariants can be found by considering the following forms of invariants
\begin{equation}
\begin{split}
    \cA_{a_1b_1c_1d_1}^{a_2b_2c_2d_2}(C_{\rm qu}^{(1,8)}) \equiv & ~\Im\[ e^{-i\thetaQCD} (X_{\rm u}^{a_1} X_{\rm d}^{b_1} X_{\rm u}^{c_1} X_{\rm d}^{d_1})_{ij}^{\vphantom{\dagger}} (X_{\rm u}^{a_2} X_{\rm d}^{b_2} X_{\rm u}^{c_2} X_{\rm d}^{d_2})_{kl}^{\vphantom{\dagger}} \right. \\
    & \qquad \left. \times Y_{{\rm u},lm}^{\vphantom{\dagger}} C_{{\rm qu},jimn}^{(1,8)} Y_{{\rm u},nk}^\dagger \det \left(Y_{\rm u} Y_{\rm d} \right) \] \,, \\
    \cB_{a_1b_1c_1d_1}^{a_2b_2c_2d_2}(C_{\rm qu}^{(1,8)}) \equiv & ~\Im\[ e^{-i\thetaQCD} (X_{\rm u}^{a_1} X_{\rm d}^{b_1} X_{\rm u}^{c_1} X_{\rm d}^{d_1})_{ij}^{\vphantom{\dagger}} (X_{\rm u}^{a_2} X_{\rm d}^{b_2} X_{\rm u}^{c_2} X_{\rm d}^{d_2})_{kl}^{\vphantom{\dagger}} \right. \\
    & \qquad \left. \times Y_{{\rm u},lm}^{\vphantom{\dagger}} C_{{\rm qu},mijn}^{(1,8)} Y_{{\rm u},nk}^\dagger \det \left( Y_{\rm u} Y_{\rm d}\right) \] \,,
\end{split}
\end{equation}
where for $\cO_{\rm qd}^{(1,8)}$ the replacement $Y_{{\rm u},lm}^{\vphantom{\dagger}} C_{{\rm qu},mijn}^{(1,8)} Y_{{\rm u},nk}^\dagger \to Y_{{\rm d},lm}^{\vphantom{\dagger}} C_{{\rm qd},mijn}^{(1,8)} Y_{{\rm d},nk}^\dagger$ has to be made.

\noindent Finally, the CP violation introduced by the operator $\cO_{\rm ud}^{(1,8)}$ can be captured 
by invariants of the form
\begin{equation}
\begin{split}
    \cA_{a_1b_1c_1d_1}^{a_2b_2c_2d_2}(C_{\rm ud}^{(1,8)}) \equiv & ~\Im\[ e^{-i\thetaQCD} (X_{\rm u}^{a_1} X_{\rm d}^{b_1} X_{\rm u}^{c_1} X_{\rm d}^{d_1})_{ij}^{\vphantom{\dagger}} (X_{\rm u}^{a_2} X_{\rm d}^{b_2} X_{\rm u}^{c_2} X_{\rm d}^{d_2})_{kl}^{\vphantom{\dagger}} \right.\\
    & \qquad \left. \times Y_{{\rm u},jm}^{\vphantom{\dagger}} Y_{{\rm d},ln}^{\vphantom{\dagger}} C_{{\rm ud},monp}^{(1,8)} Y_{{\rm u},oi}^\dagger Y_{{\rm d},pk}^\dagger \det \left(Y_{\rm u} Y_{\rm d}\right) \] \,, \\
    \cB_{a_1b_1c_1d_1}^{a_2b_2c_2d_2}(C_{\rm ud}^{(1,8)}) \equiv & ~\Im\[ e^{-i\thetaQCD} (X_{\rm u}^{a_1} X_{\rm d}^{b_1} X_{\rm u}^{c_1} X_{\rm d}^{d_1})_{ij}^{\vphantom{\dagger}} (X_{\rm u}^{a_2} X_{\rm d}^{b_2} X_{\rm u}^{c_2} X_{\rm d}^{d_2})_{kl}^{\vphantom{\dagger}} \right.\\
    & \qquad \left. \times Y_{{\rm u},jm}^{\vphantom{\dagger}} Y_{{\rm d},ln}^{\vphantom{\dagger}} C_{{\rm ud},nomp}^{(1,8)} Y_{{\rm u},oi}^\dagger Y_{{\rm d},pk}^\dagger \det \left(Y_{\rm u} Y_{\rm d}\right) \] \,.
\end{split}
\end{equation}

\subsection{Conversion of invariants with negative powers of Yukawa couplings} \label{app:RelInvYuk}
In Section~\ref{sec:basis}, all determinant-like invariants were related to the basis of trace invariants of Ref.~\cite{Bonnefoy:2021tbt}, except for invariants of the form $\cI_{0bcd}(C_{\rm uH})$ which are mapped to the trace invariants $L(C)$ and $R(C)$ defined below Eq.~\eqref{eq:relCHq}
\begin{equation} \label{eq:CuH0000wrongBasis}
         \cI_{0bcd}(C_{\rm uH})=2\,\left( J_{\theta}\,R_{(-1)bcd}(C_{\rm uH}^{\vphantom{\dagger}}Y_{\vphantom{H}\rm u}^\dagger)+K_{\theta}\,L_{(-1)bcd}(C_{\rm uH}^{\vphantom{\dagger}}Y_{\vphantom{H}\rm u}^\dagger)\right)\,,
\end{equation}
that contain inverse powers of Yukawa couplings, which is clearly not in the basis of Ref.~\cite{Bonnefoy:2021tbt}. To convert the right-hand side of Eq.~\eqref{eq:CuH0000wrongBasis} to the basis we will use the following relation
\begin{equation}
    A^{-1} = \frac{1}{\det A} \[ A^2 - A \, \Tr\,A +  \frac{1}{2} \( (\Tr A)^2 - \Tr(A^2) \) \mathbb{1} \] \,,\label{eq:Cayley_Hamilton_3x3}
\end{equation}
which directly follows from the Cayley--Hamilton theorem. Making use of this identity, we can write
\begin{equation} \label{eq:MapInverseYuk}
\begin{split}
    L_{(-1)bcd}(C_{\rm uH}^{\vphantom{\dagger}}Y_{\vphantom{H}\rm u}^\dagger) & = \Im\,\Tr \( X_{\rm u}^{-1} X_{\rm d}^b X_{\rm u}^c X_{\rm d}^d C_{\rm uH}^{\vphantom{\dagger}} Y_{\vphantom{H}\rm u}^\dagger \)\\ 
    &= \frac{1}{\det X_{\rm u}} \left( L_{2bcd}(C_{\rm uH}^{\vphantom{\dagger}}Y_{\vphantom{H}\rm u}^\dagger) - \Tr (X_{\rm u}) \, L_{1bcd}(C_{\rm uH}^{\vphantom{\dagger}}Y_{\vphantom{H}\rm u}^\dagger) \right. \\
    & \qquad\left. +  \frac{1}{2} \( (\Tr X_{\rm u})^2 - \Tr (X_{\rm u}^2)\) L_{0bcd}(C_{\rm uH}^{\vphantom{\dagger}}Y_{\vphantom{H}\rm u}^\dagger)\)\,,
\end{split}
\end{equation}
given that $\det X_{\rm u} \neq 0$. Repeating the same for $R_{(-1)000}(C_{\rm uH}^{\vphantom{\dagger}}Y_{\vphantom{H}\rm u}^\dagger)$ enables us to fully map all determinant-like invariants of $C_{\rm uH}$ to the trace invariants of Ref.~\cite{Bonnefoy:2021tbt}. The same procedure can be followed for all other operators with chirality flipping currents where the same problem occurs. In some cases further syzygies have to be imposed in order to map the invariants appearing on the right-hand side of Eq.~\eqref{eq:MapInverseYuk} to the basis of Ref.~\cite{Bonnefoy:2021tbt}. For instance, the invariant $\cI_{0100}(C_{\rm uH})$ is mapped to the CP-odd trace invariants $L_{2100}(C_{\rm uH}^{\vphantom{\dagger}}Y_{\vphantom{H}\rm u}^\dagger)$, $L_{1100}(C_{\rm uH}^{\vphantom{\dagger}}Y_{\vphantom{H}\rm u}^\dagger)$ and $L_{0100}(C_{\rm uH}^{\vphantom{\dagger}}Y_{\vphantom{H}\rm u}^\dagger)$, out of which only the last two appear in the basis of Ref.~\cite{Bonnefoy:2021tbt} without further manipulations. Therefore, more syzygies have to be applied to the invariant $L_{2100}(C_{\rm uH}^{\vphantom{\dagger}}Y_{\vphantom{H}\rm u}^\dagger)$ in order to reduce it to the basis in Ref.~\cite{Bonnefoy:2021tbt}.

\section{Basics of instanton calculations}
\label{app:basics}

In this Appendix, we briefly give an overview of instantons and their calculus. There is a vast literature on instantons, and more details can be found in the standard lectures and recent reviews of this topic such as Refs.~\cite{Coleman:1978ae,Vainshtein:1981wh,Shifman:2022shi,Vandoren:2008xg,Dorey:2002ik,Tong:2005un,Reece:2023czb,Csaki:2019vte,Csaki:2023ziz}.

\subsection{Instanton calculations: technical preliminaries}
\label{sec: Instanton: review}

The presence of instantons is necessary if the vacuum of the theory considered is degenerate in the space of fields. Within the semi-classical approximation, instantons describe the tunneling effects that connect the two distinct energy-degenerate states in the space of fields. They are localized objects in Euclidean spacetime, satisfying the Euclidean equation of motion with non-trivial topologies and therefore minimize the Euclidean action.

The instanton solutions for pure Yang--Mills theories were first discovered in 1975 by Belavin, Polyakov, Schwarz and Tyupkin (usually refered as BPST instantons)~\cite{Belavin:1975fg}. These solutions play a primary role in revealing the non-trivial vacuum structure of Yang--Mills theories, i.e. the existence of the $\theta$ vacuum, as a superposition of the so-called ``$n-$vacua'' which are degenerate but topologically distinct and characterized by the \textit{winding number} $n$ of the gauge field at infinity~\cite{tHooft:1986ooh}.

We begin with the pure Yang--Mills part of the QCD Lagrangian, including the contribution of the vacuum angle $\thetaQCD$. From the Lagrangian formalism, we then write down the Euclidean action of this theory.\footnote{When switching from Minkowski to Euclidean space, the field strength tensor components are related by $(G^A_{ij})_{M}=(G^A_{ij})_{E}$, $(G^A_{0j})_{M}=i(G^A_{4j})_{E}$. For the Euclidean path integral, each trajectory is weighted by the factor $e^{-S_{\text{YM}}}$.}
\begin{align}
    S_{\rm YM} = \int d^4x \left( \dfrac{1}{4}G_{\mu\nu}^A G^{A,\,\mu\nu} + i\thetaQCD \dfrac{g^2}{32\pi^2}G^A_{\mu\nu} \tilde{G}^{A,\,\mu\nu} \right)
    \,,
    \label{Definition: Lag-QCD-YM}
\end{align}
where $G_{\mu\nu}=G^A_{\mu\nu}T^A$ with the gluon field strength tensor given by
\begin{align}
	G^A_{\mu\nu} = \partial_{\mu}G^A_{\nu} - \partial_{\nu}G^A_{\mu} - gf^{ABC}G^B_{\mu}G^C_{\nu}
	\,.
\end{align}
Here, $A=1,\dots,8$ are gauge indices, $T^A$ and $f^{ABC}$ are $SU(3)$ generators and structure constant, respectively. The QCD gauge coupling is $g$. Our convention for the dual field strength tensor is $G^A_{\mu\nu}=\frac{1}{2}\epsilon_{\mu\nu\rho\sigma}G^{A,\rho\sigma}$, with the choice $\epsilon^{0123}=+1$. 
We define the topological charge as
\begin{align}
    \dfrac{g^2}{32\pi^2} \int d^4x  \, G^A_{\mu\nu}\tilde{G}^{A,\,\mu\nu}(x) \bigg|_{\text{inst.}} = Q
     \,,
     ~ \text{where}~ Q \in \mathbb{Z}
     \,.
     \label{Definition: Topo-charge}
\end{align}
Within the perturbative regime, we can fix the topological charge $Q=\pm 1$, because these configurations will minimize the Euclidean action and dominate over all path integral trajectories.\footnote{In the non-perturbative regime, all topological configurations contribute to the path integral equally. Thus, one needs to consider multi-instanton solutions and the interactions between (anti-)instantons~\cite{Shuryak:1981ff,Shuryak:1982dp,Diakonov:1983hh,Diakonov:1984vw,Schafer:1995pz,Schafer:2002af}.
}
Later on, the subscript $|_{\text{inst.}}$ will be replaced by $|_{\text{1-(a)-inst.}}$ for the background with the one-(anti)-instanton solution.

For the $Q=+1$ configuration, using the regular Landau gauge, an explicit form of the BPST instanton solutions for the $SU(2)$ gauge theory is given by~\cite{Belavin:1975fg},
\begin{align}
    G^a_{\mu}(x)\big|_{\rm 1-inst.} = 2\,\eta_{a\mu\nu} \dfrac{(x-x_0)_{\nu}}{(x-x_0)^2 + \rho^2}  
    \,,\quad
    \eta_{a\mu\nu} = 
    \begin{cases}
        \epsilon_{a\mu\nu}, \,& \mu,\nu \in \{1,2,3\}
        \\
        -\delta_{a\nu}, \,& \mu = 0
        \\
        +\delta_{a\mu}, \,& \nu = 0
        \\
        0, \,& \mu=\nu = 0
    \end{cases}
    \,,
    \label{Definition: BPST instantons}
\end{align}
where $a = 1,2,3$ label the $SU(2)$ gauge indices, $\mu,\nu$ are the Euclidean spacetime indices and $\eta_{a\mu\nu}$ is the group-theoretic 't~Hooft $\eta$ symbol defined in Eq.~\eqref{Definition: BPST instantons}. The relations and index contractions of the $\eta$ symbols can be found in Ref.~\cite{tHooft:1976snw}. The instanton solution in Eq.~\eqref{Definition: BPST instantons} depends on five parameters, the Euclidean four-vector $x_0^{\mu}$ and $\rho$ which describes the instanton location and size, respectively.\footnote{For an $SU(2)$ theory, $G_{\mu}=G^a_{\mu}T^a$ there are an additional three gauge parameters, for a total of eight parameters.
Different values of these parameters just lead to equivalent instanton solutions. In the language of soliton physics, these parameters and the family of equivalent solutions given by Eq.~\eqref{Definition: BPST instantons} are referred to as collective coordinates and zero modes, respectively.}

The anti-instanton solution has exactly the same form illustrated by Eq.~\eqref{Definition: BPST instantons}, with the replacement $\eta_{a\mu\nu} \rightarrow \bar{\eta}_{a\mu\nu}$, where the symbols $ \bar{\eta}_{a\mu\nu}$  are defined by the modification $\delta_{a\mu},\,\delta_{a\nu}\to -\delta_{a\mu},\,\delta_{a\nu}$ in Eq.~\eqref{Definition: BPST instantons}. 
Since we usually work with field strength tensors instead of gauge fields, it is convenient to give 
an explicit form of the field strength tensor in the presence of a one-instanton background,
\begin{align}
    G_{\mu\nu}^a \big|_{\text{1-inst.}} = -4\,\eta_{a\mu\nu}\dfrac{\rho^2}{[(x-x_0)^2+\rho^2]^2}
    \,.
    \label{Definition: G-1-instanton}
\end{align}
Furthermore, note that the instanton solutions for the $SU(N)$ gauge theory can be obtained by embedding the $SU(2)$ solutions into  $SU(N)$. Therefore, in this work, when contracting the gauge index of $G^A_{\mu\nu}\big|_{\text{1-inst.}}$ with $T^A$ or $f^{ABC}$, only $A,B,C \in \{1,2,3\}$ yields non-vanishing results.

An important property of the one-(anti-)instanton solution is that it satisfies the \textit{(anti-) self-dual} equation
\begin{equation}
	G^a_{\mu\nu} = \pm \,\tilde{G}^a_{\mu\nu}
	\,,
	\label{Definition: self-dual-eq}
\end{equation}
and thus, due to the Bianchi identity, automatically solves the gluon equation of motion $D^{\mu}G^a_{\mu\nu} = D^{\mu}\tilde{G}^a_{\mu\nu} = 0$. With all of these properties, the one-(anti)-instanton solution then yields the finite QCD classical action
\begin{align}
    S_{\rm YM}^{\text{1-inst.}} = \int d^4x \left( \dfrac{1}{4}G_{\mu\nu}^A G^{A,\,\mu\nu} + i\thetaQCD\dfrac{g^2}{32\pi^2}G_{\mu\nu}^A \tilde{G}^{A,\,\mu\nu} \right)\bigg|_{\text{1-(a.-)inst.}}
    = \dfrac{8\pi^2}{g^2} \pm i\thetaQCD
    \label{E-action: QCD-YM-1-inst}
    \,.
\end{align}

\paragraph{Fermion zero modes}
Next, we consider the $SU(N)$ gauge theory with massless fermions in the presence of an instanton background. The fermionic Euclidean action is given by
\begin{align}
	S_{\psi}= \int d^4x \, \bar{\psi}_f \big(-i\slashed{D} \big) \psi_f
	\,,
	\label{E-action: massless-fermion}
\end{align}
where $D_{\mu}=\partial_{\mu} + igG^a_{\mu}T^a$, and $f$ is fermion flavor index. The spectrum of the Dirac operator can be obtained by expanding the fermion fields into their eigenmodes,
\begin{align}
	\psi_f(x) = \sum_k \xi_f^{(k)}\psi^{(k)}(x)
	\,;\quad
	\bar{\psi}_f(x) = \sum_k \bar{\xi}_f^{(k)}\bar{\psi}^{(k)}(x)
	\,,
	\label{Inst-fermion: eigenmodes}
\end{align}
where $\xi_f^{(k)}$ and $\bar{\xi}_f^{(k)}$ are Grassmann variables. The crucial point is that the interaction of fermion with the instanton background leads to the so-called fermion zero modes which satisfy the massless Dirac equation, $-i\slashed{D}\big|_{\text{1-inst.}}\psi^{(0)}(x)=0$. 
For the $SU(2)$ gauge theory, an explicit form of $\psi^{(0)}(x)$ in the regular Landau gauge is given by~\cite{Vainshtein:1981wh, Shifman:2022shi}
\begin{equation} 
    \label{Inst-fermion: zero-modes}
    \psi^{(0)}(x)\bigg|_{\text{1-inst.}} = \begin{pmatrix} \chi_L \\ \chi_R \end{pmatrix}
    = \dfrac{1}{\pi}\dfrac{\rho}{\big[ (x-x_0)^2 + \rho^2 \big]^{3/2}} \begin{pmatrix} 0 \\ \varphi \end{pmatrix} \,, \quad \varphi_{\alpha m} = \epsilon^{\alpha m} \,,
\end{equation}
where $\alpha = 1,2$ and $m=1,2$ are spinor and $SU(2)$ gauge indices and $\epsilon$ is the anti-symmetric tensor in two dimensions. Here, the fermion zero modes given by Eq.~\eqref{Inst-fermion: zero-modes} are normalized by imposing the condition $\int d^4x \, \big[\psi^{(0)^{\dagger}}\psi^{(0)}\big](x) = 1$.
In the presence of an instanton background, only $\chi_L^{\dagger}, \chi_R$ Weyl components possess the zero mode solutions of the Dirac equations and vice versa for the anti-instanton background. Also, notice that the zero modes are independent of the flavor of the respective fermion.

\paragraph{Vacuum-to-vacuum amplitude} 
Within the semi-classical approximation, one can expand the fields around their classical configuration in the presence of the one-instanton background, up to quadratic order in the quantum fluctuations, the Euclidean action reads  
\begin{align}
S_E = S_{\rm YM}^{\text{1-inst.}} + \int d^4x \sum_i \delta\Phi^{\dagger}_i \big( M_{\Phi_i \Phi_i} \big) \delta\Phi_i
\,,
	\label{E-action: full-action}
\end{align}
where in QCD, $S_{\rm YM}^{\text{1-inst.}}$ is given by Eq.~\eqref{E-action: QCD-YM-1-inst}, $\delta\Phi$ encapsulates all quantum fluctuations of the gauge $A_{\mu}$, ghost $\eta$, scalar $\phi$ and fermion $\psi$ fields.\footnote{The fluctuations $\delta\Phi$ include both zero modes and non-zero modes.} We are interested in the vacuum-to-vacuum amplitude, and, following Refs.~\cite{tHooft:1976snw, Csaki:2019vte}, we express this amplitude in terms of path integral as follows,
\begin{align}
\braket{0}{0}\big|_{\text{1-inst.}} = 
\dfrac{
\left. \int\mathcal{D}A_{\mu} \mathcal{D}\eta \mathcal{D}\bar{\eta} \mathcal{D}\phi \mathcal{D}\phi^{\dagger} \mathcal{D}\psi\mathcal{D}\bar{\psi} \, e^{-S_E} \right|_{\text{1-inst.}}
}
{
\left. \int\mathcal{D}A_{\mu} \mathcal{D}\eta \mathcal{D}\bar{\eta} \mathcal{D}\phi \mathcal{D}\phi^{\dagger} \mathcal{D}\psi\mathcal{D}\bar{\psi} \, e^{-S_E} \right|_{A^{\rm cl}_{\mu}=0}
}
\,.
	\label{Definition: vac-to-vac-amp}
\end{align}
The computation of this amplitude requires a lot of effort. First, one has to split the path integral measure into an integration over zero modes and non-zero modes. The path integral over zero modes can be replaced by an integration over collective coordinates and the corresponding Jacobian must be computed properly. Second, the non-zero modes need to be integrated out, their contributions can be viewed as the product of the infinite non-vanishing eigenvalues of the operator $M_{\Phi_i \Phi_i}$. This product is divergent due to many large eigenvalues and needs to be regularized. The final result that will be used in our calculations is given by~\cite{tHooft:1976snw}
\begin{align}
\braket{0}{0}\big|_{\text{1-inst.}} &= e^{-i\thetaQCD} \int d^4x_0 \int\dfrac{d\rho}{\rho^5} \,  d_N(\rho) \int \prod_{f=1}^{N_f}\left( \rho \, d\xi^{(0)}_f d\bar{\xi}^{(0)}_f \right) \, e^{\int d^4x \, (-\bar{\psi} J \psi + \rm{h.c.})}
\,,
	\label{Definition: 'tHooft-result}
\end{align} 
where $J$ is a source term describing the interaction between fermions (charged under the instanton gauge groups) and other quantum fields (unrelated to instanton dynamics). An important quantity appearing in Eq.~\eqref{Definition: 'tHooft-result} is the instanton density in $SU(N)$ theory,
\begin{align}
        d_N(\rho) &= C[N] \left( \dfrac{8\pi^2}{g^2} \right)^{2N} e^{-8\pi^2/g^2(1/\rho)}
        \,.
        \label{Definition: inst-density}
\end{align}
Notice that the gauge coupling $g$ in the pre-factor $(8\pi^2/g^2)^{2N}$ is the bare coupling and only receives radiative corrections beginning at two-loop order. Furthermore, the running gauge coupling in the exponential factor, resulting from the contributions of non-zero modes, is evaluated at one-loop order and the full expression can be found in Eq.~\eqref{eq:RunningCoupling}. In presence of scalars, $\sigma$, that are charged under the gauge group, the instanton density is modified as in Eq.~\eqref{eq:sclar_inst_measure}.

The coefficient $C[N]$ includes the contributions of non-zero modes and the Jacobian factor when transforming $\int \mathcal{D}A_{\mu}^{(0)}$ to the integration over collective coordinates and it is given by~\cite{tHooft:1976snw,Bernard:1979qt,Csaki:2019vte,Csaki:2023ziz}
\begin{align}
        C[N] = \dfrac{C_1 \, e^{-C_2 N}}{(N-1)!(N-2)!}\,
        e^{0.292N_f}        
        \,,
        \label{Definition: inst-factor-CN}
 \end{align}
where $C_1\approx 0.466$, $C_2 \approx 1.678$; the contribution of fermion non-zero modes yields the factor $e^{0.292N_f}$ where $N_f$ is the number of fermions. Note that the coefficient $C_1$ is just a constant while $C_2$ and $e^{0.292N_f}$ are scheme-dependent due to the renormalization of the gauge coupling. Here, the coefficient $C[N]$ is defined in the Pauli--Villars regularization scheme. In this paper, we evaluate loop integrals using dimensional regularization and the $\overline{\text{MS}}$ scheme, therefore $d_N(\rho)$ should be converted into $\overline{\text{MS}}$ scheme. The details of this step will be discussed in Appendix~\ref{app: MS-bar}.
Eventually, the running of the gauge coupling is given by
\begin{align} 
    \label{eq:RunningCoupling}
    \dfrac{8\pi^2}{g^2(1/\rho)} = \dfrac{8\pi^2}{g_0^2(\Lambda_{\rm UV})} - b_0\log\rho\Lambda_{\rm UV}
    \,,\quad b_0 = \dfrac{11}{3}N - \dfrac{2}{3}N_f
    \,.
\end{align}

\subsection{Divergences and scheme independence of the results} \label{app: MS-bar}
The calculation of the topological susceptibility, $\chi_{_\mathcal{O}}(0)$ defined in Eq.~\eqref{eq:suscepdef}, induced by effective operators  can involve divergent loop integrals. Within the SMEFT, the standard technique to regularize these divergences is to use dimensional regularization, supplemented by the modified minimal subtraction $(\overline{\text{MS}})$ renormalization scheme. Therefore, to consistently evaluate $\chi_{_\mathcal{O}}(0)$, as well as the ratio 
$\chi_{_\mathcal{O}}(0) / \chi(0)$ in Eq.~\eqref{eq:thetaind},
 the instanton density $d_N(\rho)$, defined in Pauli--Villars (PV) renormalization scheme, must be converted to the $\overline{\text{MS}}$ renormalization scheme. The details of this conversion procedure can be found in the Appendix B of Ref.~\cite{Csaki:2019vte}.

It is clear that the computations of $\chi_{_{\rm \mathcal{O}}}(0)$ and $\chi(0)$ are scheme dependent. However, $\chi_{_{\rm \mathcal{O}}}(0) / \chi(0)$ does not depend on the renormalization scheme since it is the ratio of two topological susceptibilities and both scale in the same fashion when converted to a different renormalization scheme~\cite{Csaki:2019vte}
\begin{align}
    \chi_{_{\rm (\mathcal{O})}}^{\overline{\text{MS}}}(0) &= e^{(N-N_f)/6} \, \chi^{\rm PV}_{_{\rm (\mathcal{O})}}(0)
    \,,
    \label{Appendix: scheme-relations-inst-density}
\end{align}
where the bracket notation in the subscript of $ \chi_{_{\rm (\mathcal{O})}}(0) $ indicates either the  $ \chi (0)$ or $ \chi_{_{\rm  \mathcal{O} }}(0)$
susceptibilities. To deal with the divergences arising from the insertion of SMEFT operators that will only affect $\chi_{_\mathcal{O}}$, one has to consider the renormalized SMEFT, i.e. with the appropriate counterterms calculated in the $\overline{\text{MS}}$ scheme. 

As an example to illustrate this feature, in Appendix~\ref{app:Semilept}, we explicitly calculate the divergent part of the topological susceptibility $\chi^{(1)}_{\rm lequ}(0)$ resulting from the insertion of the semi-leptonic operator $\mathcal{O}^{(1)}_{\rm lequ}$. The divergences appearing in loop calculations are expected to be canceled in renormalized perturbation theory by appropriate counterterms. 
To obtain these counterterms we make use of the SMEFT RGEs\footnote{Inspired by previous studies~\cite{Novikov:1983uc, Novikov:1985rd, Monin:2023tjm}, the $\beta$-function of a given operator (related to instanton dynamics) can also be computed in the instanton background instead of using the traditional diagrammatic approach.}~\cite{Grojean:2013kd,Jenkins:2013zja,Jenkins:2013wua,Alonso:2013hga} that are computed from the SMEFT counterterms. It is important to note that we will require off-shell correlation functions and therefore to observe the cancellation at the level of the correlation functions, the counterterms will be calculated in an off-shell Green's basis~\cite{Gherardi:2020det}. The details of this computation will be given in Appendix~\ref{app:Semilept}.

\section{Evaluating loop and collective coordinates integrals}
\label{sec:SolveInt}

\subsection{Four-quark operator}
\label{app:Four-fermion}
With the insertion of the four-quark operator $\mathcal{O}_{\rm quqd}$, we have shown in Eq.~\eqref{result-4fermions: upto-inv} that the two-point correlation function $\chi_{\rm quqd}^{(1)}(0)$ is proportional to the determinant-like flavor invariant. However, to complete the calculation of the topological susceptibility we still need to evaluate the integral $\mathcal{I}$ and perform the final integral over the collective coordinate $x_0$ in Eq.~\eqref{result-4fermions: upto-inv}.

The integral $\cI$ has been previously calculated in the literature~\cite{Flynn:1987rs,Csaki:2019vte, Bedi:2022qrd} and reads,
\begin{align}
    \mathcal{I} &= \epsilon_{IJ}\epsilon^{IJ} \int d^4x_1 \int d^4x_2~\big(\bar{\psi}^{(0)} P_R \psi^{(0)} \big)(x_1) \Delta_H(x_1-x_2) \big(\bar{\psi}^{(0)} P_R \psi^{(0)} \big)(x_2)\,,
    \nonumber \\
    &= \int d^4x_1 \int d^4x_2 \int d^4k \, \dfrac{\rho^4}{2\pi^8} \dfrac{e^{-ikx_1}}{\big( x_1^2 + \rho^2 \big)^3} \dfrac{1}{k^2+m_H^2} \dfrac{e^{ikx_2}}{\big( x_2^2 + \rho^2 \big)^3}
    = \dfrac{\rho^2}{8\pi^4} \int d^4k \,  \dfrac{\big[ k\rho \,K_1(k\rho) \big]^2}{(k\rho)^2 + (m_H\rho)^2}\,,
    \label{Appendix: zero-modes-Higgs-int}
\end{align}
where we have substituted the zero mode profile in Eq.~\eqref{Inst-fermion: zero-modes} and the 
(Euclidean) scalar propagator into the first line of Eq.~\eqref{Appendix: zero-modes-Higgs-int}. The integrals over the Euclidean coordinates $x_1,~ x_2$ are performed using the identity
\begin{align}
    \int d^4x \dfrac{e^{-ikx}}{(x^2+\rho^2)^3} = \dfrac{\pi^2}{2}  \dfrac{k}{\rho} K_1(k\rho)
    \,,
    \label{Appendix: Bessel-K1-int}
\end{align}
where $K_1(k\rho)$ is the modified Bessel function of the second kind. In the small instanton limit, i.e., $m_H \rho \ll 1$, the integral in Eq.~\eqref{Appendix: zero-modes-Higgs-int} can be evaluated to give
\begin{align}
    \mathcal{I}^{(\rm UV)} \simeq \dfrac{1}{6\pi^2\rho^2}
    \,.
    \label{Appendix: solved-I-integral}
\end{align}

Secondly, we can evaluate the integrals over the collective coordinate ($x_0$ in Eq.~\eqref{result-4fermions: upto-inv}) resulting from the insertion of the four-quark operator:
\begin{align}
    \int d^4x_0 \big( \bar{\psi}^{(0)} P_{R(L)} \psi^{(0)} \, \bar{\psi}^{(0)} P_{R(L)} \psi^{(0)} \big)(0) \bigg|_{\rm 1-(a)-inst.}
    &= \int d^4x_0 \dfrac{4\rho^4}{\pi^4} \dfrac{1}{(x_0^2+\rho^2)^6}
    = \dfrac{1}{5\pi^2\rho^4} 
    \,.
    \label{Appendix: 4F-coordinates-x0-int}
\end{align}
Finally, substituting the results derived in Eqs.~\eqref{Appendix: zero-modes-Higgs-int} and
\eqref{Appendix: 4F-coordinates-x0-int} into  Eq.~\eqref{result-4fermions: upto-inv}, we obtain
\begin{align}
& \chi_{\rm quqd}^{(1)}(0) = \nonumber \\ 
& \frac{i}{\CPVm^2}\,\(\cA_{0000}^{0000}\(C_{\rm quqd}^{(1)}\) + \cB_{0000}^{0000}\(C_{\rm quqd}^{(1)}\)\) \int\dfrac{d\rho}{\rho^5} \, d_N(\rho) \rho^6
\left[ \dfrac{\rho^2}{8\pi^4} \int d^4k \,  \dfrac{(k\rho)^2 K_1^2(k\rho) }{(k\rho)^2 + (m_H\rho)^2}  \right]^{2} \dfrac{2}{5\pi^2\rho^4}
\,.
\end{align}

In the small instanton limit, $m_H\rho \ll 1$, the topological susceptibility induced by the operator $\mathcal{O}^{(1)}_{\rm quqd}$ is then given by
\begin{align}
\chi_{\rm quqd}^{(1)\,(\rm UV)}(0)  
&= \frac{i}{\CPVm^2}\,\(\cA_{0000}^{0000}\(C_{\rm quqd}^{(1)}\) + \cB_{0000}^{0000}\(C_{\rm quqd}^{(1)}\)\) \int\dfrac{d\rho}{\rho^5} \, d_N(\rho)
\dfrac{2!}{(6\pi^2)^2} \dfrac{2}{5\pi^2\rho^2} 
\,.
\label{Appendix: final-result-4F}
\end{align}
The integration over the instanton size, $\rho$, can be performed once the details of the UV dynamics at the small instanton scale are known. Some examples of UV models will be explored in Section~\ref{sec:Pheno}.

The results obtained in this Appendix also allow us to estimate the higher-order contribution of having an insertion of the CP-odd phase from $\cO^{(1)}_{\mathrm{quqd}}$ and a CP-even parameter from the same operator. This extra insertion of the SMEFT operator would result in one less Higgs needed to close the fermion legs in Fig.~\ref{fig:FlowerDiagramsSMEFT}. With this substitution, the final result has one less power of the integral $\cI$, Eq.~\eqref{Appendix: solved-I-integral}, which would then be substituted by one more power of the result of Eq.~\eqref{Appendix: 4F-coordinates-x0-int}. As such, the higher-dimensional contribution will be suppressed by an additional factor of $\left(\frac{\Lambda_{\mathrm{SI}}}{\CPVm}\right)^2$.

\subsection{Semi-leptonic operator}\label{app:Semilept}
Evaluating the integral associated with the insertion of the semi-leptonic operator $\mathcal{O}^{(1)}_{\rm lequ}$ is analogous to previous computations, where we begin with Eq.~\eqref{result-lequ: upto-inv} and then add the anti-instanton contribution. The topological susceptibility, $\chi_{\rm lequ}^{(1)}(0)$ reads
\begin{align}
    \chi_{\rm lequ}^{(1)}(0) &= \frac{i}{\CPVm^2}\,\cI_{0000}^0\(C_{\rm lequ}^{(1)}\) \int \dfrac{d\rho}{\rho^5} d_N(\rho)\, \big(3! \, \rho^6 \mathcal{I}^2 \big)\, \mathcal{I}_{\rm lequ}
    \,,
    \label{Appendix: lequ-upto-inv}
\end{align}
where the contribution of $\mathcal{O}^{(1)}_{\rm lequ}$ is included inside the integral $\mathcal{I}_{\rm lequ}$, defined as
\begin{align}
    \mathcal{I}_{\rm lequ} &= \epsilon_{OP}\epsilon^{OP} 
    \int d^4x_0 \int d^4x_5 \int d^4x_6
    \nonumber \\
    &\times \big( \bar{\psi}^{(0)} P_R \psi^{(0)} \big)(x_5) \Delta_H(x_5-x_6) \tr \big( P_R\, \Delta_F(x_6-0) P_L\, \Delta_F(0-x_6) \big) \big( \bar{\psi}^{(0)} P_R \psi^{(0)} \big)(0) 
    \,.
    \label{Appendix: lequ-master-integral}
\end{align}
\paragraph{Evaluating the divergent part of $\mathcal{I}_{\rm lequ}$}
Next, we substitute the (Euclidean) scalar and fermion propagators into Eq.~\eqref{Appendix: lequ-master-integral} to give
\begin{align}
    \mathcal{I}_{\rm lequ} &= 2 \int d^4x_0 \big(\bar{\psi}^{(0)} P_R \psi^{(0)}\big)(0) 
    \nonumber \\
    &\times \int d^4 x_5 \int\dfrac{d^4k}{(2\pi)^4}\int\dfrac{d^dq}{(2\pi)^d} \, \tr\left[
    P_R \dfrac{\slashed{q}}{q^2} P_L \dfrac{\slashed{q}+\slashed{k}}{(q+k)^2} \right]
    \dfrac{e^{-ikx_5}}{k^2+m_H^2} \big(\bar{\psi}^{(0)} P_R \psi^{(0)}\big)(x_5)
    \,.
    \label{eq:propIlequ}
\end{align}
Here, to obtain Eq.~\eqref{eq:propIlequ}, the integral representation of the four-dimensional Dirac delta distribution has been used to eliminate the $\int d^4x_6$ integration and simplify the four-momentum variables. To regulate the divergences appearing in the integral in Eq.~\eqref{eq:propIlequ}, we employ dimensional regularization~\cite{tHooft:1972tcz} in the $\overline{\rm MS}$ scheme as well as the semi-naive procedure~\cite{Chanowitz:1979zu} to deal with $\gamma^5$ matrices in $d=4-2\epsilon$ dimensions. Finally, we obtain the following for the divergent part of the integral $\mathcal{I}_{\rm lequ}$, 
\begin{equation} 
\label{Appendix: lequ-master-integral-div}
    \mathcal{I}_{\rm lequ}^{\rm \, div.} = \dfrac{1}{16\pi^2\epsilon} \left[ 2\int d^4x_0 d^4 x_5 \big(\bar{\psi}^{(0)} P_R \psi^{(0)}\big)(0) \int\dfrac{d^4k}{(2\pi)^4} \dfrac{k^2 \, e^{-ikx_5} }{k^2 + m_H^2} \big(\bar{\psi}^{(0)} P_R \psi^{(0)}\big)(x_5) \right]
    \,.
\end{equation}
The crucial point is that Eq.~\eqref{Appendix: lequ-master-integral-div} contains a UV divergence manifested as a $\frac{1}{\epsilon}$-pole, which can be canceled by identifying the appropriate counterterms.

\paragraph{Divergence cancellation and relation with SMEFT RGEs.}
Using the results in Ref.~\cite{Jenkins:2013wua} we can extract the appropriate counterterm needed to cancel the divergence in $\chi_{\rm lequ}^{(1)}(0)$. The SMEFT RGEs reveal that the only counterterm that can cancel the divergence in $\chi^{(1)}_{\rm lequ}(0)$ is the one responsible for the running of the on-shell operator $\cO_{\rm quqd}^{(1)}$ (all other counterterms either yield the wrong flavor structure or require additional insertions of gauge couplings). However, since we are requiring the divergence cancellation at the level of correlation functions, which are not invariant under field redefinitions~\cite{Arzt:1993gz,Manohar:2018aog}, we need to consider the counterterms in an enlarged Green's basis instead. For this particular case, we can verify that the contribution of $\mathcal{O}^{(1)}_{\rm lequ}$ to the RGE of $\mathcal{O}^{(1)}_{\rm quqd}$ is fully determined by a Green's basis operator. Considering the Green's basis of Ref.~\cite{Gherardi:2020det}, we find~\cite{Jenkins:2013wua}
\begin{equation} \label{Appendix: c.t.-quqd-offshell}
    C^{(1), \,\rm c.t.}_{{\rm quqd},mnop} \supset -Y_{{\rm d},op}^{\vphantom{t}} \, G^{\rm \, c.t.}_{{\rm uHD1},\,mn} \,,\quad G^{\rm \, c.t.}_{{\rm uHD1},\,mn} = \dfrac{1}{16\pi^2\, \epsilon} \, C^{(1)}_{{\rm lequ},stmn} \, Y^{\dagger}_{{\rm e},ts}  \,,
\end{equation}
where $G^{\rm \, c.t.}_{{\rm uHD1},\,mn}$ is the Wilson coefficient of the redundant operator $\mathcal{O}_{\rm uHD1}=\bar{Q}u D^2\tilde{H}$ that is reduced to $\cO_{\rm quqd}^{(1)}$ via field redefinitions -- or equivalently at this order, replacing $D^2 \tilde{H}$ by the Higgs equation of motion. To cancel the $\frac{1}{\epsilon}$-pole in $\chi_{\rm lequ}^{(1),\,\rm div.}(0)$, we need to compute the correlation function
\begin{equation}
    \chi_{\rm uHD1}^{\rm c.t.}(0)\big|_{\rm 1-inst.}  = -i \lim _{k \rightarrow 0} \int d^4 x e^{i k x}\left\langle 0\left|T\left\{\dfrac{1}{32 \pi^2} G \wtilde{G}(x), \dfrac{G^{\rm \, c.t.}_{\rm uHD1}}{\CPVm^2}\mathcal{O}_{\rm uHD1} (0)\right\}\right| 0\right\rangle_{\rm 1-inst.} \,,
\end{equation}
using similar steps to those used previously in Section~\ref{app:Four-fermion}. Eventually, we obtain
\begin{equation} \label{Appendix: c.t.-uHD1-master-formula}
    \chi_{\rm uHD1}^{\rm c.t.}(0) = - \frac{i}{\CPVm^2}\,\Im\big( I_{\rm uHD1} \big) \int \dfrac{d\rho}{\rho^5} d_N(\rho) \big(3! \, \rho^6 \mathcal{I}^2 \big)\, \mathcal{I}_{\rm uHD1}
    \,,
\end{equation}
where the invariant $I_{\rm uHD1}$, supplemented by the counterterm in Eq.~\eqref{Appendix: c.t.-quqd-offshell}, yields
\begin{equation}
    \Im\big( I_{\rm uHD1} \big) = \Im\left[ 
    e^{-i\thetaQCD} \epsilon^{i_1i_2m} \epsilon^{j_1j_2n} Y_{{\rm u},i_1j_1} Y_{{\rm u},i_2j_2} G^{\rm \, c.t.}_{{\rm uHD1},\,mn} \det Y_{\rm d} 
    \right]
    = \dfrac{1}{16\pi^2 \,\epsilon}\cI_{0000}^0\(C_{\rm lequ}^{(1)}\)
    \,.
    \label{Appendix: c.t.-uHD1-invariant}
\end{equation}
The explicit form of the integral $\mathcal{I}_{\rm uHD1}$ reads
\begin{equation} \label{Appendix: c.t.-uHD1-integral}
    \mathcal{I}_{\rm uHD1} = 2\int d^4x_0 \big(\bar{\psi}^{(0)} P_R \psi^{(0)}\big)(0) 
    \int d^4 x_5 \int\dfrac{d^4k}{(2\pi)^4} \dfrac{k^2 \, e^{-ikx_5} }{k^2 + m_H^2} 
    \big(\bar{\psi}^{(0)} P_R \psi^{(0)}\big)(x_5)
    \,,
\end{equation}
where we have used the fact that the Green's basis operator contains derivatives (in Euclidean space) acting on the Higgs, hence the path integral over the Higgs fields yields
\begin{align}
    \int \cD H \cD H^\dagger e^{-S_0\[H,H^\dagger\]} H_I^{\vphantom{\dagger}}(x_1) \partial^2 H_J^\dagger(x_2) 
    = \partial_{x_2}^2 \Delta_H(x_1-x_2) \delta_{IJ} 
    &= -\int \dfrac{d^4k}{(2\pi)^4} \dfrac{k^2 e^{-i k(x_1-x_2)}}{k^2+m_H^2} \delta_{IJ} \,.
\end{align}
At this point, we have found that the integral $\cI_{\rm uHD1}$ in Eq.~\eqref{Appendix: c.t.-uHD1-integral} obtained from including the counterterm is the same as the integral $\mathcal{I}_{\rm lequ}^{\rm \, div.}$ in Eq.~\eqref{Appendix: lequ-master-integral-div} up to the overall factor $1/(16\pi^2\epsilon)$. Thus, substituting Eqs.~\eqref{Appendix: c.t.-uHD1-invariant} and \eqref{Appendix: c.t.-uHD1-integral} into Eq.~\eqref{Appendix: c.t.-uHD1-master-formula}, one can easily observe that $\chi_{\rm uHD1}^{\rm c.t.}(0)$ precisely cancels the $\frac{1}{\epsilon}$-pole in $\chi_{\rm lequ}^{(1)}(0)$. 

\paragraph{Evaluating the finite part of $\mathcal{I}_{\rm lequ}$} Starting from Eq.~\eqref{eq:propIlequ}, we can also extract the finite contribution of the integration over the loop momentum $q$, 
\begin{align}
\mathcal{I}_{\rm lequ}^{\rm (finite)} 
    &= 2\int d^4x_0 \big(\bar{\psi}^{(0)} P_R \psi^{(0)}\big)(0) 
    \nonumber \\
    &\times \int d^4 x_5 \int\dfrac{d^4k}{(2\pi)^4} \dfrac{k^2 \, e^{-ikx_5} }{k^2 + m_H^2} 
    \left[ \dfrac{1}{8\pi^2} + \dfrac{1}{16\pi^2}\log\dfrac{\mu^2}{k^2} 
    \right]  \big(\bar{\psi}^{(0)} P_R \psi^{(0)}\big)(x_5)
    \,,
    \label{Appendix: lequ-master-integral-finite}
\end{align}
where $\mu$ is the renormalization scale. To evaluate Eq~\eqref{Appendix: lequ-master-integral-finite}, we follow similar steps to the previous computations by first substituting the fermion zero mode solutions in Eq.~\eqref{Inst-fermion: zero-modes}, then integrating over all locations $x_5$ and the collective coordinates $x_0$. The final integral over the momentum $k$ can be performed in the limit of $m_H \rightarrow 0$, to explicitly give
\begin{align}
\mathcal{I}_{\rm lequ}^{\rm (finite,UV)}
    &\simeq \int\dfrac{d^4k}{(2\pi)^4}\dfrac{1}{4\pi^2}\left[1+\dfrac{1}{2}\log\dfrac{\mu^2}{k^2} \right] \dfrac{4\rho^4}{\pi^4} \int d^4x_0 \dfrac{e^{-ikx_0}}{\big(x_0^2+\rho^2\big)^3}\int d^4x_5 \dfrac{e^{-ik(x_5-x_0)}}{\big((x_5-x_0)^2 + \rho^2 \big)^3}\,,
    \nonumber \\
    &= \int\dfrac{d^4k}{(2\pi)^4}\dfrac{1}{4\pi^2}\left[1+\dfrac{1}{2}\log\dfrac{\mu^2}{k^2} \right] \big[k\rho\, K_1(k\rho) \big]^2\,,
    \nonumber \\
    &= \dfrac{1}{20\pi^4\rho^4}\left( \frac{11}{30} + \log\mu\rho + \gamma_{\rm E} -\log2\right)
    \,.
    \label{Appendix: solved-lequ-master-integral-finite}
\end{align}
Finally, substituting Eq.~\eqref{Appendix: solved-I-integral} and Eq.~\eqref{Appendix: solved-lequ-master-integral-finite} into Eq.~\eqref{Appendix: lequ-upto-inv}, the topological susceptibility $\chi^{(1)}_{\rm lequ}(0)$ induced by the operator $\mathcal{O}^{(1)}_{\rm lequ}$ becomes
\begin{align}
\chi^{(1)(\rm finite,UV)}_{\rm lequ}(0)
    &= \dfrac{i}{\CPVm^2}\, \cI_{0000}^0\(C_{\rm lequ}^{(1)}\) \int \dfrac{d\rho}{\rho^5} d_N(\rho)\dfrac{3!}{(6\pi^2)^2} \dfrac{11+30\(\log\(\rho\CPVm\) +\gamma_{\rm E}-\log2 \)}{600\pi^4\rho^2} 
    \,.\label{eq:susc_lequ_finite}
\end{align}
The dependence on the renormalization scale $\mu$ in Eq.~\eqref{eq:susc_lequ_finite} has already been removed by performing the RG evolution induced by $C_{\rm lequ}^{(1)}$, rendering the final result of $\theta_{\text{ind}}$ independent of the renormalization scale as expected. The result in Eq.~\eqref{eq:susc_lequ_finite} will be used in Section~\ref{sec:semileptop} to place bounds on the scale $\CPVm$.

\subsection{Gluon dipole operator} 
\label{app:GluonDip}
The calculation for the insertion of the gluon dipole operator $\mathcal{O}_{\rm dG} = \bar{Q} \sigma^{\mu\nu} T^A d \, H  G_{\mu\nu}^A $ in the correlation function works similarly to those previously presented in Section~\ref{sec:4Fermi_op}. The field strength tensor in $\mathcal{O}_{\rm dG}$ is set to its instanton background value, while the rest of the calculation proceeds in a similar fashion to the other effective operators:
we assume all fermions only contribute zero modes at leading order and subsequently contract all Higgses. As discussed in Appendix~\ref{app:MoreInvs}, the insertion of the gluon dipole operator yields a similar flavor invariant structure as given in Eq.~\eqref{eq:CuHinv}. Following the previous calculations, combining both the instanton and anti-instanton contributions, the topological susceptibility $\chi_{\rm dG}(0)$ can be written as a flavor invariant times a complicated integral
\begin{align}
    \chi_{\rm dG}(0) &= \frac{i}{\CPVm^2}\,\mathcal{I}_{0000}(C_{\rm dG}) \int\dfrac{d\rho}{\rho^5} d_N(\rho)~3! \, \rho^6  \mathcal{I}^2~\, 
    \mathcal{I}_{\rm dG}\,,
    \label{Appendix: cEDM-upto-inv}
\end{align}
where the $\mathcal{O}_{\rm dG}$ operator is included inside the integral $\mathcal{I}_{\rm dG}$, and defined as
\begin{align}
    \mathcal{I}_{\rm dG} &= \epsilon_{IJ}\epsilon^{IJ} \int d^4x_0 \int d^4x_3 \big( \bar{\psi}^{(0)}P_R \psi^{(0)}  \big)(x_3) \Delta_H(x_3) \big( \bar{\psi}^{(0)} \sigma^{\mu\nu}T^A P_R \psi^{(0)} G^A_{\mu\nu}\big|_{\rm 1-inst.} \big)(0)
    \,.
    \label{Appendix: cEDM-Higgs-zero-modes-int}
\end{align}
The computation of $\chi_{\rm dG}(0)$ proceeds in the same way as the integral of $\chi_{\rm quqd}^{(1)}(0)$. Since most of the computations have already been performed in Appendix~\ref{app:Four-fermion}, we only need to evaluate the remaining integral $\mathcal{I}_{\rm dG}$. Substituting the zero modes of fermions~\eqref{Inst-fermion: zero-modes} and gauge fields~\eqref{Definition: G-1-instanton} into Eq.~\eqref{Appendix: cEDM-Higgs-zero-modes-int}, then contracting spinor and color indices\footnote{As mentioned in Appendix~\ref{sec: Instanton: review}, $T^A G^A_{\mu\nu}\big|_{\rm 1-inst.}$ only receives contributions from $A=1,2,3$.}, the integral $\mathcal{I}_{\rm dG}$ becomes
\begin{align}
    \mathcal{I}_{\rm dG} 
    &= \dfrac{192\rho^6}{\pi^4} \int\dfrac{d^4k}{(2\pi)^4} \int d^4x_0 \int d^4x_3 \dfrac{e^{ikx_3}}{(x_3^2 + \rho^2)^3} \dfrac{1}{k^2 + m_H^2} \dfrac{e^{-ikx_0}}{(x_0^2 + \rho^2)^5}\,,
    \nonumber \\
    &= \dfrac{96\rho^6}{\pi^2} \int\dfrac{d^4k}{(2\pi)^4} \dfrac{k\rho~K_1(k\rho)}{(k\rho)^2 + (m_H\rho)^2} \int d^4x_0 \dfrac{e^{-ikx_0}}{(x_0^2 + \rho^2)^5}\,,
    \nonumber \\
    &= \dfrac{1}{16\pi^4}\int d^4k \, \dfrac{(k\rho)^4 \, K_1(k\rho) K_3(k\rho)}{(k\rho)^2 + (m_H\rho)^2}
    \,.
    \label{Appendix: cEDM-dG-integral}
\end{align}
Here, the only extra computation is to evaluate the integral over collective coordinate $\int d^4x_0$ and express the result in terms of the Bessel function, $K_n$. This step can be easily performed by $\rho$ differentiation and using the identity in Eq.~\eqref{Appendix: Bessel-K1-int}, to give
\begin{align}
    \int d^4x_0 \dfrac{e^{-ikx_0}}{(x_0^2 + \rho^2)^5}
    &= \dfrac{1}{12} \left(\dfrac{\partial}{\partial \rho^2} \right)^2 \int d^4x_0 \dfrac{e^{-ikx_0}}{(x_0^2 + \rho^2)^3}
    = \dfrac{\pi^2}{24} \left(\dfrac{\partial}{\partial \rho^2} \right)^2 \bigg[ \dfrac{k}{\rho}K_1(k\rho) \bigg]\,,
    \nonumber\\
    &= \dfrac{\pi^2}{96}\left( \dfrac{k}{\rho} \right)^3 K_3(k\rho)
    \,,
\end{align}
where differentiation of $K_n$ satisfies the following identity
\begin{align}
\dfrac{\partial}{\partial \rho^2}\bigg[ \dfrac{1}{\rho^n} K_n(k\rho) \bigg] &= -\dfrac{k}{2\rho^{n+1}}K_{n+1}(k\rho)\,.
\end{align}
Analogously to the previous section, the last integral in Eq.~\eqref{Appendix: cEDM-dG-integral} can be evaluated in the small instanton limit, i.e., $m_H \rho \ll 1$, to give
\begin{align}
\mathcal{I}_{\rm dG}^{(\rm UV)}  
\simeq \dfrac{6}{5\pi^2\rho^4}\,.
    \label{Appendix: solved-IdG-integral}
\end{align}
Substituting Eqs.~\eqref{Appendix: solved-I-integral} and ~\eqref{Appendix: solved-IdG-integral} into Eq.~\eqref{Appendix: cEDM-upto-inv}, we obtain
\begin{align}
\chi_{\rm dG}^{(\rm UV)}(0)
&\simeq \frac {i}{\CPVm^2}\,\mathcal{I}_{0000}(C_{\rm dG}) \int \dfrac{d\rho}{\rho^5}\, d_N(\rho)
\dfrac{3!}{(6\pi^2)^2}\dfrac{6}{5\pi^2\rho^2}
\,.\label{eq:cEDM_op}
\end{align}
This result will be used in Section~\ref{sec:gluondipop} to place bounds on $\CPVm$ for various UV models.

\clearpage
\bibliographystyle{apsrev4-1_title}
\bibliography{biblio.bib}

\begin{thebibliography}{99}%
\makeatletter
\providecommand \@ifxundefined [1]{%
 \@ifx{#1\undefined}
}%
\providecommand \@ifnum [1]{%
 \ifnum #1\expandafter \@firstoftwo
 \else \expandafter \@secondoftwo
 \fi
}%
\providecommand \@ifx [1]{%
 \ifx #1\expandafter \@firstoftwo
 \else \expandafter \@secondoftwo
 \fi
}%
\providecommand \natexlab [1]{#1}%
\providecommand \enquote  [1]{``#1''}%
\providecommand \bibnamefont  [1]{#1}%
\providecommand \bibfnamefont [1]{#1}%
\providecommand \citenamefont [1]{#1}%
\providecommand \href@noop [0]{\@secondoftwo}%
\providecommand \href [0]{\begingroup \@sanitize@url \@href}%
\providecommand \@href[1]{\@@startlink{#1}\@@href}%
\providecommand \@@href[1]{\endgroup#1\@@endlink}%
\providecommand \@sanitize@url [0]{\catcode `\\12\catcode `\$12\catcode
  `\&12\catcode `\#12\catcode `\^12\catcode `\_12\catcode `\%12\relax}%
\providecommand \@@startlink[1]{}%
\providecommand \@@endlink[0]{}%
\providecommand \url  [0]{\begingroup\@sanitize@url \@url }%
\providecommand \@url [1]{\endgroup\@href {#1}{\urlprefix }}%
\providecommand \urlprefix  [0]{URL }%
\providecommand \Eprint [0]{\href }%
\providecommand \doibase [0]{http://dx.doi.org/}%
\providecommand \selectlanguage [0]{\@gobble}%
\providecommand \bibinfo [0]{\@secondoftwo}%
\providecommand \bibfield [0]{\@secondoftwo}%
\providecommand \translation [1]{[#1]}%
\providecommand \BibitemOpen [0]{}%
\providecommand \bibitemStop [0]{}%
\providecommand \bibitemNoStop [0]{.\EOS\space}%
\providecommand \EOS [0]{\spacefactor3000\relax}%
\providecommand \BibitemShut  [1]{\csname bibitem#1\endcsname}%
\let\auto@bib@innerbib\@empty
\bibitem [{\citenamefont{Abel} \emph {et\,al.}(2020)}]{Abel:2020pzs}%
  \BibitemOpen
  \bibfield{author}{\bibinfo{author}{\bibfnamefont{C.}\,\bibnamefont{Abel}}
  \emph {et\,al.}, }\bibfield{title}{\emph {\bibinfo{title}{{Measurement of the
  Permanent Electric Dipole Moment of the Neutron}}}, }\href {\doibase
  10.1103/PhysRevLett.124.081803} {\bibfield{journal}{\bibinfo{journal}{Phys.
  Rev.
  Lett.}\,}\textbf{\bibinfo{volume}{124}}\,(\bibinfo{year}{2020})\,\bibinfo{pages}{081803}},
  \Eprint {http://arxiv.org/abs/2001.11966}{arXiv:2001.11966
  [hep-ex]}\BibitemShut {NoStop}%
\bibitem [{\citenamefont{Ellis} and
  \citenamefont{Gaillard}(1979)}]{ELLIS1979141}%
  \BibitemOpen
  \bibfield{author}{\bibinfo{author}{\bibfnamefont{J.}\,\bibnamefont{Ellis}}
  and \bibinfo{author}{\bibfnamefont{M.~K.} \bibnamefont{Gaillard}},
  }\bibfield{title}{\emph {\bibinfo{title}{Strong and weak {CP} violation}},
  }\href {\doibase https://doi.org/10.1016/0550-3213(79)90297-9}
  {\bibfield{journal}{\bibinfo{journal}{Nucl. Phys.
  B}\,}\textbf{\bibinfo{volume}{150}}\,(\bibinfo{year}{1979})\,\bibinfo{pages}{141}}\BibitemShut
  {NoStop}%
\bibitem [{\citenamefont{Khriplovich} and
  \citenamefont{Vainshtein}(1994)}]{Khriplovich:1993pf}%
  \BibitemOpen
  \bibfield{author}{\bibinfo{author}{\bibfnamefont{I.~B.}
  \bibnamefont{Khriplovich}} and \bibinfo{author}{\bibfnamefont{A.~I.}
  \bibnamefont{Vainshtein}}, }\bibfield{title}{\emph {\bibinfo{title}{{Infinite
  renormalization of Theta term and Jarlskog invariant for CP violation}}},
  }\href {\doibase 10.1016/0550-3213(94)90419-7}
  {\bibfield{journal}{\bibinfo{journal}{Nucl. Phys.
  B}\,}\textbf{\bibinfo{volume}{414}}\,(\bibinfo{year}{1994})\,\bibinfo{pages}{27}},
  \Eprint
  {http://arxiv.org/abs/hep-ph/9308334}{arXiv:hep-ph/9308334}\BibitemShut
  {NoStop}%
\bibitem [{\citenamefont{Khriplovich}(1986)}]{Khriplovich:1985jr}%
  \BibitemOpen
  \bibfield{author}{\bibinfo{author}{\bibfnamefont{I.~B.}
  \bibnamefont{Khriplovich}}, }\bibfield{title}{\emph {\bibinfo{title}{{Quark
  Electric Dipole Moment and Induced $\theta$ Term in the {Kobayashi--Maskawa}
  Model}}}, }\href {\doibase 10.1016/0370-2693(86)90245-5}
  {\bibfield{journal}{\bibinfo{journal}{Phys. Lett.
  B}\,}\textbf{\bibinfo{volume}{173}}\,(\bibinfo{year}{1986})\,\bibinfo{pages}{193}}\BibitemShut
  {NoStop}%
\bibitem [{\citenamefont{Bigi} and
  \citenamefont{Uraltsev}(1991{\natexlab{a}})}]{Bigi:1990kz}%
  \BibitemOpen
  \bibfield{author}{\bibinfo{author}{\bibfnamefont{I.~I.~Y.}
  \bibnamefont{Bigi}} and \bibinfo{author}{\bibfnamefont{N.~G.}
  \bibnamefont{Uraltsev}}, }\bibfield{title}{\emph {\bibinfo{title}{{Induced
  Multi - Gluon Couplings and the Neutron Electric Dipole Moment}}}, }\href
  {\doibase 10.1016/0550-3213(91)90339-Y}
  {\bibfield{journal}{\bibinfo{journal}{Nucl. Phys.
  B}\,}\textbf{\bibinfo{volume}{353}}\,(\bibinfo{year}{1991}{\natexlab{a}})\,\bibinfo{pages}{321}}\BibitemShut
  {NoStop}%
\bibitem [{\citenamefont{Peccei} and
  \citenamefont{Quinn}(1977)}]{Peccei:1977hh}%
  \BibitemOpen
  \bibfield{author}{\bibinfo{author}{\bibfnamefont{R.~D.} \bibnamefont{Peccei}}
  and \bibinfo{author}{\bibfnamefont{H.~R.} \bibnamefont{Quinn}},
  }\bibfield{title}{\emph {\bibinfo{title}{{CP Conservation in the Presence of
  Instantons}}}, }\href {\doibase 10.1103/PhysRevLett.38.1440}
  {\bibfield{journal}{\bibinfo{journal}{Phys. Rev.
  Lett.}\,}\textbf{\bibinfo{volume}{38}}\,(\bibinfo{year}{1977})\,\bibinfo{pages}{1440}}\BibitemShut
  {NoStop}%
\bibitem [{\citenamefont{Weinberg}(1978)}]{Weinberg:1977ma}%
  \BibitemOpen
  \bibfield{author}{\bibinfo{author}{\bibfnamefont{S.}\,\bibnamefont{Weinberg}},
  }\bibfield{title}{\emph {\bibinfo{title}{{A New Light Boson?}}}, }\href
  {\doibase 10.1103/PhysRevLett.40.223}
  {\bibfield{journal}{\bibinfo{journal}{Phys. Rev.
  Lett.}\,}\textbf{\bibinfo{volume}{40}}\,(\bibinfo{year}{1978})\,\bibinfo{pages}{223}}\BibitemShut
  {NoStop}%
\bibitem [{\citenamefont{Wilczek}(1978)}]{Wilczek:1977pj}%
  \BibitemOpen
  \bibfield{author}{\bibinfo{author}{\bibfnamefont{F.}\,\bibnamefont{Wilczek}},
  }\bibfield{title}{\emph {\bibinfo{title}{{Problem of Strong $P$ and $T$
  Invariance in the Presence of Instantons}}}, }\href {\doibase
  10.1103/PhysRevLett.40.279} {\bibfield{journal}{\bibinfo{journal}{Phys. Rev.
  Lett.}\,}\textbf{\bibinfo{volume}{40}}\,(\bibinfo{year}{1978})\,\bibinfo{pages}{279}}\BibitemShut
  {NoStop}%
\bibitem [{\citenamefont{Witten}(1984)}]{Witten:1984dg}%
  \BibitemOpen
  \bibfield{author}{\bibinfo{author}{\bibfnamefont{E.}\,\bibnamefont{Witten}},
  }\bibfield{title}{\emph {\bibinfo{title}{{Some Properties of O(32)
  Superstrings}}}, }\href {\doibase 10.1016/0370-2693(84)90422-2}
  {\bibfield{journal}{\bibinfo{journal}{Phys. Lett.
  B}\,}\textbf{\bibinfo{volume}{149}}\,(\bibinfo{year}{1984})\,\bibinfo{pages}{351}}\BibitemShut
  {NoStop}%
\bibitem [{\citenamefont{Randall}(1992)}]{Randall:1992ut}%
  \BibitemOpen
  \bibfield{author}{\bibinfo{author}{\bibfnamefont{L.}\,\bibnamefont{Randall}},
  }\bibfield{title}{\emph {\bibinfo{title}{{Composite axion models and Planck
  scale physics}}}, }\href {\doibase 10.1016/0370-2693(92)91928-3}
  {\bibfield{journal}{\bibinfo{journal}{Phys. Lett.
  B}\,}\textbf{\bibinfo{volume}{284}}\,(\bibinfo{year}{1992})\,\bibinfo{pages}{77}}\BibitemShut
  {NoStop}%
\bibitem [{\citenamefont{Choi}(2004)}]{Choi:2003wr}%
  \BibitemOpen
  \bibfield{author}{\bibinfo{author}{\bibfnamefont{K.-w.} \bibnamefont{Choi}},
  }\bibfield{title}{\emph {\bibinfo{title}{{A QCD axion from higher dimensional
  gauge field}}}, }\href {\doibase 10.1103/PhysRevLett.92.101602}
  {\bibfield{journal}{\bibinfo{journal}{Phys. Rev.
  Lett.}\,}\textbf{\bibinfo{volume}{92}}\,(\bibinfo{year}{2004})\,\bibinfo{pages}{101602}},
  \Eprint
  {http://arxiv.org/abs/hep-ph/0308024}{arXiv:hep-ph/0308024}\BibitemShut
  {NoStop}%
\bibitem [{\citenamefont{Lillard} and
  \citenamefont{Tait}(2018)}]{Lillard:2018fdt}%
  \BibitemOpen
  \bibfield{author}{\bibinfo{author}{\bibfnamefont{B.}\,\bibnamefont{Lillard}}
  and \bibinfo{author}{\bibfnamefont{T.~M.~P.} \bibnamefont{Tait}},
  }\bibfield{title}{\emph {\bibinfo{title}{{A High Quality Composite Axion}}},
  }\href {\doibase 10.1007/JHEP11(2018)199}
  {\bibfield{journal}{\bibinfo{journal}{JHEP}\,}\textbf{\bibinfo{volume}{11}}\,(\bibinfo{year}{2018})\,\bibinfo{pages}{199}},
  \Eprint {http://arxiv.org/abs/1811.03089}{arXiv:1811.03089
  [hep-ph]}\BibitemShut {NoStop}%
\bibitem [{\citenamefont{Gavela} \emph {et\,al.}(2019)\citenamefont{Gavela},
  \citenamefont{Ibe}, \citenamefont{Quilez}, and
  \citenamefont{Yanagida}}]{Gavela:2018paw}%
  \BibitemOpen
  \bibfield{author}{\bibinfo{author}{\bibfnamefont{M.~B.}
  \bibnamefont{Gavela}},
  \bibinfo{author}{\bibfnamefont{M.}\,\bibnamefont{Ibe}},
  \bibinfo{author}{\bibfnamefont{P.}\,\bibnamefont{Quilez}},  and
  \bibinfo{author}{\bibfnamefont{T.~T.} \bibnamefont{Yanagida}},
  }\bibfield{title}{\emph {\bibinfo{title}{{Automatic Peccei\textendash{}Quinn
  symmetry}}}, }\href {\doibase 10.1140/epjc/s10052-019-7046-3}
  {\bibfield{journal}{\bibinfo{journal}{Eur. Phys. J.
  C}\,}\textbf{\bibinfo{volume}{79}}\,(\bibinfo{year}{2019})\,\bibinfo{pages}{542}},
  \Eprint {http://arxiv.org/abs/1812.08174}{arXiv:1812.08174
  [hep-ph]}\BibitemShut {NoStop}%
\bibitem [{\citenamefont{Hook} \emph {et\,al.}(2020)\citenamefont{Hook},
  \citenamefont{Kumar}, \citenamefont{Liu}, and
  \citenamefont{Sundrum}}]{Hook:2019qoh}%
  \BibitemOpen
  \bibfield{author}{\bibinfo{author}{\bibfnamefont{A.}\,\bibnamefont{Hook}},
  \bibinfo{author}{\bibfnamefont{S.}\,\bibnamefont{Kumar}},
  \bibinfo{author}{\bibfnamefont{Z.}\,\bibnamefont{Liu}},  and
  \bibinfo{author}{\bibfnamefont{R.}\,\bibnamefont{Sundrum}},
  }\bibfield{title}{\emph {\bibinfo{title}{{High Quality QCD Axion and the
  LHC}}}, }\href {\doibase 10.1103/PhysRevLett.124.221801}
  {\bibfield{journal}{\bibinfo{journal}{Phys. Rev.
  Lett.}\,}\textbf{\bibinfo{volume}{124}}\,(\bibinfo{year}{2020})\,\bibinfo{pages}{221801}},
  \Eprint {http://arxiv.org/abs/1911.12364}{arXiv:1911.12364
  [hep-ph]}\BibitemShut {NoStop}%
\bibitem [{\citenamefont{Contino} \emph {et\,al.}(2022)\citenamefont{Contino},
  \citenamefont{Podo}, and \citenamefont{Revello}}]{Contino:2021ayn}%
  \BibitemOpen
  \bibfield{author}{\bibinfo{author}{\bibfnamefont{R.}\,\bibnamefont{Contino}},
  \bibinfo{author}{\bibfnamefont{A.}\,\bibnamefont{Podo}},  and
  \bibinfo{author}{\bibfnamefont{F.}\,\bibnamefont{Revello}},
  }\bibfield{title}{\emph {\bibinfo{title}{{Chiral models of composite axions
  and accidental Peccei-Quinn symmetry}}}, }\href {\doibase
  10.1007/JHEP04(2022)180}
  {\bibfield{journal}{\bibinfo{journal}{JHEP}\,}\textbf{\bibinfo{volume}{04}}\,(\bibinfo{year}{2022})\,\bibinfo{pages}{180}},
  \Eprint {http://arxiv.org/abs/2112.09635}{arXiv:2112.09635
  [hep-ph]}\BibitemShut {NoStop}%
\bibitem [{\citenamefont{Cox} \emph {et\,al.}(2023)\citenamefont{Cox},
  \citenamefont{Gherghetta}, and \citenamefont{Paul}}]{Cox:2023dou}%
  \BibitemOpen
  \bibfield{author}{\bibinfo{author}{\bibfnamefont{P.}\,\bibnamefont{Cox}},
  \bibinfo{author}{\bibfnamefont{T.}\,\bibnamefont{Gherghetta}},  and
  \bibinfo{author}{\bibfnamefont{A.}\,\bibnamefont{Paul}},
  }\bibfield{title}{\emph {\bibinfo{title}{{A common origin for the QCD axion
  and sterile neutrinos from SU(5) strong dynamics}}}, }\href {\doibase
  10.1007/JHEP12(2023)180}
  {\bibfield{journal}{\bibinfo{journal}{JHEP}\,}\textbf{\bibinfo{volume}{12}}\,(\bibinfo{year}{2023})\,\bibinfo{pages}{180}},
  \Eprint {http://arxiv.org/abs/2310.08557}{arXiv:2310.08557
  [hep-ph]}\BibitemShut {NoStop}%
\bibitem [{\citenamefont{Dine} and \citenamefont{Seiberg}(1986)}]{Dine:1986bg}%
  \BibitemOpen
  \bibfield{author}{\bibinfo{author}{\bibfnamefont{M.}\,\bibnamefont{Dine}} and
  \bibinfo{author}{\bibfnamefont{N.}\,\bibnamefont{Seiberg}},
  }\bibfield{title}{\emph {\bibinfo{title}{{String Theory and the Strong {CP}
  Problem}}}, }\href {\doibase 10.1016/0550-3213(86)90043-X}
  {\bibfield{journal}{\bibinfo{journal}{Nucl. Phys.
  B}\,}\textbf{\bibinfo{volume}{273}}\,(\bibinfo{year}{1986})\,\bibinfo{pages}{109}}\BibitemShut
  {NoStop}%
\bibitem [{\citenamefont{Dine}(2022)}]{Dine:2022mjw}%
  \BibitemOpen
  \bibfield{author}{\bibinfo{author}{\bibfnamefont{M.}\,\bibnamefont{Dine}},
  }\bibfield{title}{\emph {\bibinfo{title}{{The Problem of Axion Quality: A Low
  Energy Effective Action Perspective}}}, }\href@noop {}
  {\,(\bibinfo{year}{2022})}, \Eprint
  {http://arxiv.org/abs/2207.01068}{arXiv:2207.01068 [hep-ph]}\BibitemShut
  {NoStop}%
\bibitem [{\citenamefont{Bedi} \emph {et\,al.}(2022)\citenamefont{Bedi},
  \citenamefont{Gherghetta}, and \citenamefont{Pospelov}}]{Bedi:2022qrd}%
  \BibitemOpen
  \bibfield{author}{\bibinfo{author}{\bibfnamefont{R.~S.} \bibnamefont{Bedi}},
  \bibinfo{author}{\bibfnamefont{T.}\,\bibnamefont{Gherghetta}},  and
  \bibinfo{author}{\bibfnamefont{M.}\,\bibnamefont{Pospelov}},
  }\bibfield{title}{\emph {\bibinfo{title}{{Enhanced EDMs from small
  instantons}}}, }\href {\doibase 10.1103/PhysRevD.106.015030}
  {\bibfield{journal}{\bibinfo{journal}{Phys. Rev.
  D}\,}\textbf{\bibinfo{volume}{106}}\,(\bibinfo{year}{2022})\,\bibinfo{pages}{015030}},
  \Eprint {http://arxiv.org/abs/2205.07948}{arXiv:2205.07948
  [hep-ph]}\BibitemShut {NoStop}%
\bibitem [{\citenamefont{Dimopoulos}(1979)}]{Dimopoulos:1979pp}%
  \BibitemOpen
  \bibfield{author}{\bibinfo{author}{\bibfnamefont{S.}\,\bibnamefont{Dimopoulos}},
  }\bibfield{title}{\emph {\bibinfo{title}{{A Solution of the Strong {CP}
  Problem in Models With Scalars}}}, }\href {\doibase
  10.1016/0370-2693(79)91233-4} {\bibfield{journal}{\bibinfo{journal}{Phys.
  Lett.
  B}\,}\textbf{\bibinfo{volume}{84}}\,(\bibinfo{year}{1979})\,\bibinfo{pages}{435}}\BibitemShut
  {NoStop}%
\bibitem [{\citenamefont{Holdom} and
  \citenamefont{Peskin}(1982)}]{Holdom:1982ex}%
  \BibitemOpen
  \bibfield{author}{\bibinfo{author}{\bibfnamefont{B.}\,\bibnamefont{Holdom}}
  and \bibinfo{author}{\bibfnamefont{M.~E.} \bibnamefont{Peskin}},
  }\bibfield{title}{\emph {\bibinfo{title}{{Raising the Axion Mass}}}, }\href
  {\doibase 10.1016/0550-3213(82)90228-0}
  {\bibfield{journal}{\bibinfo{journal}{Nucl. Phys.
  B}\,}\textbf{\bibinfo{volume}{208}}\,(\bibinfo{year}{1982})\,\bibinfo{pages}{397}}\BibitemShut
  {NoStop}%
\bibitem [{\citenamefont{Holdom}(1985)}]{Holdom:1985vx}%
  \BibitemOpen
  \bibfield{author}{\bibinfo{author}{\bibfnamefont{B.}\,\bibnamefont{Holdom}},
  }\bibfield{title}{\emph {\bibinfo{title}{{Strong QCD at High-energies and a
  Heavy Axion}}}, }\href {\doibase 10.1016/0370-2693(85)90371-5}
  {\bibfield{journal}{\bibinfo{journal}{Phys.
  Lett.}\,}\textbf{\bibinfo{volume}{154B}}\,(\bibinfo{year}{1985})\,\bibinfo{pages}{316}},
  \bibinfo{note}{[Erratum: Phys. Lett.156B,452(1985)]}\BibitemShut {NoStop}%
\bibitem [{\citenamefont{Flynn} and
  \citenamefont{Randall}(1987)}]{Flynn:1987rs}%
  \BibitemOpen
  \bibfield{author}{\bibinfo{author}{\bibfnamefont{J.~M.} \bibnamefont{Flynn}}
  and \bibinfo{author}{\bibfnamefont{L.}\,\bibnamefont{Randall}},
  }\bibfield{title}{\emph {\bibinfo{title}{{A Computation of the Small
  Instanton Contribution to the Axion Potential}}}, }\href {\doibase
  10.1016/0550-3213(87)90089-7} {\bibfield{journal}{\bibinfo{journal}{Nucl.
  Phys.
  B}\,}\textbf{\bibinfo{volume}{293}}\,(\bibinfo{year}{1987})\,\bibinfo{pages}{731}}\BibitemShut
  {NoStop}%
\bibitem [{\citenamefont{Rubakov}(1997)}]{Rubakov:1997vp}%
  \BibitemOpen
  \bibfield{author}{\bibinfo{author}{\bibfnamefont{V.~A.}
  \bibnamefont{Rubakov}}, }\bibfield{title}{\emph {\bibinfo{title}{{Grand
  unification and heavy axion}}}, }\href {\doibase 10.1134/1.567390}
  {\bibfield{journal}{\bibinfo{journal}{JETP
  Lett.}\,}\textbf{\bibinfo{volume}{65}}\,(\bibinfo{year}{1997})\,\bibinfo{pages}{621}},
  \Eprint
  {http://arxiv.org/abs/hep-ph/9703409}{arXiv:hep-ph/9703409}\BibitemShut
  {NoStop}%
\bibitem [{\citenamefont{Fukuda} \emph {et\,al.}(2015)\citenamefont{Fukuda},
  \citenamefont{Harigaya}, \citenamefont{Ibe}, and
  \citenamefont{Yanagida}}]{Fukuda:2015ana}%
  \BibitemOpen
  \bibfield{author}{\bibinfo{author}{\bibfnamefont{H.}\,\bibnamefont{Fukuda}},
  \bibinfo{author}{\bibfnamefont{K.}\,\bibnamefont{Harigaya}},
  \bibinfo{author}{\bibfnamefont{M.}\,\bibnamefont{Ibe}},  and
  \bibinfo{author}{\bibfnamefont{T.~T.} \bibnamefont{Yanagida}},
  }\bibfield{title}{\emph {\bibinfo{title}{{Model of visible QCD axion}}},
  }\href {\doibase 10.1103/PhysRevD.92.015021}
  {\bibfield{journal}{\bibinfo{journal}{Phys.
  Rev.}\,}\textbf{\bibinfo{volume}{D92}}\,(\bibinfo{year}{2015})\,\bibinfo{pages}{015021}},
  \Eprint {http://arxiv.org/abs/1504.06084}{arXiv:1504.06084
  [hep-ph]}\BibitemShut {NoStop}%
\bibitem [{\citenamefont{Gherghetta} \emph
  {et\,al.}(2016)\citenamefont{Gherghetta}, \citenamefont{Nagata}, and
  \citenamefont{Shifman}}]{Gherghetta:2016fhp}%
  \BibitemOpen
  \bibfield{author}{\bibinfo{author}{\bibfnamefont{T.}\,\bibnamefont{Gherghetta}},
  \bibinfo{author}{\bibfnamefont{N.}\,\bibnamefont{Nagata}},  and
  \bibinfo{author}{\bibfnamefont{M.}\,\bibnamefont{Shifman}},
  }\bibfield{title}{\emph {\bibinfo{title}{{A Visible QCD Axion from an
  Enlarged Color Group}}}, }\href {\doibase 10.1103/PhysRevD.93.115010}
  {\bibfield{journal}{\bibinfo{journal}{Phys.
  Rev.}\,}\textbf{\bibinfo{volume}{D93}}\,(\bibinfo{year}{2016})\,\bibinfo{pages}{115010}},
  \Eprint {http://arxiv.org/abs/1604.01127}{arXiv:1604.01127
  [hep-ph]}\BibitemShut {NoStop}%
\bibitem [{\citenamefont{Agrawal} and
  \citenamefont{Howe}(2018)}]{Agrawal:2017ksf}%
  \BibitemOpen
  \bibfield{author}{\bibinfo{author}{\bibfnamefont{P.}\,\bibnamefont{Agrawal}}
  and \bibinfo{author}{\bibfnamefont{K.}\,\bibnamefont{Howe}},
  }\bibfield{title}{\emph {\bibinfo{title}{{Factoring the Strong CP Problem}}},
  }\href {\doibase 10.1007/JHEP12(2018)029}
  {\bibfield{journal}{\bibinfo{journal}{JHEP}\,}\textbf{\bibinfo{volume}{12}}\,(\bibinfo{year}{2018})\,\bibinfo{pages}{029}},
  \Eprint {http://arxiv.org/abs/1710.04213}{arXiv:1710.04213
  [hep-ph]}\BibitemShut {NoStop}%
\bibitem [{\citenamefont{Cs\'aki} \emph {et\,al.}(2020)\citenamefont{Cs\'aki},
  \citenamefont{Ruhdorfer}, and \citenamefont{Shirman}}]{Csaki:2019vte}%
  \BibitemOpen
  \bibfield{author}{\bibinfo{author}{\bibfnamefont{C.}\,\bibnamefont{Cs\'aki}},
  \bibinfo{author}{\bibfnamefont{M.}\,\bibnamefont{Ruhdorfer}},  and
  \bibinfo{author}{\bibfnamefont{Y.}\,\bibnamefont{Shirman}},
  }\bibfield{title}{\emph {\bibinfo{title}{{UV Sensitivity of the Axion Mass
  from Instantons in Partially Broken Gauge Groups}}}, }\href {\doibase
  10.1007/JHEP04(2020)031}
  {\bibfield{journal}{\bibinfo{journal}{JHEP}\,}\textbf{\bibinfo{volume}{04}}\,(\bibinfo{year}{2020})\,\bibinfo{pages}{031}},
  \Eprint {http://arxiv.org/abs/1912.02197}{arXiv:1912.02197
  [hep-ph]}\BibitemShut {NoStop}%
\bibitem [{\citenamefont{Gherghetta} and
  \citenamefont{Nguyen}(2020)}]{Gherghetta:2020ofz}%
  \BibitemOpen
  \bibfield{author}{\bibinfo{author}{\bibfnamefont{T.}\,\bibnamefont{Gherghetta}}
  and \bibinfo{author}{\bibfnamefont{M.~D.} \bibnamefont{Nguyen}},
  }\bibfield{title}{\emph {\bibinfo{title}{{A Composite Higgs with a Heavy
  Composite Axion}}}, }\href {\doibase 10.1007/JHEP12(2020)094}
  {\bibfield{journal}{\bibinfo{journal}{JHEP}\,}\textbf{\bibinfo{volume}{12}}\,(\bibinfo{year}{2020})\,\bibinfo{pages}{094}},
  \Eprint {http://arxiv.org/abs/2007.10875}{arXiv:2007.10875
  [hep-ph]}\BibitemShut {NoStop}%
\bibitem [{\citenamefont{Kitano} and
  \citenamefont{Yin}(2021)}]{Kitano:2021fdl}%
  \BibitemOpen
  \bibfield{author}{\bibinfo{author}{\bibfnamefont{R.}\,\bibnamefont{Kitano}}
  and \bibinfo{author}{\bibfnamefont{W.}\,\bibnamefont{Yin}},
  }\bibfield{title}{\emph {\bibinfo{title}{{Strong CP problem and axion dark
  matter with small instantons}}}, }\href {\doibase 10.1007/JHEP07(2021)078}
  {\bibfield{journal}{\bibinfo{journal}{JHEP}\,}\textbf{\bibinfo{volume}{07}}\,(\bibinfo{year}{2021})\,\bibinfo{pages}{078}},
  \Eprint {http://arxiv.org/abs/2103.08598}{arXiv:2103.08598
  [hep-ph]}\BibitemShut {NoStop}%
\bibitem [{\citenamefont{Demirtas} \emph
  {et\,al.}(2023)\citenamefont{Demirtas}, \citenamefont{Gendler},
  \citenamefont{Long}, \citenamefont{McAllister}, and
  \citenamefont{Moritz}}]{Demirtas:2021gsq}%
  \BibitemOpen
  \bibfield{author}{\bibinfo{author}{\bibfnamefont{M.}\,\bibnamefont{Demirtas}},
  \bibinfo{author}{\bibfnamefont{N.}\,\bibnamefont{Gendler}},
  \bibinfo{author}{\bibfnamefont{C.}\,\bibnamefont{Long}},
  \bibinfo{author}{\bibfnamefont{L.}\,\bibnamefont{McAllister}},  and
  \bibinfo{author}{\bibfnamefont{J.}\,\bibnamefont{Moritz}},
  }\bibfield{title}{\emph {\bibinfo{title}{{PQ axiverse}}}, }\href {\doibase
  10.1007/JHEP06(2023)092}
  {\bibfield{journal}{\bibinfo{journal}{JHEP}\,}\textbf{\bibinfo{volume}{06}}\,(\bibinfo{year}{2023})\,\bibinfo{pages}{092}},
  \Eprint {http://arxiv.org/abs/2112.04503}{arXiv:2112.04503
  [hep-th]}\BibitemShut {NoStop}%
\bibitem [{\citenamefont{Cs\'aki} \emph {et\,al.}(2024)\citenamefont{Cs\'aki},
  \citenamefont{D'Agnolo}, \citenamefont{Kuflik}, and
  \citenamefont{Ruhdorfer}}]{Csaki:2023ziz}%
  \BibitemOpen
  \bibfield{author}{\bibinfo{author}{\bibfnamefont{C.}\,\bibnamefont{Cs\'aki}},
  \bibinfo{author}{\bibfnamefont{R.~T.} \bibnamefont{D'Agnolo}},
  \bibinfo{author}{\bibfnamefont{E.}\,\bibnamefont{Kuflik}},  and
  \bibinfo{author}{\bibfnamefont{M.}\,\bibnamefont{Ruhdorfer}},
  }\bibfield{title}{\emph {\bibinfo{title}{{Instanton NDA and applications to
  axion models}}}, }\href {\doibase 10.1007/JHEP04(2024)074}
  {\bibfield{journal}{\bibinfo{journal}{JHEP}\,}\textbf{\bibinfo{volume}{04}}\,(\bibinfo{year}{2024})\,\bibinfo{pages}{074}},
  \Eprint {http://arxiv.org/abs/2311.09285}{arXiv:2311.09285
  [hep-ph]}\BibitemShut {NoStop}%
\bibitem [{\citenamefont{de\,Blas} \emph
  {et\,al.}(2018)\citenamefont{de\,Blas}, \citenamefont{Criado},
  \citenamefont{Perez-Victoria}, and \citenamefont{Santiago}}]{deBlas:2017xtg}%
  \BibitemOpen
  \bibfield{author}{\bibinfo{author}{\bibfnamefont{J.}\,\bibnamefont{de\,Blas}},
  \bibinfo{author}{\bibfnamefont{J.~C.} \bibnamefont{Criado}},
  \bibinfo{author}{\bibfnamefont{M.}\,\bibnamefont{Perez-Victoria}},  and
  \bibinfo{author}{\bibfnamefont{J.}\,\bibnamefont{Santiago}},
  }\bibfield{title}{\emph {\bibinfo{title}{{Effective description of general
  extensions of the Standard Model: the complete tree-level dictionary}}},
  }\href {\doibase 10.1007/JHEP03(2018)109}
  {\bibfield{journal}{\bibinfo{journal}{JHEP}\,}\textbf{\bibinfo{volume}{03}}\,(\bibinfo{year}{2018})\,\bibinfo{pages}{109}},
  \Eprint {http://arxiv.org/abs/1711.10391}{arXiv:1711.10391
  [hep-ph]}\BibitemShut {NoStop}%
\bibitem [{\citenamefont{Guedes} \emph {et\,al.}(2023)\citenamefont{Guedes},
  \citenamefont{Olgoso}, and \citenamefont{Santiago}}]{Guedes:2023azv}%
  \BibitemOpen
  \bibfield{author}{\bibinfo{author}{\bibfnamefont{G.}\,\bibnamefont{Guedes}},
  \bibinfo{author}{\bibfnamefont{P.}\,\bibnamefont{Olgoso}},  and
  \bibinfo{author}{\bibfnamefont{J.}\,\bibnamefont{Santiago}},
  }\bibfield{title}{\emph {\bibinfo{title}{{Towards the one loop IR/UV
  dictionary in the SMEFT: One loop generated operators from new scalars and
  fermions}}}, }\href {\doibase 10.21468/SciPostPhys.15.4.143}
  {\bibfield{journal}{\bibinfo{journal}{SciPost
  Phys.}\,}\textbf{\bibinfo{volume}{15}}\,(\bibinfo{year}{2023})\,\bibinfo{pages}{143}},
  \Eprint {http://arxiv.org/abs/2303.16965}{arXiv:2303.16965
  [hep-ph]}\BibitemShut {NoStop}%
\bibitem [{\citenamefont{Farrar} and
  \citenamefont{Shaposhnikov}(1994)}]{Farrar:1993hn}%
  \BibitemOpen
  \bibfield{author}{\bibinfo{author}{\bibfnamefont{G.~R.} \bibnamefont{Farrar}}
  and \bibinfo{author}{\bibfnamefont{M.~E.} \bibnamefont{Shaposhnikov}},
  }\bibfield{title}{\emph {\bibinfo{title}{{Baryon asymmetry of the universe in
  the standard electroweak theory}}}, }\href {\doibase 10.1103/PhysRevD.50.774}
  {\bibfield{journal}{\bibinfo{journal}{Phys. Rev.
  D}\,}\textbf{\bibinfo{volume}{50}}\,(\bibinfo{year}{1994})\,\bibinfo{pages}{774}},
  \Eprint
  {http://arxiv.org/abs/hep-ph/9305275}{arXiv:hep-ph/9305275}\BibitemShut
  {NoStop}%
\bibitem [{\citenamefont{Smith} and
  \citenamefont{Touati}(2017)}]{Smith:2017dtz}%
  \BibitemOpen
  \bibfield{author}{\bibinfo{author}{\bibfnamefont{C.}\,\bibnamefont{Smith}}
  and \bibinfo{author}{\bibfnamefont{S.}\,\bibnamefont{Touati}},
  }\bibfield{title}{\emph {\bibinfo{title}{{Electric dipole moments with and
  beyond flavor invariants}}}, }\href {\doibase
  10.1016/j.nuclphysb.2017.09.013} {\bibfield{journal}{\bibinfo{journal}{Nucl.
  Phys.
  B}\,}\textbf{\bibinfo{volume}{924}}\,(\bibinfo{year}{2017})\,\bibinfo{pages}{417}},
  \Eprint {http://arxiv.org/abs/1707.06805}{arXiv:1707.06805
  [hep-ph]}\BibitemShut {NoStop}%
\bibitem [{\citenamefont{Denton} and
  \citenamefont{Parke}(2019)}]{Denton:2019yiw}%
  \BibitemOpen
  \bibfield{author}{\bibinfo{author}{\bibfnamefont{P.~B.} \bibnamefont{Denton}}
  and \bibinfo{author}{\bibfnamefont{S.~J.} \bibnamefont{Parke}},
  }\bibfield{title}{\emph {\bibinfo{title}{{Simple and Precise Factorization of
  the Jarlskog Invariant for Neutrino Oscillations in Matter}}}, }\href
  {\doibase 10.1103/PhysRevD.100.053004}
  {\bibfield{journal}{\bibinfo{journal}{Phys. Rev.
  D}\,}\textbf{\bibinfo{volume}{100}}\,(\bibinfo{year}{2019})\,\bibinfo{pages}{053004}},
  \Eprint {http://arxiv.org/abs/1902.07185}{arXiv:1902.07185
  [hep-ph]}\BibitemShut {NoStop}%
\bibitem [{\citenamefont{Yu} and \citenamefont{Zhou}(2022)}]{Yu:2022ttm}%
  \BibitemOpen
  \bibfield{author}{\bibinfo{author}{\bibfnamefont{B.}\,\bibnamefont{Yu}} and
  \bibinfo{author}{\bibfnamefont{S.}\,\bibnamefont{Zhou}},
  }\bibfield{title}{\emph {\bibinfo{title}{{CP violation and flavor invariants
  in the seesaw effective field theory}}}, }\href {\doibase
  10.1007/JHEP08(2022)017}
  {\bibfield{journal}{\bibinfo{journal}{JHEP}\,}\textbf{\bibinfo{volume}{08}}\,(\bibinfo{year}{2022})\,\bibinfo{pages}{017}},
  \Eprint {http://arxiv.org/abs/2203.10121}{arXiv:2203.10121
  [hep-ph]}\BibitemShut {NoStop}%
\bibitem [{\citenamefont{Bonnefoy} \emph
  {et\,al.}(2022)\citenamefont{Bonnefoy}, \citenamefont{Gendy},
  \citenamefont{Grojean}, and \citenamefont{Ruderman}}]{Bonnefoy:2021tbt}%
  \BibitemOpen
  \bibfield{author}{\bibinfo{author}{\bibfnamefont{Q.}\,\bibnamefont{Bonnefoy}},
  \bibinfo{author}{\bibfnamefont{E.}\,\bibnamefont{Gendy}},
  \bibinfo{author}{\bibfnamefont{C.}\,\bibnamefont{Grojean}},  and
  \bibinfo{author}{\bibfnamefont{J.~T.} \bibnamefont{Ruderman}},
  }\bibfield{title}{\emph {\bibinfo{title}{{Beyond Jarlskog: 699 invariants for
  CP violation in SMEFT}}}, }\href {\doibase 10.1007/JHEP08(2022)032}
  {\bibfield{journal}{\bibinfo{journal}{JHEP}\,}\textbf{\bibinfo{volume}{08}}\,(\bibinfo{year}{2022})\,\bibinfo{pages}{032}},
  \Eprint {http://arxiv.org/abs/2112.03889}{arXiv:2112.03889
  [hep-ph]}\BibitemShut {NoStop}%
\bibitem [{\citenamefont{Bonnefoy} \emph
  {et\,al.}(2023{\natexlab{a}})\citenamefont{Bonnefoy}, \citenamefont{Gendy},
  \citenamefont{Grojean}, and \citenamefont{Ruderman}}]{Bonnefoy:2023bzx}%
  \BibitemOpen
  \bibfield{author}{\bibinfo{author}{\bibfnamefont{Q.}\,\bibnamefont{Bonnefoy}},
  \bibinfo{author}{\bibfnamefont{E.}\,\bibnamefont{Gendy}},
  \bibinfo{author}{\bibfnamefont{C.}\,\bibnamefont{Grojean}},  and
  \bibinfo{author}{\bibfnamefont{J.~T.} \bibnamefont{Ruderman}},
  }\bibfield{title}{\emph {\bibinfo{title}{{Opportunistic CP violation}}},
  }\href {\doibase 10.1007/JHEP06(2023)141}
  {\bibfield{journal}{\bibinfo{journal}{JHEP}\,}\textbf{\bibinfo{volume}{06}}\,(\bibinfo{year}{2023}{\natexlab{a}})\,\bibinfo{pages}{141}},
  \Eprint {http://arxiv.org/abs/2302.07288}{arXiv:2302.07288
  [hep-ph]}\BibitemShut {NoStop}%
\bibitem [{\citenamefont{Jarlskog}(1985{\natexlab{a}})}]{Jarlskog:1985ht}%
  \BibitemOpen
  \bibfield{author}{\bibinfo{author}{\bibfnamefont{C.}\,\bibnamefont{Jarlskog}},
  }\bibfield{title}{\emph {\bibinfo{title}{{Commutator of the Quark Mass
  Matrices in the Standard Electroweak Model and a Measure of Maximal CP
  Nonconservation}}}, }\href {\doibase 10.1103/PhysRevLett.55.1039}
  {\bibfield{journal}{\bibinfo{journal}{Phys. Rev.
  Lett.}\,}\textbf{\bibinfo{volume}{55}}\,(\bibinfo{year}{1985}{\natexlab{a}})\,\bibinfo{pages}{1039}}\BibitemShut
  {NoStop}%
\bibitem [{\citenamefont{Jarlskog}(1985{\natexlab{b}})}]{Jarlskog:1985cw}%
  \BibitemOpen
  \bibfield{author}{\bibinfo{author}{\bibfnamefont{C.}\,\bibnamefont{Jarlskog}},
  }\bibfield{title}{\emph {\bibinfo{title}{{A Basis Independent Formulation of
  the Connection Between Quark Mass Matrices, CP Violation and Experiment}}},
  }\href {\doibase 10.1007/BF01565198} {\bibfield{journal}{\bibinfo{journal}{Z.
  Phys.
  C}\,}\textbf{\bibinfo{volume}{29}}\,(\bibinfo{year}{1985}{\natexlab{b}})\,\bibinfo{pages}{491}}\BibitemShut
  {NoStop}%
\bibitem [{\citenamefont{Bernabeu} \emph
  {et\,al.}(1986)\citenamefont{Bernabeu}, \citenamefont{Branco}, and
  \citenamefont{Gronau}}]{Bernabeu:1986fc}%
  \BibitemOpen
  \bibfield{author}{\bibinfo{author}{\bibfnamefont{J.}\,\bibnamefont{Bernabeu}},
  \bibinfo{author}{\bibfnamefont{G.~C.} \bibnamefont{Branco}},  and
  \bibinfo{author}{\bibfnamefont{M.}\,\bibnamefont{Gronau}},
  }\bibfield{title}{\emph {\bibinfo{title}{{CP Restrictions on Quark Mass
  Matrices}}}, }\href {\doibase 10.1016/0370-2693(86)90659-3}
  {\bibfield{journal}{\bibinfo{journal}{Phys. Lett.
  B}\,}\textbf{\bibinfo{volume}{169}}\,(\bibinfo{year}{1986})\,\bibinfo{pages}{243}}\BibitemShut
  {NoStop}%
\bibitem [{\citenamefont{Marsh}(2016)}]{Marsh:2015xka}%
  \BibitemOpen
  \bibfield{author}{\bibinfo{author}{\bibfnamefont{D.~J.~E.}
  \bibnamefont{Marsh}}, }\bibfield{title}{\emph {\bibinfo{title}{{Axion
  Cosmology}}}, }\href {\doibase 10.1016/j.physrep.2016.06.005}
  {\bibfield{journal}{\bibinfo{journal}{Phys.
  Rept.}\,}\textbf{\bibinfo{volume}{643}}\,(\bibinfo{year}{2016})\,\bibinfo{pages}{1}},
  \Eprint {http://arxiv.org/abs/1510.07633}{arXiv:1510.07633
  [astro-ph.CO]}\BibitemShut {NoStop}%
\bibitem [{\citenamefont{Di~Luzio} \emph
  {et\,al.}(2020)\citenamefont{Di~Luzio}, \citenamefont{Giannotti},
  \citenamefont{Nardi}, and \citenamefont{Visinelli}}]{DiLuzio:2020wdo}%
  \BibitemOpen
  \bibfield{author}{\bibinfo{author}{\bibfnamefont{L.}\,\bibnamefont{Di~Luzio}},
  \bibinfo{author}{\bibfnamefont{M.}\,\bibnamefont{Giannotti}},
  \bibinfo{author}{\bibfnamefont{E.}\,\bibnamefont{Nardi}},  and
  \bibinfo{author}{\bibfnamefont{L.}\,\bibnamefont{Visinelli}},
  }\bibfield{title}{\emph {\bibinfo{title}{{The landscape of QCD axion
  models}}}, }\href {\doibase 10.1016/j.physrep.2020.06.002}
  {\bibfield{journal}{\bibinfo{journal}{Phys.
  Rept.}\,}\textbf{\bibinfo{volume}{870}}\,(\bibinfo{year}{2020})\,\bibinfo{pages}{1}},
  \Eprint {http://arxiv.org/abs/2003.01100}{arXiv:2003.01100
  [hep-ph]}\BibitemShut {NoStop}%
\bibitem [{\citenamefont{'t\,Hooft}(1976)}]{tHooft:1976snw}%
  \BibitemOpen
  \bibfield{author}{\bibinfo{author}{\bibfnamefont{G.}\,\bibnamefont{'t\,Hooft}},
  }\bibfield{title}{\emph {\bibinfo{title}{{Computation of the Quantum Effects
  Due to a Four-Dimensional Pseudoparticle}}}, }\href {\doibase
  10.1103/PhysRevD.14.3432} {\bibfield{journal}{\bibinfo{journal}{Phys. Rev.
  D}\,}\textbf{\bibinfo{volume}{14}}\,(\bibinfo{year}{1976})\,\bibinfo{pages}{3432}},
  \bibinfo{note}{[Erratum: Phys.Rev.D 18, 2199 (1978)]}\BibitemShut {NoStop}%
\bibitem [{\citenamefont{Grzadkowski} \emph
  {et\,al.}(2010)\citenamefont{Grzadkowski}, \citenamefont{Iskrzynski},
  \citenamefont{Misiak}, and \citenamefont{Rosiek}}]{Grzadkowski:2010es}%
  \BibitemOpen
  \bibfield{author}{\bibinfo{author}{\bibfnamefont{B.}\,\bibnamefont{Grzadkowski}},
  \bibinfo{author}{\bibfnamefont{M.}\,\bibnamefont{Iskrzynski}},
  \bibinfo{author}{\bibfnamefont{M.}\,\bibnamefont{Misiak}},  and
  \bibinfo{author}{\bibfnamefont{J.}\,\bibnamefont{Rosiek}},
  }\bibfield{title}{\emph {\bibinfo{title}{{Dimension-Six Terms in the Standard
  Model Lagrangian}}}, }\href {\doibase 10.1007/JHEP10(2010)085}
  {\bibfield{journal}{\bibinfo{journal}{JHEP}\,}\textbf{\bibinfo{volume}{10}}\,(\bibinfo{year}{2010})\,\bibinfo{pages}{085}},
  \Eprint {http://arxiv.org/abs/1008.4884}{arXiv:1008.4884
  [hep-ph]}\BibitemShut {NoStop}%
\bibitem [{\citenamefont{Bigi} and
  \citenamefont{Uraltsev}(1991{\natexlab{b}})}]{Bigi:1991rh}%
  \BibitemOpen
  \bibfield{author}{\bibinfo{author}{\bibfnamefont{I.~I.~Y.}
  \bibnamefont{Bigi}} and \bibinfo{author}{\bibfnamefont{N.~G.}
  \bibnamefont{Uraltsev}}, }\bibfield{title}{\emph {\bibinfo{title}{{Effective
  gluon operators and the dipole moment of the neutron}}}, }\href@noop {}
  {\bibfield{journal}{\bibinfo{journal}{Sov. Phys.
  JETP}\,}\textbf{\bibinfo{volume}{73}}\,(\bibinfo{year}{1991}{\natexlab{b}})\,\bibinfo{pages}{198}}\BibitemShut
  {NoStop}%
\bibitem [{\citenamefont{Witten}(1979)}]{Witten:1979vv}%
  \BibitemOpen
  \bibfield{author}{\bibinfo{author}{\bibfnamefont{E.}\,\bibnamefont{Witten}},
  }\bibfield{title}{\emph {\bibinfo{title}{{Current Algebra Theorems for the
  U(1) Goldstone Boson}}}, }\href {\doibase 10.1016/0550-3213(79)90031-2}
  {\bibfield{journal}{\bibinfo{journal}{Nucl. Phys.
  B}\,}\textbf{\bibinfo{volume}{156}}\,(\bibinfo{year}{1979})\,\bibinfo{pages}{269}}\BibitemShut
  {NoStop}%
\bibitem [{\citenamefont{Shifman} \emph {et\,al.}(1980)\citenamefont{Shifman},
  \citenamefont{Vainshtein}, and \citenamefont{Zakharov}}]{Shifman:1979if}%
  \BibitemOpen
  \bibfield{author}{\bibinfo{author}{\bibfnamefont{M.~A.}
  \bibnamefont{Shifman}}, \bibinfo{author}{\bibfnamefont{A.~I.}
  \bibnamefont{Vainshtein}},  and \bibinfo{author}{\bibfnamefont{V.~I.}
  \bibnamefont{Zakharov}}, }\bibfield{title}{\emph {\bibinfo{title}{{Can
  Confinement Ensure Natural CP Invariance of Strong Interactions?}}}, }\href
  {\doibase 10.1016/0550-3213(80)90209-6}
  {\bibfield{journal}{\bibinfo{journal}{Nucl. Phys.
  B}\,}\textbf{\bibinfo{volume}{166}}\,(\bibinfo{year}{1980})\,\bibinfo{pages}{493}}\BibitemShut
  {NoStop}%
\bibitem [{\citenamefont{Pospelov}(1998)}]{Pospelov:1997uv}%
  \BibitemOpen
  \bibfield{author}{\bibinfo{author}{\bibfnamefont{M.}\,\bibnamefont{Pospelov}},
  }\bibfield{title}{\emph {\bibinfo{title}{{CP odd interaction of axion with
  matter}}}, }\href {\doibase 10.1103/PhysRevD.58.097703}
  {\bibfield{journal}{\bibinfo{journal}{Phys. Rev.
  D}\,}\textbf{\bibinfo{volume}{58}}\,(\bibinfo{year}{1998})\,\bibinfo{pages}{097703}},
  \Eprint
  {http://arxiv.org/abs/hep-ph/9707431}{arXiv:hep-ph/9707431}\BibitemShut
  {NoStop}%
\bibitem [{\citenamefont{Pospelov} and
  \citenamefont{Ritz}(2005)}]{Pospelov:2005pr}%
  \BibitemOpen
  \bibfield{author}{\bibinfo{author}{\bibfnamefont{M.}\,\bibnamefont{Pospelov}}
  and \bibinfo{author}{\bibfnamefont{A.}\,\bibnamefont{Ritz}},
  }\bibfield{title}{\emph {\bibinfo{title}{{Electric dipole moments as probes
  of new physics}}}, }\href {\doibase 10.1016/j.aop.2005.04.002}
  {\bibfield{journal}{\bibinfo{journal}{Annals
  Phys.}\,}\textbf{\bibinfo{volume}{318}}\,(\bibinfo{year}{2005})\,\bibinfo{pages}{119}},
  \Eprint
  {http://arxiv.org/abs/hep-ph/0504231}{arXiv:hep-ph/0504231}\BibitemShut
  {NoStop}%
\bibitem [{\citenamefont{Georgi} \emph {et\,al.}(1981)\citenamefont{Georgi},
  \citenamefont{Hall}, and \citenamefont{Wise}}]{Georgi:1981pu}%
  \BibitemOpen
  \bibfield{author}{\bibinfo{author}{\bibfnamefont{H.~M.}
  \bibnamefont{Georgi}}, \bibinfo{author}{\bibfnamefont{L.~J.}
  \bibnamefont{Hall}},  and \bibinfo{author}{\bibfnamefont{M.~B.}
  \bibnamefont{Wise}}, }\bibfield{title}{\emph {\bibinfo{title}{{Grand Unified
  Models With an Automatic {Peccei-Quinn} Symmetry}}}, }\href {\doibase
  10.1016/0550-3213(81)90433-8} {\bibfield{journal}{\bibinfo{journal}{Nucl.
  Phys.
  B}\,}\textbf{\bibinfo{volume}{192}}\,(\bibinfo{year}{1981})\,\bibinfo{pages}{409}}\BibitemShut
  {NoStop}%
\bibitem [{\citenamefont{Holman} \emph {et\,al.}(1992)\citenamefont{Holman},
  \citenamefont{Hsu}, \citenamefont{Kephart}, \citenamefont{Kolb},
  \citenamefont{Watkins}, and \citenamefont{Widrow}}]{Holman:1992us}%
  \BibitemOpen
  \bibfield{author}{\bibinfo{author}{\bibfnamefont{R.}\,\bibnamefont{Holman}},
  \bibinfo{author}{\bibfnamefont{S.~D.~H.} \bibnamefont{Hsu}},
  \bibinfo{author}{\bibfnamefont{T.~W.} \bibnamefont{Kephart}},
  \bibinfo{author}{\bibfnamefont{E.~W.} \bibnamefont{Kolb}},
  \bibinfo{author}{\bibfnamefont{R.}\,\bibnamefont{Watkins}},  and
  \bibinfo{author}{\bibfnamefont{L.~M.} \bibnamefont{Widrow}},
  }\bibfield{title}{\emph {\bibinfo{title}{{Solutions to the strong CP problem
  in a world with gravity}}}, }\href {\doibase 10.1016/0370-2693(92)90491-L}
  {\bibfield{journal}{\bibinfo{journal}{Phys.
  Lett.}\,}\textbf{\bibinfo{volume}{B282}}\,(\bibinfo{year}{1992})\,\bibinfo{pages}{132}},
  \Eprint
  {http://arxiv.org/abs/hep-ph/9203206}{arXiv:hep-ph/9203206}\BibitemShut
  {NoStop}%
\bibitem [{\citenamefont{Kamionkowski} and
  \citenamefont{March-Russell}(1992)}]{Kamionkowski:1992mf}%
  \BibitemOpen
  \bibfield{author}{\bibinfo{author}{\bibfnamefont{M.}\,\bibnamefont{Kamionkowski}}
  and \bibinfo{author}{\bibfnamefont{J.}\,\bibnamefont{March-Russell}},
  }\bibfield{title}{\emph {\bibinfo{title}{{Planck scale physics and the
  Peccei-Quinn mechanism}}}, }\href {\doibase 10.1016/0370-2693(92)90492-M}
  {\bibfield{journal}{\bibinfo{journal}{Phys. Lett.
  B}\,}\textbf{\bibinfo{volume}{282}}\,(\bibinfo{year}{1992})\,\bibinfo{pages}{137}},
  \Eprint
  {http://arxiv.org/abs/hep-th/9202003}{arXiv:hep-th/9202003}\BibitemShut
  {NoStop}%
\bibitem [{\citenamefont{Barr} and \citenamefont{Seckel}(1992)}]{Barr:1992qq}%
  \BibitemOpen
  \bibfield{author}{\bibinfo{author}{\bibfnamefont{S.~M.} \bibnamefont{Barr}}
  and \bibinfo{author}{\bibfnamefont{D.}\,\bibnamefont{Seckel}},
  }\bibfield{title}{\emph {\bibinfo{title}{{Planck scale corrections to axion
  models}}}, }\href {\doibase 10.1103/PhysRevD.46.539}
  {\bibfield{journal}{\bibinfo{journal}{Phys.
  Rev.}\,}\textbf{\bibinfo{volume}{D46}}\,(\bibinfo{year}{1992})\,\bibinfo{pages}{539}}\BibitemShut
  {NoStop}%
\bibitem [{\citenamefont{Ghigna} \emph {et\,al.}(1992)\citenamefont{Ghigna},
  \citenamefont{Lusignoli}, and \citenamefont{Roncadelli}}]{Ghigna:1992iv}%
  \BibitemOpen
  \bibfield{author}{\bibinfo{author}{\bibfnamefont{S.}\,\bibnamefont{Ghigna}},
  \bibinfo{author}{\bibfnamefont{M.}\,\bibnamefont{Lusignoli}},  and
  \bibinfo{author}{\bibfnamefont{M.}\,\bibnamefont{Roncadelli}},
  }\bibfield{title}{\emph {\bibinfo{title}{{Instability of the invisible
  axion}}}, }\href {\doibase 10.1016/0370-2693(92)90019-Z}
  {\bibfield{journal}{\bibinfo{journal}{Phys.
  Lett.}\,}\textbf{\bibinfo{volume}{B283}}\,(\bibinfo{year}{1992})\,\bibinfo{pages}{278}}\BibitemShut
  {NoStop}%
\bibitem [{\citenamefont{Bonnefoy} \emph
  {et\,al.}(2023{\natexlab{b}})\citenamefont{Bonnefoy}, \citenamefont{Grojean},
  and \citenamefont{Kley}}]{Bonnefoy:2022rik}%
  \BibitemOpen
  \bibfield{author}{\bibinfo{author}{\bibfnamefont{Q.}\,\bibnamefont{Bonnefoy}},
  \bibinfo{author}{\bibfnamefont{C.}\,\bibnamefont{Grojean}},  and
  \bibinfo{author}{\bibfnamefont{J.}\,\bibnamefont{Kley}},
  }\bibfield{title}{\emph {\bibinfo{title}{{Shift-Invariant Orders of an
  Axionlike Particle}}}, }\href {\doibase 10.1103/PhysRevLett.130.111803}
  {\bibfield{journal}{\bibinfo{journal}{Phys. Rev.
  Lett.}\,}\textbf{\bibinfo{volume}{130}}\,(\bibinfo{year}{2023}{\natexlab{b}})\,\bibinfo{pages}{111803}},
  \Eprint {http://arxiv.org/abs/2206.04182}{arXiv:2206.04182
  [hep-ph]}\BibitemShut {NoStop}%
\bibitem [{\citenamefont{Bonnefoy}(2023)}]{Bonnefoy:2022vop}%
  \BibitemOpen
  \bibfield{author}{\bibinfo{author}{\bibfnamefont{Q.}\,\bibnamefont{Bonnefoy}},
  }\bibfield{title}{\emph {\bibinfo{title}{{Heavy fields and the axion quality
  problem}}}, }\href {\doibase 10.1103/PhysRevD.108.035023}
  {\bibfield{journal}{\bibinfo{journal}{Phys. Rev.
  D}\,}\textbf{\bibinfo{volume}{108}}\,(\bibinfo{year}{2023})\,\bibinfo{pages}{035023}},
  \Eprint {http://arxiv.org/abs/2212.00102}{arXiv:2212.00102
  [hep-ph]}\BibitemShut {NoStop}%
\bibitem [{\citenamefont{Callan} \emph {et\,al.}(1978)\citenamefont{Callan},
  \citenamefont{Dashen}, and \citenamefont{Gross}}]{Callan:1977gz}%
  \BibitemOpen
  \bibfield{author}{\bibinfo{author}{\bibfnamefont{C.~G.} \bibnamefont{Callan},
  \bibfnamefont{Jr.}}, \bibinfo{author}{\bibfnamefont{R.~F.}
  \bibnamefont{Dashen}},  and \bibinfo{author}{\bibfnamefont{D.~J.}
  \bibnamefont{Gross}}, }\bibfield{title}{\emph {\bibinfo{title}{{Toward a
  Theory of the Strong Interactions}}}, }\href {\doibase
  10.1103/PhysRevD.17.2717} {\bibfield{journal}{\bibinfo{journal}{Phys. Rev.
  D}\,}\textbf{\bibinfo{volume}{17}}\,(\bibinfo{year}{1978})\,\bibinfo{pages}{2717}}\BibitemShut
  {NoStop}%
\bibitem [{\citenamefont{Gherghetta} \emph
  {et\,al.}(2020)\citenamefont{Gherghetta}, \citenamefont{Khoze},
  \citenamefont{Pomarol}, and \citenamefont{Shirman}}]{Gherghetta:2020keg}%
  \BibitemOpen
  \bibfield{author}{\bibinfo{author}{\bibfnamefont{T.}\,\bibnamefont{Gherghetta}},
  \bibinfo{author}{\bibfnamefont{V.~V.} \bibnamefont{Khoze}},
  \bibinfo{author}{\bibfnamefont{A.}\,\bibnamefont{Pomarol}},  and
  \bibinfo{author}{\bibfnamefont{Y.}\,\bibnamefont{Shirman}},
  }\bibfield{title}{\emph {\bibinfo{title}{{The Axion Mass from 5D Small
  Instantons}}}, }\href {\doibase 10.1007/JHEP03(2020)063}
  {\bibfield{journal}{\bibinfo{journal}{JHEP}\,}\textbf{\bibinfo{volume}{03}}\,(\bibinfo{year}{2020})\,\bibinfo{pages}{063}},
  \Eprint {http://arxiv.org/abs/2001.05610}{arXiv:2001.05610
  [hep-ph]}\BibitemShut {NoStop}%
\bibitem [{\citenamefont{Buras} \emph {et\,al.}(2001)\citenamefont{Buras},
  \citenamefont{Gambino}, \citenamefont{Gorbahn}, \citenamefont{Jager}, and
  \citenamefont{Silvestrini}}]{Buras:2000dm}%
  \BibitemOpen
  \bibfield{author}{\bibinfo{author}{\bibfnamefont{A.~J.} \bibnamefont{Buras}},
  \bibinfo{author}{\bibfnamefont{P.}\,\bibnamefont{Gambino}},
  \bibinfo{author}{\bibfnamefont{M.}\,\bibnamefont{Gorbahn}},
  \bibinfo{author}{\bibfnamefont{S.}\,\bibnamefont{Jager}},  and
  \bibinfo{author}{\bibfnamefont{L.}\,\bibnamefont{Silvestrini}},
  }\bibfield{title}{\emph {\bibinfo{title}{{Universal unitarity triangle and
  physics beyond the standard model}}}, }\href {\doibase
  10.1016/S0370-2693(01)00061-2} {\bibfield{journal}{\bibinfo{journal}{Phys.
  Lett.
  B}\,}\textbf{\bibinfo{volume}{500}}\,(\bibinfo{year}{2001})\,\bibinfo{pages}{161}},
  \Eprint
  {http://arxiv.org/abs/hep-ph/0007085}{arXiv:hep-ph/0007085}\BibitemShut
  {NoStop}%
\bibitem [{\citenamefont{D'Ambrosio} \emph
  {et\,al.}(2002)\citenamefont{D'Ambrosio}, \citenamefont{Giudice},
  \citenamefont{Isidori}, and \citenamefont{Strumia}}]{DAmbrosio:2002vsn}%
  \BibitemOpen
  \bibfield{author}{\bibinfo{author}{\bibfnamefont{G.}\,\bibnamefont{D'Ambrosio}},
  \bibinfo{author}{\bibfnamefont{G.~F.} \bibnamefont{Giudice}},
  \bibinfo{author}{\bibfnamefont{G.}\,\bibnamefont{Isidori}},  and
  \bibinfo{author}{\bibfnamefont{A.}\,\bibnamefont{Strumia}},
  }\bibfield{title}{\emph {\bibinfo{title}{{Minimal flavor violation: An
  Effective field theory approach}}}, }\href {\doibase
  10.1016/S0550-3213(02)00836-2} {\bibfield{journal}{\bibinfo{journal}{Nucl.
  Phys.
  B}\,}\textbf{\bibinfo{volume}{645}}\,(\bibinfo{year}{2002})\,\bibinfo{pages}{155}},
  \Eprint
  {http://arxiv.org/abs/hep-ph/0207036}{arXiv:hep-ph/0207036}\BibitemShut
  {NoStop}%
\bibitem [{\citenamefont{Isidori} and
  \citenamefont{Straub}(2012)}]{Isidori:2012ts}%
  \BibitemOpen
  \bibfield{author}{\bibinfo{author}{\bibfnamefont{G.}\,\bibnamefont{Isidori}}
  and \bibinfo{author}{\bibfnamefont{D.~M.} \bibnamefont{Straub}},
  }\bibfield{title}{\emph {\bibinfo{title}{{Minimal Flavour Violation and
  Beyond}}}, }\href {\doibase 10.1140/epjc/s10052-012-2103-1}
  {\bibfield{journal}{\bibinfo{journal}{Eur. Phys. J.
  C}\,}\textbf{\bibinfo{volume}{72}}\,(\bibinfo{year}{2012})\,\bibinfo{pages}{2103}},
  \Eprint {http://arxiv.org/abs/1202.0464}{arXiv:1202.0464
  [hep-ph]}\BibitemShut {NoStop}%
\bibitem [{\citenamefont{Froggatt} and
  \citenamefont{Nielsen}(1979)}]{Froggatt:1978nt}%
  \BibitemOpen
  \bibfield{author}{\bibinfo{author}{\bibfnamefont{C.~D.}
  \bibnamefont{Froggatt}} and \bibinfo{author}{\bibfnamefont{H.~B.}
  \bibnamefont{Nielsen}}, }\bibfield{title}{\emph {\bibinfo{title}{{Hierarchy
  of Quark Masses, Cabibbo Angles and CP Violation}}}, }\href {\doibase
  10.1016/0550-3213(79)90316-X} {\bibfield{journal}{\bibinfo{journal}{Nucl.
  Phys.
  B}\,}\textbf{\bibinfo{volume}{147}}\,(\bibinfo{year}{1979})\,\bibinfo{pages}{277}}\BibitemShut
  {NoStop}%
\bibitem [{\citenamefont{Bordone} \emph {et\,al.}(2020)\citenamefont{Bordone},
  \citenamefont{Cat\`a}, and \citenamefont{Feldmann}}]{Bordone:2019uzc}%
  \BibitemOpen
  \bibfield{author}{\bibinfo{author}{\bibfnamefont{M.}\,\bibnamefont{Bordone}},
  \bibinfo{author}{\bibfnamefont{O.}\,\bibnamefont{Cat\`a}},  and
  \bibinfo{author}{\bibfnamefont{T.}\,\bibnamefont{Feldmann}},
  }\bibfield{title}{\emph {\bibinfo{title}{{Effective Theory Approach to New
  Physics with Flavour: General Framework and a Leptoquark Example}}}, }\href
  {\doibase 10.1007/JHEP01(2020)067}
  {\bibfield{journal}{\bibinfo{journal}{JHEP}\,}\textbf{\bibinfo{volume}{01}}\,(\bibinfo{year}{2020})\,\bibinfo{pages}{067}},
  \Eprint {http://arxiv.org/abs/1910.02641}{arXiv:1910.02641
  [hep-ph]}\BibitemShut {NoStop}%
\bibitem [{\citenamefont{Dekens} and
  \citenamefont{de\,Vries}(2013)}]{Dekens:2013zca}%
  \BibitemOpen
  \bibfield{author}{\bibinfo{author}{\bibfnamefont{W.}\,\bibnamefont{Dekens}}
  and \bibinfo{author}{\bibfnamefont{J.}\,\bibnamefont{de\,Vries}},
  }\bibfield{title}{\emph {\bibinfo{title}{{Renormalization Group Running of
  Dimension-Six Sources of Parity and Time-Reversal Violation}}}, }\href
  {\doibase 10.1007/JHEP05(2013)149}
  {\bibfield{journal}{\bibinfo{journal}{JHEP}\,}\textbf{\bibinfo{volume}{05}}\,(\bibinfo{year}{2013})\,\bibinfo{pages}{149}},
  \Eprint {http://arxiv.org/abs/1303.3156}{arXiv:1303.3156
  [hep-ph]}\BibitemShut {NoStop}%
\bibitem [{\citenamefont{Cirigliano} \emph
  {et\,al.}(2016)\citenamefont{Cirigliano}, \citenamefont{Dekens},
  \citenamefont{de\,Vries}, and
  \citenamefont{Mereghetti}}]{Cirigliano:2016nyn}%
  \BibitemOpen
  \bibfield{author}{\bibinfo{author}{\bibfnamefont{V.}\,\bibnamefont{Cirigliano}},
  \bibinfo{author}{\bibfnamefont{W.}\,\bibnamefont{Dekens}},
  \bibinfo{author}{\bibfnamefont{J.}\,\bibnamefont{de\,Vries}},  and
  \bibinfo{author}{\bibfnamefont{E.}\,\bibnamefont{Mereghetti}},
  }\bibfield{title}{\emph {\bibinfo{title}{{Constraining the top-Higgs sector
  of the Standard Model Effective Field Theory}}}, }\href {\doibase
  10.1103/PhysRevD.94.034031} {\bibfield{journal}{\bibinfo{journal}{Phys. Rev.
  D}\,}\textbf{\bibinfo{volume}{94}}\,(\bibinfo{year}{2016})\,\bibinfo{pages}{034031}},
  \Eprint {http://arxiv.org/abs/1605.04311}{arXiv:1605.04311
  [hep-ph]}\BibitemShut {NoStop}%
\bibitem [{\citenamefont{Alioli} \emph {et\,al.}(2017)\citenamefont{Alioli},
  \citenamefont{Cirigliano}, \citenamefont{Dekens}, \citenamefont{de\,Vries},
  and \citenamefont{Mereghetti}}]{Alioli:2017ces}%
  \BibitemOpen
  \bibfield{author}{\bibinfo{author}{\bibfnamefont{S.}\,\bibnamefont{Alioli}},
  \bibinfo{author}{\bibfnamefont{V.}\,\bibnamefont{Cirigliano}},
  \bibinfo{author}{\bibfnamefont{W.}\,\bibnamefont{Dekens}},
  \bibinfo{author}{\bibfnamefont{J.}\,\bibnamefont{de\,Vries}},  and
  \bibinfo{author}{\bibfnamefont{E.}\,\bibnamefont{Mereghetti}},
  }\bibfield{title}{\emph {\bibinfo{title}{{Right-handed charged currents in
  the era of the Large Hadron Collider}}}, }\href {\doibase
  10.1007/JHEP05(2017)086}
  {\bibfield{journal}{\bibinfo{journal}{JHEP}\,}\textbf{\bibinfo{volume}{05}}\,(\bibinfo{year}{2017})\,\bibinfo{pages}{086}},
  \Eprint {http://arxiv.org/abs/1703.04751}{arXiv:1703.04751
  [hep-ph]}\BibitemShut {NoStop}%
\bibitem [{\citenamefont{de\,Vries} \emph
  {et\,al.}(2019)\citenamefont{de\,Vries}, \citenamefont{Draper},
  \citenamefont{Fuyuto}, \citenamefont{Kozaczuk}, and
  \citenamefont{Sutherland}}]{deVries:2018mgf}%
  \BibitemOpen
  \bibfield{author}{\bibinfo{author}{\bibfnamefont{J.}\,\bibnamefont{de\,Vries}},
  \bibinfo{author}{\bibfnamefont{P.}\,\bibnamefont{Draper}},
  \bibinfo{author}{\bibfnamefont{K.}\,\bibnamefont{Fuyuto}},
  \bibinfo{author}{\bibfnamefont{J.}\,\bibnamefont{Kozaczuk}},  and
  \bibinfo{author}{\bibfnamefont{D.}\,\bibnamefont{Sutherland}},
  }\bibfield{title}{\emph {\bibinfo{title}{{Indirect Signs of the Peccei-Quinn
  Mechanism}}}, }\href {\doibase 10.1103/PhysRevD.99.015042}
  {\bibfield{journal}{\bibinfo{journal}{Phys. Rev.
  D}\,}\textbf{\bibinfo{volume}{99}}\,(\bibinfo{year}{2019})\,\bibinfo{pages}{015042}},
  \Eprint {http://arxiv.org/abs/1809.10143}{arXiv:1809.10143
  [hep-ph]}\BibitemShut {NoStop}%
\bibitem [{\citenamefont{Kley} \emph {et\,al.}(2022)\citenamefont{Kley},
  \citenamefont{Theil}, \citenamefont{Venturini}, and
  \citenamefont{Weiler}}]{Kley:2021yhn}%
  \BibitemOpen
  \bibfield{author}{\bibinfo{author}{\bibfnamefont{J.}\,\bibnamefont{Kley}},
  \bibinfo{author}{\bibfnamefont{T.}\,\bibnamefont{Theil}},
  \bibinfo{author}{\bibfnamefont{E.}\,\bibnamefont{Venturini}},  and
  \bibinfo{author}{\bibfnamefont{A.}\,\bibnamefont{Weiler}},
  }\bibfield{title}{\emph {\bibinfo{title}{{Electric dipole moments at one-loop
  in the dimension-6 SMEFT}}}, }\href {\doibase
  10.1140/epjc/s10052-022-10861-5} {\bibfield{journal}{\bibinfo{journal}{Eur.
  Phys. J.
  C}\,}\textbf{\bibinfo{volume}{82}}\,(\bibinfo{year}{2022})\,\bibinfo{pages}{926}},
  \Eprint {http://arxiv.org/abs/2109.15085}{arXiv:2109.15085
  [hep-ph]}\BibitemShut {NoStop}%
\bibitem [{\citenamefont{Coleman}(1979)}]{Coleman:1978ae}%
  \BibitemOpen
  \bibfield{author}{\bibinfo{author}{\bibfnamefont{S.~R.}
  \bibnamefont{Coleman}}, }\bibfield{title}{\emph {\bibinfo{title}{{The Uses of
  Instantons}}}, }\href@noop {} {\bibfield{journal}{\bibinfo{journal}{Subnucl.
  Ser.}\,}\textbf{\bibinfo{volume}{15}}\,(\bibinfo{year}{1979})\,\bibinfo{pages}{805}}\BibitemShut
  {NoStop}%
\bibitem [{\citenamefont{Vainshtein} \emph
  {et\,al.}(1982)\citenamefont{Vainshtein}, \citenamefont{Zakharov},
  \citenamefont{Novikov}, and \citenamefont{Shifman}}]{Vainshtein:1981wh}%
  \BibitemOpen
  \bibfield{author}{\bibinfo{author}{\bibfnamefont{A.~I.}
  \bibnamefont{Vainshtein}}, \bibinfo{author}{\bibfnamefont{V.~I.}
  \bibnamefont{Zakharov}}, \bibinfo{author}{\bibfnamefont{V.~A.}
  \bibnamefont{Novikov}},  and \bibinfo{author}{\bibfnamefont{M.~A.}
  \bibnamefont{Shifman}}, }\bibfield{title}{\emph {\bibinfo{title}{{ABC's of
  Instantons}}}, }\href {\doibase 10.1070/PU1982v025n04ABEH004533}
  {\bibfield{journal}{\bibinfo{journal}{Sov. Phys.
  Usp.}\,}\textbf{\bibinfo{volume}{25}}\,(\bibinfo{year}{1982})\,\bibinfo{pages}{195}}\BibitemShut
  {NoStop}%
\bibitem [{\citenamefont{Shifman}(2022)}]{Shifman:2022shi}%
  \BibitemOpen
  \bibfield{author}{\bibinfo{author}{\bibfnamefont{M.}\,\bibnamefont{Shifman}},
  }\href {\doibase 10.1017/9781108885911} {\emph {\bibinfo{title}{{Advanced
  Topics in Quantum Field Theory}}}}\,(\bibinfo{publisher}{Cambridge University
  Press}, \bibinfo{year}{2022})\BibitemShut {NoStop}%
\bibitem [{\citenamefont{Vandoren} and
  \citenamefont{van\,Nieuwenhuizen}(2008)}]{Vandoren:2008xg}%
  \BibitemOpen
  \bibfield{author}{\bibinfo{author}{\bibfnamefont{S.}\,\bibnamefont{Vandoren}}
  and \bibinfo{author}{\bibfnamefont{P.}\,\bibnamefont{van\,Nieuwenhuizen}},
  }\bibfield{title}{\emph {\bibinfo{title}{{Lectures on instantons}}},
  }\href@noop {} {\,(\bibinfo{year}{2008})}, \Eprint
  {http://arxiv.org/abs/0802.1862}{arXiv:0802.1862 [hep-th]}\BibitemShut
  {NoStop}%
\bibitem [{\citenamefont{Dorey} \emph {et\,al.}(2002)\citenamefont{Dorey},
  \citenamefont{Hollowood}, \citenamefont{Khoze}, and
  \citenamefont{Mattis}}]{Dorey:2002ik}%
  \BibitemOpen
  \bibfield{author}{\bibinfo{author}{\bibfnamefont{N.}\,\bibnamefont{Dorey}},
  \bibinfo{author}{\bibfnamefont{T.~J.} \bibnamefont{Hollowood}},
  \bibinfo{author}{\bibfnamefont{V.~V.} \bibnamefont{Khoze}},  and
  \bibinfo{author}{\bibfnamefont{M.~P.} \bibnamefont{Mattis}},
  }\bibfield{title}{\emph {\bibinfo{title}{{The Calculus of many instantons}}},
  }\href {\doibase 10.1016/S0370-1573(02)00301-0}
  {\bibfield{journal}{\bibinfo{journal}{Phys.
  Rept.}\,}\textbf{\bibinfo{volume}{371}}\,(\bibinfo{year}{2002})\,\bibinfo{pages}{231}},
  \Eprint
  {http://arxiv.org/abs/hep-th/0206063}{arXiv:hep-th/0206063}\BibitemShut
  {NoStop}%
\bibitem [{\citenamefont{Tong}(2005)}]{Tong:2005un}%
  \BibitemOpen
  \bibfield{author}{\bibinfo{author}{\bibfnamefont{D.}\,\bibnamefont{Tong}},
  }\bibfield{title}{\emph {\bibinfo{title}{{TASI lectures on solitons:
  Instantons, monopoles, vortices and kinks}}}, }in \href@noop {} {\emph
  {\bibinfo{booktitle}{{Theoretical Advanced Study Institute in Elementary
  Particle Physics}: {Many Dimensions of String
  Theory}}}}\,(\bibinfo{year}{2005})\,\Eprint
  {http://arxiv.org/abs/hep-th/0509216}{arXiv:hep-th/0509216}\BibitemShut
  {NoStop}%
\bibitem [{\citenamefont{Reece}(2023)}]{Reece:2023czb}%
  \BibitemOpen
  \bibfield{author}{\bibinfo{author}{\bibfnamefont{M.}\,\bibnamefont{Reece}},
  }\bibfield{title}{\emph {\bibinfo{title}{{TASI Lectures: (No) Global
  Symmetries to Axion Physics}}}, }\href@noop {} {\,(\bibinfo{year}{2023})},
  \Eprint {http://arxiv.org/abs/2304.08512}{arXiv:2304.08512
  [hep-ph]}\BibitemShut {NoStop}%
\bibitem [{\citenamefont{Belavin} \emph {et\,al.}(1975)\citenamefont{Belavin},
  \citenamefont{Polyakov}, \citenamefont{Schwartz}, and
  \citenamefont{Tyupkin}}]{Belavin:1975fg}%
  \BibitemOpen
  \bibfield{author}{\bibinfo{author}{\bibfnamefont{A.~A.}
  \bibnamefont{Belavin}}, \bibinfo{author}{\bibfnamefont{A.~M.}
  \bibnamefont{Polyakov}}, \bibinfo{author}{\bibfnamefont{A.~S.}
  \bibnamefont{Schwartz}},  and \bibinfo{author}{\bibfnamefont{Y.~S.}
  \bibnamefont{Tyupkin}}, }\bibfield{title}{\emph
  {\bibinfo{title}{{Pseudoparticle Solutions of the Yang-Mills Equations}}},
  }\href {\doibase 10.1016/0370-2693(75)90163-X}
  {\bibfield{journal}{\bibinfo{journal}{Phys. Lett.
  B}\,}\textbf{\bibinfo{volume}{59}}\,(\bibinfo{year}{1975})\,\bibinfo{pages}{85}}\BibitemShut
  {NoStop}%
\bibitem [{\citenamefont{'t\,Hooft}(1986)}]{tHooft:1986ooh}%
  \BibitemOpen
  \bibfield{author}{\bibinfo{author}{\bibfnamefont{G.}\,\bibnamefont{'t\,Hooft}},
  }\bibfield{title}{\emph {\bibinfo{title}{{How Instantons Solve the U(1)
  Problem}}}, }\href {\doibase 10.1016/0370-1573(86)90117-1}
  {\bibfield{journal}{\bibinfo{journal}{Phys.
  Rept.}\,}\textbf{\bibinfo{volume}{142}}\,(\bibinfo{year}{1986})\,\bibinfo{pages}{357}}\BibitemShut
  {NoStop}%
\bibitem [{\citenamefont{Shuryak}(1982{\natexlab{a}})}]{Shuryak:1981ff}%
  \BibitemOpen
  \bibfield{author}{\bibinfo{author}{\bibfnamefont{E.~V.}
  \bibnamefont{Shuryak}}, }\bibfield{title}{\emph {\bibinfo{title}{{The Role of
  Instantons in Quantum Chromodynamics. 1. Physical Vacuum}}}, }\href {\doibase
  10.1016/0550-3213(82)90478-3} {\bibfield{journal}{\bibinfo{journal}{Nucl.
  Phys.
  B}\,}\textbf{\bibinfo{volume}{203}}\,(\bibinfo{year}{1982}{\natexlab{a}})\,\bibinfo{pages}{93}}\BibitemShut
  {NoStop}%
\bibitem [{\citenamefont{Shuryak}(1982{\natexlab{b}})}]{Shuryak:1982dp}%
  \BibitemOpen
  \bibfield{author}{\bibinfo{author}{\bibfnamefont{E.~V.}
  \bibnamefont{Shuryak}}, }\bibfield{title}{\emph {\bibinfo{title}{{The Role of
  Instantons in Quantum Chromodynamics. 2. Hadronic Structure}}}, }\href
  {\doibase 10.1016/0550-3213(82)90479-5}
  {\bibfield{journal}{\bibinfo{journal}{Nucl. Phys.
  B}\,}\textbf{\bibinfo{volume}{203}}\,(\bibinfo{year}{1982}{\natexlab{b}})\,\bibinfo{pages}{116}}\BibitemShut
  {NoStop}%
\bibitem [{\citenamefont{Diakonov} and
  \citenamefont{Petrov}(1984{\natexlab{a}})}]{Diakonov:1983hh}%
  \BibitemOpen
  \bibfield{author}{\bibinfo{author}{\bibfnamefont{D.}\,\bibnamefont{Diakonov}}
  and \bibinfo{author}{\bibfnamefont{V.~Y.} \bibnamefont{Petrov}},
  }\bibfield{title}{\emph {\bibinfo{title}{{Instanton Based Vacuum from Feynman
  Variational Principle}}}, }\href {\doibase 10.1016/0550-3213(84)90432-2}
  {\bibfield{journal}{\bibinfo{journal}{Nucl. Phys.
  B}\,}\textbf{\bibinfo{volume}{245}}\,(\bibinfo{year}{1984}{\natexlab{a}})\,\bibinfo{pages}{259}}\BibitemShut
  {NoStop}%
\bibitem [{\citenamefont{Diakonov} and
  \citenamefont{Petrov}(1984{\natexlab{b}})}]{Diakonov:1984vw}%
  \BibitemOpen
  \bibfield{author}{\bibinfo{author}{\bibfnamefont{D.}\,\bibnamefont{Diakonov}}
  and \bibinfo{author}{\bibfnamefont{V.~Y.} \bibnamefont{Petrov}},
  }\bibfield{title}{\emph {\bibinfo{title}{{Chiral Condensate in the Instanton
  Vacuum}}}, }\href {\doibase 10.1016/0370-2693(84)90132-1}
  {\bibfield{journal}{\bibinfo{journal}{Phys. Lett.
  B}\,}\textbf{\bibinfo{volume}{147}}\,(\bibinfo{year}{1984}{\natexlab{b}})\,\bibinfo{pages}{351}}\BibitemShut
  {NoStop}%
\bibitem [{\citenamefont{Sch\"afer} and
  \citenamefont{Shuryak}(1996)}]{Schafer:1995pz}%
  \BibitemOpen
  \bibfield{author}{\bibinfo{author}{\bibfnamefont{T.}\,\bibnamefont{Sch\"afer}}
  and \bibinfo{author}{\bibfnamefont{E.~V.} \bibnamefont{Shuryak}},
  }\bibfield{title}{\emph {\bibinfo{title}{{The Interacting instanton liquid in
  QCD at zero and finite temperature}}}, }\href {\doibase
  10.1103/PhysRevD.53.6522} {\bibfield{journal}{\bibinfo{journal}{Phys. Rev.
  D}\,}\textbf{\bibinfo{volume}{53}}\,(\bibinfo{year}{1996})\,\bibinfo{pages}{6522}},
  \Eprint
  {http://arxiv.org/abs/hep-ph/9509337}{arXiv:hep-ph/9509337}\BibitemShut
  {NoStop}%
\bibitem [{\citenamefont{Sch\"afer}(2002)}]{Schafer:2002af}%
  \BibitemOpen
  \bibfield{author}{\bibinfo{author}{\bibfnamefont{T.}\,\bibnamefont{Sch\"afer}},
  }\bibfield{title}{\emph {\bibinfo{title}{{Instantons in QCD with many
  colors}}}, }\href {\doibase 10.1103/PhysRevD.66.076009}
  {\bibfield{journal}{\bibinfo{journal}{Phys. Rev.
  D}\,}\textbf{\bibinfo{volume}{66}}\,(\bibinfo{year}{2002})\,\bibinfo{pages}{076009}},
  \Eprint
  {http://arxiv.org/abs/hep-ph/0206062}{arXiv:hep-ph/0206062}\BibitemShut
  {NoStop}%
\bibitem [{\citenamefont{Bernard}(1979)}]{Bernard:1979qt}%
  \BibitemOpen
  \bibfield{author}{\bibinfo{author}{\bibfnamefont{C.~W.}
  \bibnamefont{Bernard}}, }\bibfield{title}{\emph {\bibinfo{title}{{Gauge Zero
  Modes, Instanton Determinants, and QCD Calculations}}}, }\href {\doibase
  10.1103/PhysRevD.19.3013} {\bibfield{journal}{\bibinfo{journal}{Phys. Rev.
  D}\,}\textbf{\bibinfo{volume}{19}}\,(\bibinfo{year}{1979})\,\bibinfo{pages}{3013}}\BibitemShut
  {NoStop}%
\bibitem [{\citenamefont{Novikov} \emph {et\,al.}(1983)\citenamefont{Novikov},
  \citenamefont{Shifman}, \citenamefont{Vainshtein}, and
  \citenamefont{Zakharov}}]{Novikov:1983uc}%
  \BibitemOpen
  \bibfield{author}{\bibinfo{author}{\bibfnamefont{V.~A.}
  \bibnamefont{Novikov}}, \bibinfo{author}{\bibfnamefont{M.~A.}
  \bibnamefont{Shifman}}, \bibinfo{author}{\bibfnamefont{A.~I.}
  \bibnamefont{Vainshtein}},  and \bibinfo{author}{\bibfnamefont{V.~I.}
  \bibnamefont{Zakharov}}, }\bibfield{title}{\emph {\bibinfo{title}{{Exact
  Gell-Mann-Low Function of Supersymmetric Yang-Mills Theories from Instanton
  Calculus}}}, }\href {\doibase 10.1016/0550-3213(83)90338-3}
  {\bibfield{journal}{\bibinfo{journal}{Nucl. Phys.
  B}\,}\textbf{\bibinfo{volume}{229}}\,(\bibinfo{year}{1983})\,\bibinfo{pages}{381}}\BibitemShut
  {NoStop}%
\bibitem [{\citenamefont{Novikov} \emph {et\,al.}(1986)\citenamefont{Novikov},
  \citenamefont{Shifman}, \citenamefont{Vainshtein}, and
  \citenamefont{Zakharov}}]{Novikov:1985rd}%
  \BibitemOpen
  \bibfield{author}{\bibinfo{author}{\bibfnamefont{V.~A.}
  \bibnamefont{Novikov}}, \bibinfo{author}{\bibfnamefont{M.~A.}
  \bibnamefont{Shifman}}, \bibinfo{author}{\bibfnamefont{A.~I.}
  \bibnamefont{Vainshtein}},  and \bibinfo{author}{\bibfnamefont{V.~I.}
  \bibnamefont{Zakharov}}, }\bibfield{title}{\emph {\bibinfo{title}{{The beta
  function in supersymmetric gauge theories. Instantons versus traditional
  approach}}}, }\href {\doibase 10.1016/0370-2693(86)90810-5}
  {\bibfield{journal}{\bibinfo{journal}{Phys. Lett.
  B}\,}\textbf{\bibinfo{volume}{166}}\,(\bibinfo{year}{1986})\,\bibinfo{pages}{329}}\BibitemShut
  {NoStop}%
\bibitem [{\citenamefont{Monin} \emph {et\,al.}(2023)\citenamefont{Monin},
  \citenamefont{Shifman}, and \citenamefont{Vainshtein}}]{Monin:2023tjm}%
  \BibitemOpen
  \bibfield{author}{\bibinfo{author}{\bibfnamefont{A.}\,\bibnamefont{Monin}},
  \bibinfo{author}{\bibfnamefont{M.}\,\bibnamefont{Shifman}},  and
  \bibinfo{author}{\bibfnamefont{A.}\,\bibnamefont{Vainshtein}},
  }\bibfield{title}{\emph {\bibinfo{title}{{Spectral Flow in Instanton
  Computations and the $\beta$ functions}}}, }\href {\doibase
  10.1103/PhysRevD.108.105002} {\bibfield{journal}{\bibinfo{journal}{Phys. Rev.
  D}\,}\textbf{\bibinfo{volume}{108}}\,(\bibinfo{year}{2023})\,\bibinfo{pages}{105002}},
  \Eprint {http://arxiv.org/abs/2307.09119}{arXiv:2307.09119
  [hep-th]}\BibitemShut {NoStop}%
\bibitem [{\citenamefont{Grojean} \emph {et\,al.}(2013)\citenamefont{Grojean},
  \citenamefont{Jenkins}, \citenamefont{Manohar}, and
  \citenamefont{Trott}}]{Grojean:2013kd}%
  \BibitemOpen
  \bibfield{author}{\bibinfo{author}{\bibfnamefont{C.}\,\bibnamefont{Grojean}},
  \bibinfo{author}{\bibfnamefont{E.~E.} \bibnamefont{Jenkins}},
  \bibinfo{author}{\bibfnamefont{A.~V.} \bibnamefont{Manohar}},  and
  \bibinfo{author}{\bibfnamefont{M.}\,\bibnamefont{Trott}},
  }\bibfield{title}{\emph {\bibinfo{title}{{Renormalization Group Scaling of
  Higgs Operators and $\Gamma(h \to \gamma \gamma)$}}}, }\href {\doibase
  10.1007/JHEP04(2013)016}
  {\bibfield{journal}{\bibinfo{journal}{JHEP}\,}\textbf{\bibinfo{volume}{04}}\,(\bibinfo{year}{2013})\,\bibinfo{pages}{016}},
  \Eprint {http://arxiv.org/abs/1301.2588}{arXiv:1301.2588
  [hep-ph]}\BibitemShut {NoStop}%
\bibitem [{\citenamefont{Jenkins} \emph {et\,al.}(2013)\citenamefont{Jenkins},
  \citenamefont{Manohar}, and \citenamefont{Trott}}]{Jenkins:2013zja}%
  \BibitemOpen
  \bibfield{author}{\bibinfo{author}{\bibfnamefont{E.~E.}
  \bibnamefont{Jenkins}}, \bibinfo{author}{\bibfnamefont{A.~V.}
  \bibnamefont{Manohar}},  and
  \bibinfo{author}{\bibfnamefont{M.}\,\bibnamefont{Trott}},
  }\bibfield{title}{\emph {\bibinfo{title}{{Renormalization Group Evolution of
  the Standard Model Dimension Six Operators I: Formalism and lambda
  Dependence}}}, }\href {\doibase 10.1007/JHEP10(2013)087}
  {\bibfield{journal}{\bibinfo{journal}{JHEP}\,}\textbf{\bibinfo{volume}{10}}\,(\bibinfo{year}{2013})\,\bibinfo{pages}{087}},
  \Eprint {http://arxiv.org/abs/1308.2627}{arXiv:1308.2627
  [hep-ph]}\BibitemShut {NoStop}%
\bibitem [{\citenamefont{Jenkins} \emph {et\,al.}(2014)\citenamefont{Jenkins},
  \citenamefont{Manohar}, and \citenamefont{Trott}}]{Jenkins:2013wua}%
  \BibitemOpen
  \bibfield{author}{\bibinfo{author}{\bibfnamefont{E.~E.}
  \bibnamefont{Jenkins}}, \bibinfo{author}{\bibfnamefont{A.~V.}
  \bibnamefont{Manohar}},  and
  \bibinfo{author}{\bibfnamefont{M.}\,\bibnamefont{Trott}},
  }\bibfield{title}{\emph {\bibinfo{title}{{Renormalization Group Evolution of
  the Standard Model Dimension Six Operators II: Yukawa Dependence}}}, }\href
  {\doibase 10.1007/JHEP01(2014)035}
  {\bibfield{journal}{\bibinfo{journal}{JHEP}\,}\textbf{\bibinfo{volume}{01}}\,(\bibinfo{year}{2014})\,\bibinfo{pages}{035}},
  \Eprint {http://arxiv.org/abs/1310.4838}{arXiv:1310.4838
  [hep-ph]}\BibitemShut {NoStop}%
\bibitem [{\citenamefont{Alonso} \emph {et\,al.}(2014)\citenamefont{Alonso},
  \citenamefont{Jenkins}, \citenamefont{Manohar}, and
  \citenamefont{Trott}}]{Alonso:2013hga}%
  \BibitemOpen
  \bibfield{author}{\bibinfo{author}{\bibfnamefont{R.}\,\bibnamefont{Alonso}},
  \bibinfo{author}{\bibfnamefont{E.~E.} \bibnamefont{Jenkins}},
  \bibinfo{author}{\bibfnamefont{A.~V.} \bibnamefont{Manohar}},  and
  \bibinfo{author}{\bibfnamefont{M.}\,\bibnamefont{Trott}},
  }\bibfield{title}{\emph {\bibinfo{title}{{Renormalization Group Evolution of
  the Standard Model Dimension Six Operators III: Gauge Coupling Dependence and
  Phenomenology}}}, }\href {\doibase 10.1007/JHEP04(2014)159}
  {\bibfield{journal}{\bibinfo{journal}{JHEP}\,}\textbf{\bibinfo{volume}{04}}\,(\bibinfo{year}{2014})\,\bibinfo{pages}{159}},
  \Eprint {http://arxiv.org/abs/1312.2014}{arXiv:1312.2014
  [hep-ph]}\BibitemShut {NoStop}%
\bibitem [{\citenamefont{Gherardi} \emph
  {et\,al.}(2020)\citenamefont{Gherardi}, \citenamefont{Marzocca}, and
  \citenamefont{Venturini}}]{Gherardi:2020det}%
  \BibitemOpen
  \bibfield{author}{\bibinfo{author}{\bibfnamefont{V.}\,\bibnamefont{Gherardi}},
  \bibinfo{author}{\bibfnamefont{D.}\,\bibnamefont{Marzocca}},  and
  \bibinfo{author}{\bibfnamefont{E.}\,\bibnamefont{Venturini}},
  }\bibfield{title}{\emph {\bibinfo{title}{{Matching scalar leptoquarks to the
  SMEFT at one loop}}}, }\href {\doibase 10.1007/JHEP07(2020)225}
  {\bibfield{journal}{\bibinfo{journal}{JHEP}\,}\textbf{\bibinfo{volume}{07}}\,(\bibinfo{year}{2020})\,\bibinfo{pages}{225}},
  \bibinfo{note}{[Erratum: JHEP 01, 006 (2021)]}, \Eprint
  {http://arxiv.org/abs/2003.12525}{arXiv:2003.12525 [hep-ph]}\BibitemShut
  {NoStop}%
\bibitem [{\citenamefont{'t\,Hooft} and
  \citenamefont{Veltman}(1972)}]{tHooft:1972tcz}%
  \BibitemOpen
  \bibfield{author}{\bibinfo{author}{\bibfnamefont{G.}\,\bibnamefont{'t\,Hooft}}
  and \bibinfo{author}{\bibfnamefont{M.~J.~G.} \bibnamefont{Veltman}},
  }\bibfield{title}{\emph {\bibinfo{title}{{Regularization and Renormalization
  of Gauge Fields}}}, }\href {\doibase 10.1016/0550-3213(72)90279-9}
  {\bibfield{journal}{\bibinfo{journal}{Nucl. Phys.
  B}\,}\textbf{\bibinfo{volume}{44}}\,(\bibinfo{year}{1972})\,\bibinfo{pages}{189}}\BibitemShut
  {NoStop}%
\bibitem [{\citenamefont{Chanowitz} \emph
  {et\,al.}(1979)\citenamefont{Chanowitz}, \citenamefont{Furman}, and
  \citenamefont{Hinchliffe}}]{Chanowitz:1979zu}%
  \BibitemOpen
  \bibfield{author}{\bibinfo{author}{\bibfnamefont{M.~S.}
  \bibnamefont{Chanowitz}},
  \bibinfo{author}{\bibfnamefont{M.}\,\bibnamefont{Furman}},  and
  \bibinfo{author}{\bibfnamefont{I.}\,\bibnamefont{Hinchliffe}},
  }\bibfield{title}{\emph {\bibinfo{title}{{The Axial Current in Dimensional
  Regularization}}}, }\href {\doibase 10.1016/0550-3213(79)90333-X}
  {\bibfield{journal}{\bibinfo{journal}{Nucl. Phys.
  B}\,}\textbf{\bibinfo{volume}{159}}\,(\bibinfo{year}{1979})\,\bibinfo{pages}{225}}\BibitemShut
  {NoStop}%
\bibitem [{\citenamefont{Arzt}(1995)}]{Arzt:1993gz}%
  \BibitemOpen
  \bibfield{author}{\bibinfo{author}{\bibfnamefont{C.}\,\bibnamefont{Arzt}},
  }\bibfield{title}{\emph {\bibinfo{title}{{Reduced effective Lagrangians}}},
  }\href {\doibase 10.1016/0370-2693(94)01419-D}
  {\bibfield{journal}{\bibinfo{journal}{Phys. Lett.
  B}\,}\textbf{\bibinfo{volume}{342}}\,(\bibinfo{year}{1995})\,\bibinfo{pages}{189}},
  \Eprint
  {http://arxiv.org/abs/hep-ph/9304230}{arXiv:hep-ph/9304230}\BibitemShut
  {NoStop}%
\bibitem [{\citenamefont{Manohar}(2018)}]{Manohar:2018aog}%
  \BibitemOpen
  \bibfield{author}{\bibinfo{author}{\bibfnamefont{A.~V.}
  \bibnamefont{Manohar}}, }\bibfield{title}{\emph
  {\bibinfo{title}{{Introduction to Effective Field Theories}}}, }\href@noop {}
  {\bibfield{journal}{\bibinfo{journal}{Les Houches Lect.Notes
  108}}\,(\bibinfo{year}{2018})}, \Eprint
  {http://arxiv.org/abs/1804.05863}{arXiv:1804.05863 [hep-ph]}\BibitemShut
  {NoStop}%
\end{thebibliography}%

\end{document}